\documentclass[journal, onecolumn]{IEEEtran}
\usepackage{setspace}
\usepackage[english]{babel}

\usepackage[letterpaper,top=2cm,bottom=2cm,left=3cm,right=3cm,marginparwidth=1.75cm]{geometry}

\usepackage[pdftex]{graphicx}
\graphicspath{{figures/}}
\DeclareGraphicsExtensions{.pdf,.jpeg,.png}
\usepackage{cite}
\usepackage{amsmath}
\usepackage{array}

\usepackage{caption, subcaption}
\usepackage{algorithm, algpseudocode}
\usepackage{bbm}
\usepackage[normalem]{ulem}
\usepackage[colorlinks=true, allcolors=blue]{hyperref}

\newcommand{\add}[1]{#1} 
\newcommand{\remove}[1]{\textcolor{blue}{\ifmmode\text{\sout{\ensuremath{#1}}}\else\sout{#1}\fi}}

\newcommand{\conjt}[1]{{#1}^\dag}
\newcommand{\conj}[1]{{#1}^{*}}
\newcommand{\abs}[1]{\vert #1 \vert}
\newcommand{\Abs}[1]{\left\vert #1 \right\vert}
\newcommand{\norm}[1]{\lVert #1 \rVert}
\newcommand{\Norm}[1]{\left\lVert #1 \right\rVert}
\newcommand{\evalgrad}[1]{\biggr\rvert_{#1}}
\newcommand{\diag}[1]{\mathrm{Diag}(#1)}
\newcommand{\Diag}[1]{\mathrm{Diag}\left(#1\right)}
\newcommand{\dset}[1]{\{1, ..., #1\}}

\newcommand{\R}{\mathbbm{R}}
\newcommand{\C}{\mathbbm{C}}
\newcommand{\E}{\mathbbm{E}}
\newcommand{\F}{\mathrm{F}}
\newcommand{\VM}{\mathcal{VM}}
\newcommand{\CN}{\mathcal{CN}}
\newcommand{\cov}{\mathrm{Cov}}

\newcommand{\iid}{\overset{\mathrm{i.i.d.}}{\sim}}
\newcommand{\Rx}{\mathrm{Rx}}
\newcommand{\Tx}{\mathrm{Tx}}
\newcommand{\Hhat}{\Hat{H}}
\newcommand{\G}{\Hat{G}}
\newcommand{\AoD}{\Phi^{\mathrm{AoD}}}
\newcommand{\AoA}{\Phi^{\mathrm{AoA}}}
\newcommand{\AoDc}{\Phi^{c, \mathrm{AoD}}}
\newcommand{\AoAc}{\Phi^{c, \mathrm{AoA}}}
\newcommand{\angleDomain}{[0, \pi) \times [0, 2\pi)}
\newcommand{\kPrior}{\kappa_0}
\newcommand{\vmDomain}{[-\pi, \pi)}
\newcommand{\allOnes}{\mathbbm{1}}
\newcommand{\true}{\mathrm{true}}

\newcommand{\hatPowProfile}{\Hat{\mathrm{P}}}
\newcommand{\powProfile}{\mathrm{P}}
\newcommand{\pow}{\mathrm{P}_{\mathrm{sig}}}
\newcommand{\hatpow}{\Hat{\mathrm{P}}_{\mathrm{sig}}}
\newcommand{\snr}{\mathrm{SNR}}
\newcommand{\new}{\mathrm{new}}
\newcommand{\lossPEOC}{\mathcal{L}^{\mathrm{PEOC}}}
\newcommand{\lossUPEC}{\mathcal{L}^{\mathrm{UPEC}}}
\newcommand{\lossPEAC}{\mathcal{L}^{\mathrm{PEAC}}}
\newcommand{\rmcal}{\mathrm{cal}}

\DeclareMathOperator*{\argmin}{arg\,min}

\DeclareMathOperator{\Tr}{Tr}

\title{Calibrating Wireless Ray Tracing for Digital Twinning using Local Phase Error Estimates}
\author{
\IEEEauthorblockN{
    Clement~Ruah,~\IEEEmembership{Student Member,~IEEE,}
    Osvaldo~Simeone,~\IEEEmembership{Fellow,~IEEE,}
    Jakob~Hoydis,~\IEEEmembership{Fellow,~IEEE,} and
    Bashir~Al-Hashimi,~\IEEEmembership{Fellow,~IEEE}
}

\thanks{
Clement Ruah and Osvaldo Simeone are with the King’s Communications, Learning \& Information Processing (KCLIP) lab within the Centre for Intelligent Information Processing Systems (CIIPS), Department of Engineering, King’s College London, WC2R 2LS London, U.K. (e-mail: clement.ruah@kcl.ac.uk; osvaldo.simeone@kcl.ac.uk).
Jakob Hoydis is with NVIDIA Corporation.
Bashir M. Al-Hashimi is with the Department of Engineering, King’s College London, WC2R 2LS London, U.K.
The work of C. Ruah was supported by the Faculty of Natural, Mathematical, and Engineering Sciences, King’s College London.
The work of O. Simeone and of J. Hoydis was partially supported by the European Union’s Horizon Europe project CENTRIC (101096379).
O. Simeone was also supported by the Open Fellowships of the EPSRC (EP/W024101/1) by the EPSRC project (EP/X011852/1), and by Project REASON, a UK Government funded project under the Future Open Networks Research Challenge (FONRC) sponsored by the Department of Science Innovation and Technology (DSIT).
}
}

\begin{document}

\maketitle

\begin{abstract}
Embodying the principle of simulation intelligence, digital twin (DT) systems construct and maintain a high-fidelity virtual model of a physical system.
This paper focuses on ray tracing (RT), which is widely seen as an enabling technology for DTs of the radio access network (RAN) segment of next-generation disaggregated wireless systems.
RT makes it possible to simulate channel conditions, enabling data augmentation and prediction-based transmission. However, the effectiveness of RT hinges on the adaptation of the electromagnetic properties assumed by the RT to actual channel conditions, a process known as calibration.
The main challenge of RT calibration is the fact that  small discrepancies in the geometric model fed to the RT software hinder the accuracy of the predicted phases of the simulated propagation paths.
Existing solutions to this problem either rely on the channel power profile, hence disregarding phase information, or they operate on the channel responses by assuming the simulated phases to be sufficiently accurate for calibration.
This paper proposes a novel channel response-based scheme that, unlike the state of the art, estimates and compensates for the phase errors in the RT-generated channel responses.
The proposed approach builds on the variational expectation maximization algorithm with a flexible choice of the prior phase-error distribution that bridges between a deterministic model with no phase errors and a stochastic model with uniform phase errors.
The algorithm is computationally efficient, and is demonstrated, by leveraging the open-source differentiable RT software available within the Sionna library, to outperform existing methods in terms of the accuracy of RT predictions.
\end{abstract}

\section{Introduction} \label{sec:introduction}

\subsection{Simulation Intelligence for Wireless Networks}

A general trend in the sciences and engineering is the use of the unprecedented processing power afforded by modern computing clusters to implement \emph{physics-driven simulators} of the real world, with the aim of facilitating scientific discovery, providing design insights, and reducing the need for data collection.
An important embodiment of this trend in engineering, referred to as \emph{simulation intelligence} \cite{lavin2021simulation}, is the adoption of \emph{digital twin} (DT) systems.
A DT software constructs and maintains a high-fidelity virtual model of a physical system through two-way communications between the real and the virtual worlds \cite{kritzinger2018digital, errandonea2020digital, liu2021review, thelen2022comprehensive, thelen2023comprehensive}.
DTs provide a \emph{virtual sandbox} in which operational policies can be designed and tested before being deployed to the physical twin (PT). 

DTs offer a natural framework for the deployment of software-based disaggregated wireless networks that follow the open radio access network (RAN) paradigm \cite{polese2023understanding, khan2022digital, jagannath2022digital, hui2022digital, salehi2023multiverse}.
Among the key advantages of DTs in this context is their capacity to synthesize data that can be leveraged to train, as well as to feed at run time, artificial intelligence (AI) models.
For example, in the O-RAN architecture, real-time RAN intelligent controllers run AI models that need to be tailored to the operating conditions of the underlying physical network \cite{polese2023understanding}.
Data are generated at the DT by simulating relevant conditions in the RAN.
For instance, a DT may be used to estimate key performance indicators (KPIs) on the RAN, such as throughput, on the basis of information about the topology of the network and traffic conditions \cite{villa2023colosseum}.
The benefits of DT-generated data hinge on the adherence of the virtual model to the physical system, and hence a key problem is the \emph{calibration} of the DT using real-world data from the physical twin \cite{hoydis2023sionna}.

\subsection{Ray Tracing at the Digital Twin}

\begin{figure*}
    \centering
    \includegraphics[width=\textwidth]{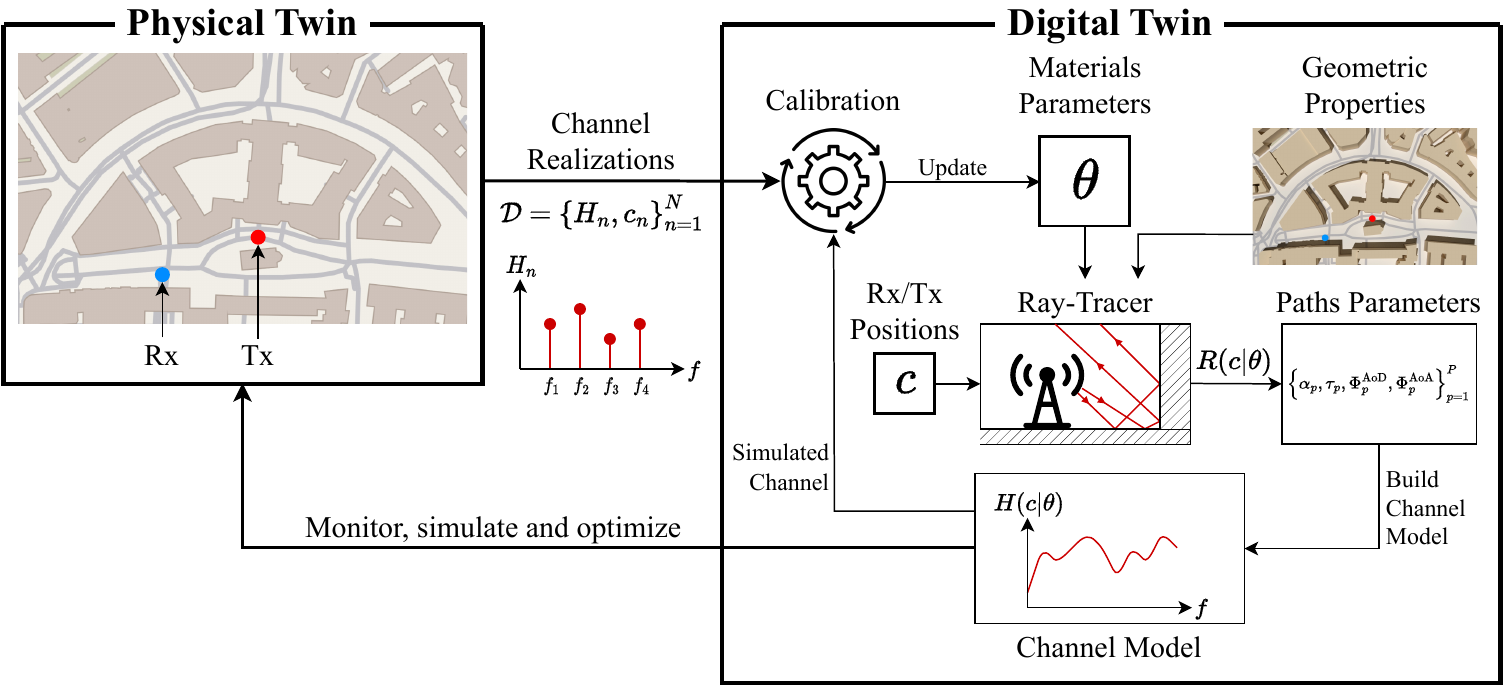}
    \caption{
    Taking as input the geometric properties of the scene, the electromagnetic material parameters $\theta$, and the coordinates $c$ of transmitter (Tx) and receiver (Rx), the ray tracer (RT) produces the features $R(c|\theta)$ of a number $P$ of propagation paths. Based on this information, the DT
    can obtain a model $H(c | \theta)$ of the channel conditions between transmitter and receiver.  
    To keep a faithful representation of its physical twin (PT), during a calibration phase, the DT compares its model predictions $H(c | \theta)$ to measured channel realizations $\mathcal{D} = \{H_n, c_n\}_{n=1}^{N}$ in order to optimize the material parameters $\theta$.
    An accurate estimate of the ground-truth channel conditions can then be used to monitor the PT, simulate alternate scenarios (counterfactual analysis), or to optimize PT operations such as beamforming, with reduced requirements on pilot transmissions and channel measurements \cite{jiang2023digital}.
    (The map and the 3D model in the the figure are built using geospatial data from the OpenStreetMap database \cite{OpenStreetMap}.) 
    }
    \label{fig:DT_RT_workflow}
\end{figure*}

\emph{Ray tracing} (RT) is widely seen as an enabling technology for DTs of the RAN \cite{yun2015ray, pengnoo2020digital, zeulin2022ml, villa2023colosseum, jiang2023digital, alkhateeb2023real, salehi2023multiverse, nguyen2023probabilistic}.
As illustrated in Fig. \ref{fig:DT_RT_workflow}, given a three-dimensional model specifying the shapes, positions, and electromagnetic properties of the objects in the propagation environment, the RT simulates multiple propagation paths between any given transmitter (Tx) and receiver (Rx) through specular and diffuse reflection, refraction, and diffraction \cite[Ch.~4]{molisch2012wireless}, \cite{itu2019propagation, degli2007measurement, degli2011analysis}.
By leveraging the predicted path components, a DT can simulate the propagation conditions for a given network deployment.
The synthesized channels can then be used to train AI models that carry out tasks such as beamforming \cite{zeulin2022ml, jiang2023digital, salehi2023multiverse, vuckovic2023paramount}, user positioning \cite{nguyen2023probabilistic}, and channel charting \cite{viswanathan2020communications, ferrand2023wireless} by simulating relevant traffic conditions and user distributions.
Given specific user locations, they may be also used to directly optimize transmission strategies with minimal requirements in terms of pilot overhead \cite{jiang2023digital, ruah2023bayesian}.

\add{Compared to purely data-driven designs \cite{zhang2020cellular, ferrand2023wireless}, RT holds the promise of more accurate and explainable predictions of channel conditions, which hinge on prior physics-based knowledge of electromagnetic dynamics.  
Nonetheless,} the fidelity of the virtual simulation depends on the precision of the geometric and electromagnetic properties, including permittivity, conductivity and permeability \cite{molisch2012wireless, itu2021effects}, fed to the DT as inputs.  
While specialized reports \cite{itu2021effects} can provide approximate values for the electromagnetic properties of generic materials, a more precise and bespoke selection is generally necessary to ensure sufficiently reliable simulations.
This optimization requires the estimation of the material parameters of each object in the scene based on measurements obtained from the ground-truth deployment scenario, in a process known as \emph{calibration} \cite{jemai2009calibration, priebe2012calibrated, he2018design, charbonnier2020calibration, kanhere2023calibration, bhatia2023tuning, hoydis2023sionna}.
\add{Through data-driven calibration, inflexible modeling choices regarding prior assumptions of the scene materials are relaxed, enabling the RT simulation to adapt to a wider range of scenarios.}

\subsection{Calibration of Ray Tracing}

The geometric properties of the deployment scenario of interest can be generally determined up to a certain degree of precision based on in-situ measurements, e.g., using LiDAR \cite{xue2020lidar, thelen2022comprehensive, lin20236g} or neural surface reconstruction \cite{li2023neuralangelo}, or based-on geo-spatial data, e.g., via the Open Street Maps database \cite{OpenStreetMap}.
However, in practical scenarios, it is not uncommon to observe small discrepancies, of the scale of a fraction of the carrier wavelength, between the ground-truth geometric properties and their virtual counterpart.
These discrepancies hinder the accuracy of the predicted phases of the simulated propagation paths, which are usually deemed to be unreliable \cite[Ch.~7.5]{molisch2012wireless}.
\add{Errors in the predicted phases can provide inaccurate interference patterns at the receiver when the temporal and/or spatial resolution of the system is not high enough to resolve each path individually, i.e., when the bandwidth and/or the antenna arrays are too small.}

We distinguish two classes of existing calibration procedures, which make distinct assumptions on the \emph{phase error} caused by imprecise geometric information.
\begin{itemize}
    \item \emph{Power profile-based} schemes aim to find the material parameters that yield the best match between the power-delay -- or more generally power-angle-delay -- profiles simulated by the RT and the corresponding power profiles measured from the real environment \cite{jemai2009calibration, priebe2012calibrated, he2018design, charbonnier2020calibration, kanhere2023calibration, bhatia2023tuning}.
    Accordingly, power profile-based schemes disregard phase information, retaining solely the amplitudes of the measured and simulated channels.
    \item \emph{Channel response-based} schemes find the material parameters that yield the best match between simulated and recorded channel \add{frequency} responses \cite{hoydis2023sionna}.
    The state of the art work \cite{hoydis2023sionna} compares RT-generated and measured channel responses by relying on the phase provided by the RT to be sufficiently accurate to enable calibration.
    We refer to this class of channel response-based methods as \emph{phase error-oblivious}.
    Such an approach adopts an opposite modelling assumption as compared to power profile-based methods, disregarding the presence of phase errors in the RT simulation.
\end{itemize}

\begin{figure*}
    \centering
    \begin{subfigure}[T]{0.46\textwidth}
        \centering
        \includegraphics[keepaspectratio=true, height=3cm]{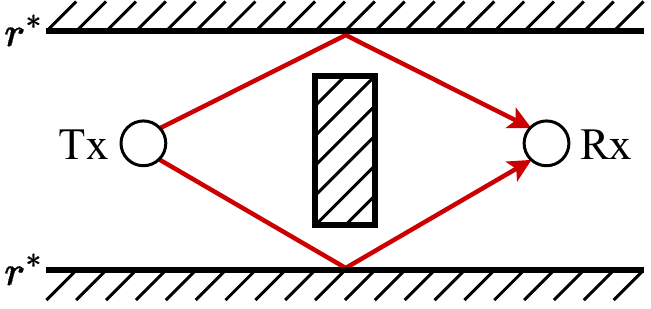}
        \vspace{0.50cm}
        \caption{Ground-truth scenario}
        \label{fig:scenario_ground_truth}
    \end{subfigure}
    \hfill
    \begin{subfigure}[T]{0.46\textwidth}
        \centering
        \includegraphics[keepaspectratio=true, height=3.75cm]{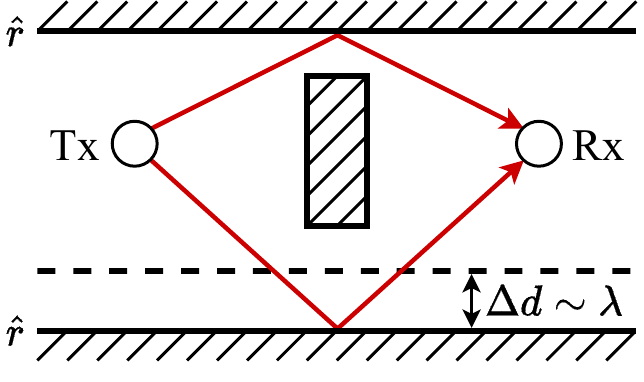}
        \vspace{-0.25cm}
        \caption{Simulated scenario}
        \label{fig:scenario_simulation}
    \end{subfigure}
    
    \vspace{0.25cm} \hrule \vspace{0.25cm}
    
    \begin{subfigure}{0.46\textwidth}
        \centering
        \includegraphics[keepaspectratio=true, height=3cm]{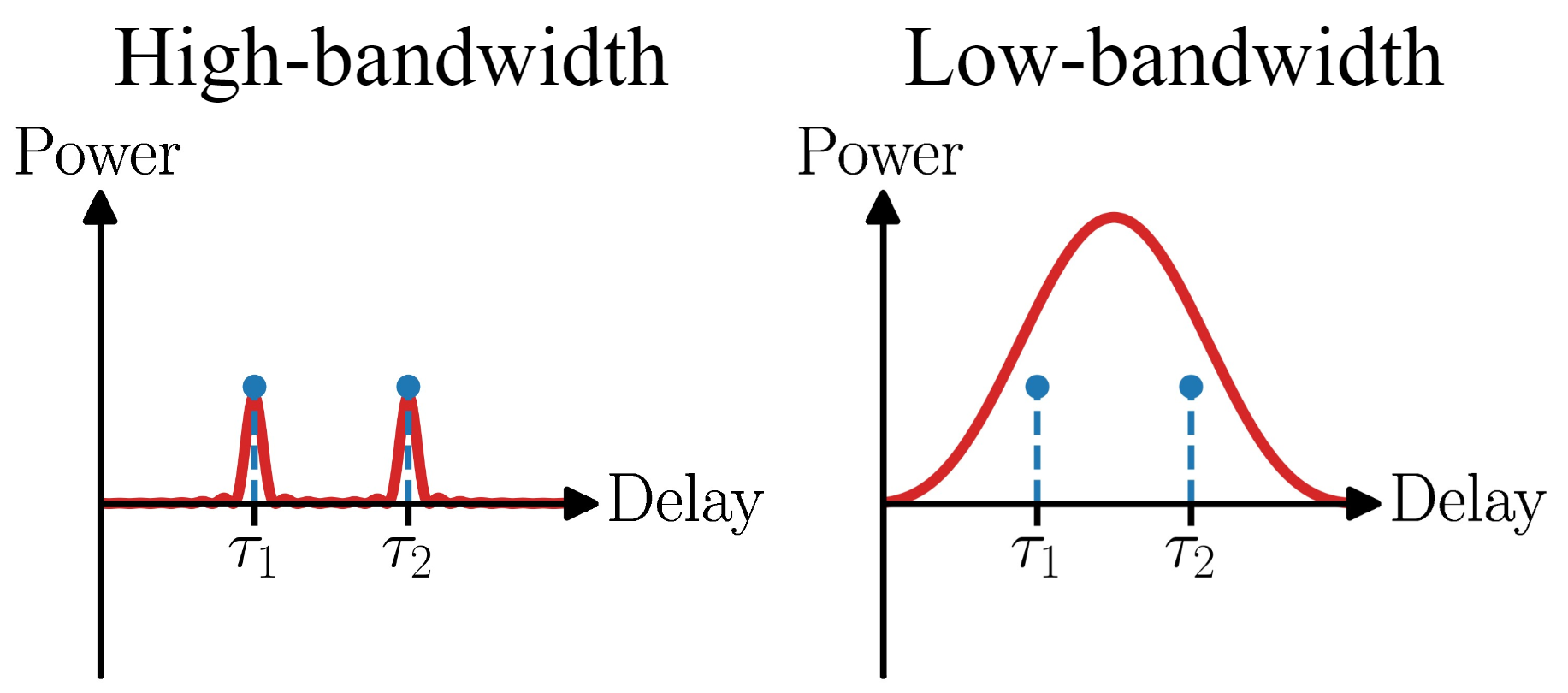}
        \vspace{-0.1cm}
        \caption{Ground-truth power-delay profiles}
        \label{fig:power_profile_ground_truth}
    \end{subfigure}
    \hfill
    \begin{subfigure}{0.46\textwidth}
        \centering
        \includegraphics[keepaspectratio=true, height=3cm]{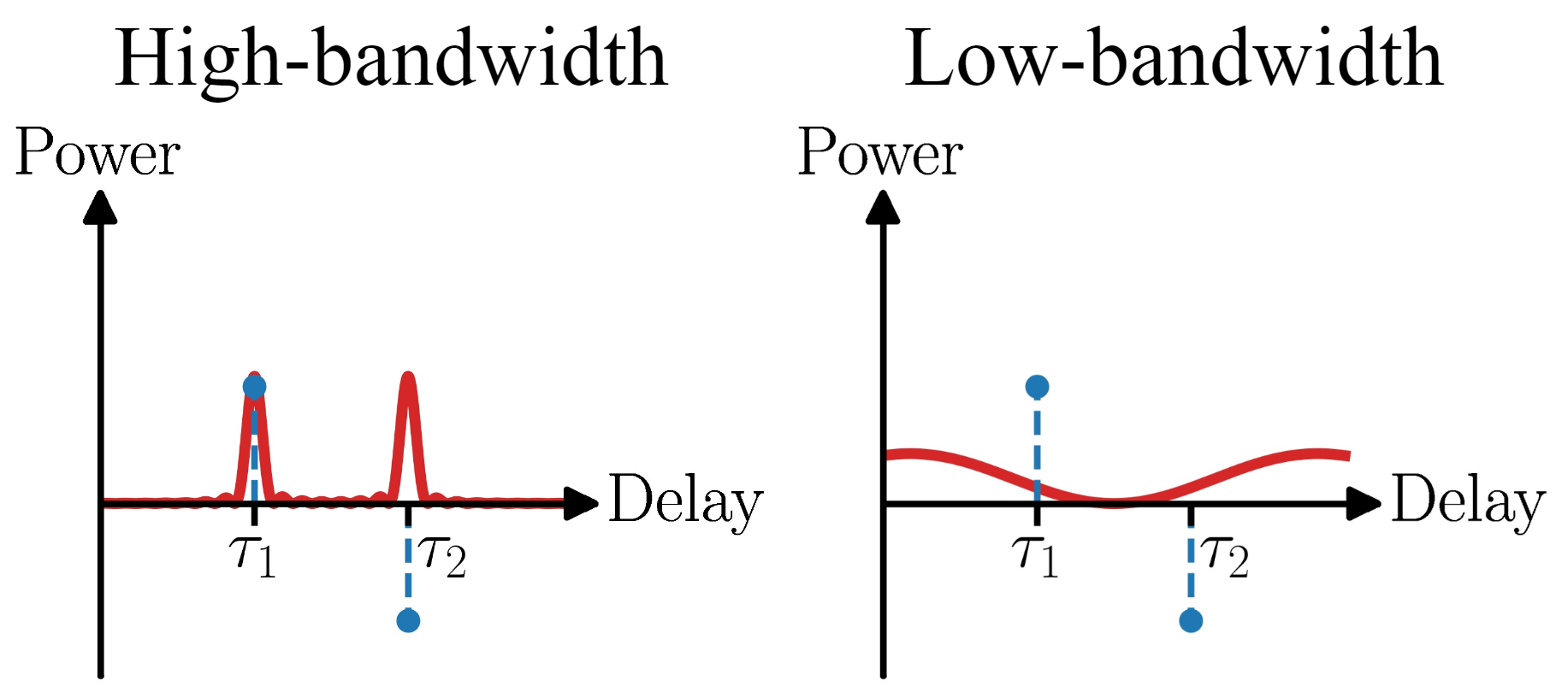}
        \vspace{-0.1cm}
        \caption{Simulated power-delay profiles}
        \label{fig:power_profile_simulation}
    \end{subfigure}
    
    \vspace{0.25cm} \hrule \vspace{0.25cm}
    
    \begin{subfigure}{0.42\textwidth}
        \centering
        \includegraphics[keepaspectratio=true, height=3cm]{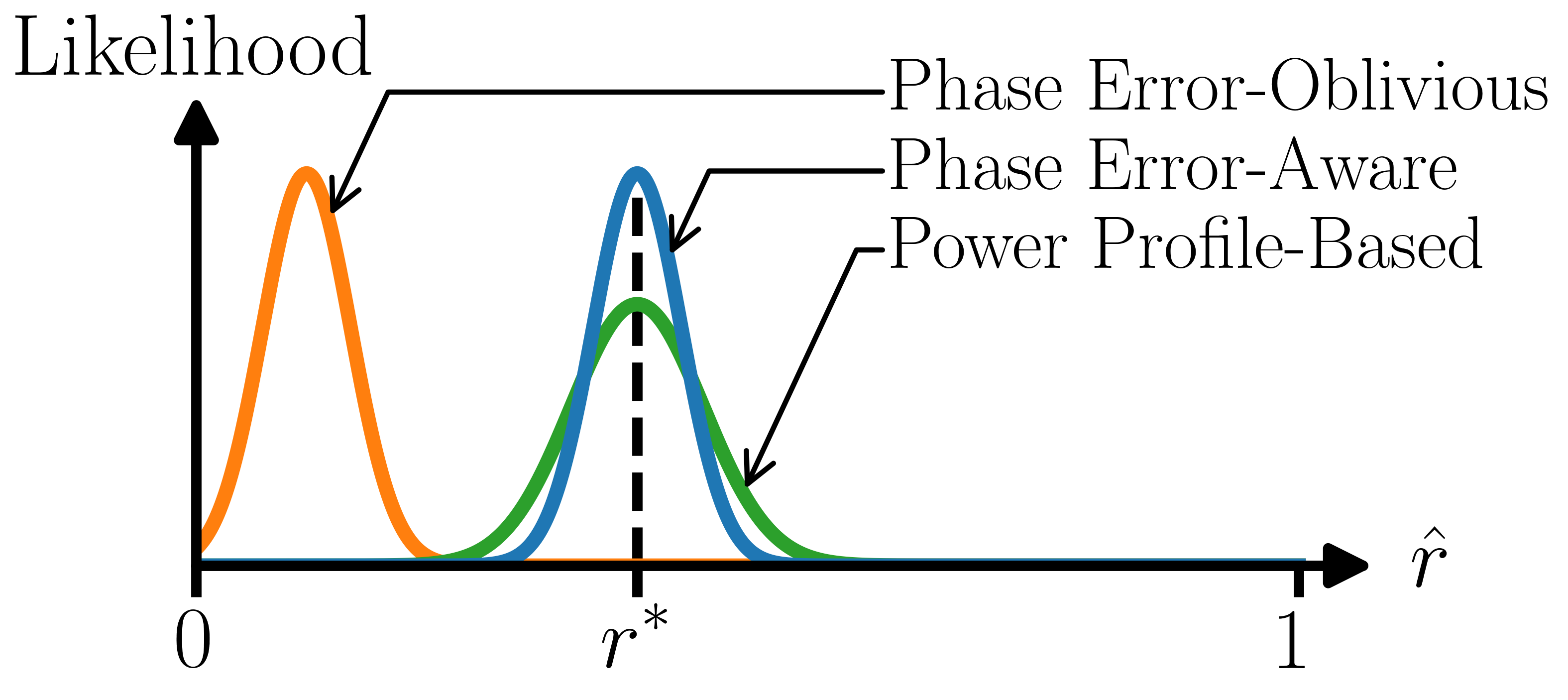}
        \caption{Calibration under high-bandwidth conditions}
        \label{fig:reflectance_likelihood_high_bandwidth}
    \end{subfigure}
    \hfill
    \begin{subfigure}{0.52\textwidth}
        \centering
        \includegraphics[keepaspectratio=true, height=3cm]{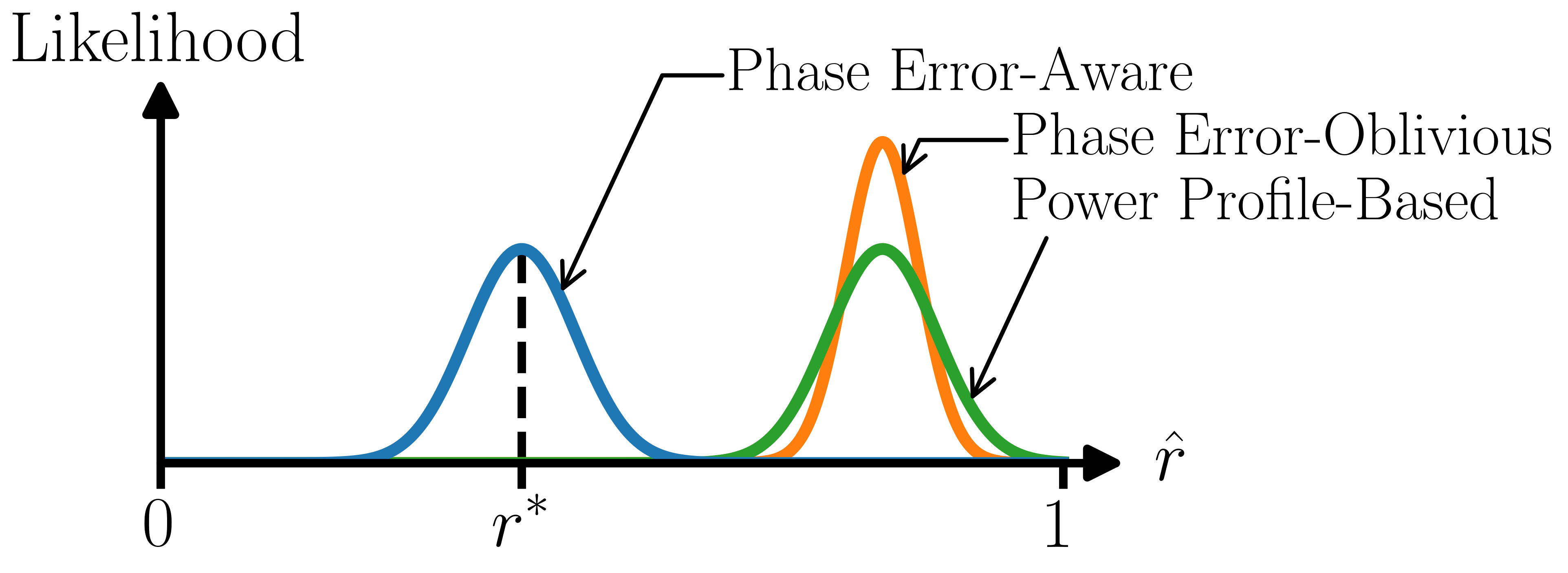}
        \caption{Calibration under low-bandwidth conditions}
        \label{fig:reflectance_likelihood_low_bandwidth}
    \end{subfigure}
    \caption{
    Toy example that illustrates the limitations of existing power profile-based RT calibration methods, as well as of channel response-based schemes that disregard path phase errors. 
    In this example, the two propagation paths between the transmitter (Tx) and the receiver (Rx) reflect on  surfaces with ground-truth reflectance $r^{*}$ in (a), and calibrated reflectance $\hat{r}$ in the simulated scenario (b). 
    Though the paths interfere constructively at the receiver under ground-truth conditions, they are predicted to interfere destructively in the simulated scenario due to an inaccuracy $\Delta d$ of the order of the carrier's wavelength $\lambda$ in the geometric model.
    \add{This difference is illustrated as the blue dashed lines in (c) and (d), which represent the signed amplitudes of each path.}
    Due to this model error, phase error-oblivious schemes provide inaccurate calibration results in both high and low-bandwidth regimes (e)-(f). Furthermore, the model error causes the RT-generated  power-delay profiles in the low-bandwidth regime (d) to diverge from the ground-truth power-delay profile (c), causing inaccurate calibration also for power profile-based schemes.
    }
    \label{fig:calibration_among_geometric_errors}
\end{figure*}

This paper proposes a novel channel response-based scheme that, unlike the state of the art, compensates for the phase errors in the RT-generated channel responses prior to carrying out a comparison with measured channel responses during RT calibration.
The main idea is to use measured channel responses to estimate the phase error locally for each path and for each measurement.
To illustrate the potential benefits of the approach, we provide the following toy example, which is depicted in Fig. \ref{fig:calibration_among_geometric_errors}.

In this example, as shown in Fig. \ref{fig:scenario_ground_truth} and Fig. \ref{fig:scenario_simulation}, the goal of calibration is to produce an estimate $\hat{r}$ of the ground-truth reflectance value \add{$r^{*} \in [0, 1]$}, which represents the fraction of an incoming ray's energy that is not absorbed by the given surfaces upon reflection.
\add{It is assumed here that the upper and lower paths, with respective times of arrivals $\tau^{*}_1$ and $\tau^{*}_2$, reach the receiver with the same phase in the ground-truth scenario in Fig.~\ref{fig:scenario_ground_truth}.}
\add{As illustrated in Fig.~\ref{fig:scenario_simulation}, the position of the lower wall in the simulated scenario differs from the ground-truth} by a small amount of the order of the carrier frequency's wavelength $\lambda$.
\add{Owing to this small geometric discrepancy, the slightly longer predicted time of arrival $\hat{\tau}_2 > \tau^{*}_2$ of the lower path causes the two paths to arrive at the receiver with opposite phases in the simulation.}

In an ideal situation in which the bandwidth, and hence the temporal resolution, is large enough to separate the contributions of each individual path, the power-delay profiles of the received signal (Fig. \ref{fig:power_profile_ground_truth}) and of the simulated paths (Fig. \ref{fig:power_profile_simulation}) match perfectly by choosing the correct reflectance value $\hat{r}=r^*$.
Accordingly, as \add{qualitatively illustrated} in Fig. \ref{fig:reflectance_likelihood_high_bandwidth}, the likelihood function maximized by a power profile-based scheme peaks at the correct value $\hat{r}=r^*$.

In a potentially more relevant low-bandwidth regime, the temporal resolution is too low to separate the two propagation paths, and the measured power profile in Fig.~\ref{fig:power_profile_ground_truth} cannot be reproduced by the power profile generated by the RT depicted in Fig. \ref{fig:power_profile_simulation}.
In fact, as mentioned, the power profile \add{simulated via} RT is characterized by a destructive interference between the two paths irrespective of the reflectance value $\hat{r}$, causing a mismatch with the constructive interference exhibited by the measured power profile.
This results in an incorrectly estimated reflectance value $\hat{r}$ (Fig. \ref{fig:reflectance_likelihood_low_bandwidth}), \add{where the power profile-based scheme over-estimates the reflectance $\hat{r}$ of the walls in order to make up for the incorrectly predicted interference pattern.}

In both high and low-bandwidth regimes, a phase error-oblivious channel response-based scheme would be unable to match measured and received responses due to its reliance on the opposite phases for the two paths predicted by the RT, \add{as illustrated by the blue dashed lines in Fig.~\ref{fig:power_profile_ground_truth} and Fig.~\ref{fig:power_profile_simulation}.}
\add{Accordingly}, this approach would return an incorrect value for the estimated reflectance $\hat{r}$, as represented by the incorrect peak value for the likelihood in Fig. \ref{fig:reflectance_likelihood_high_bandwidth} and Fig. \ref{fig:reflectance_likelihood_low_bandwidth}.
In contrast, as we will demonstrate in this work, a channel response-based strategy that attempts to estimate and mitigate simulated phase errors can provide well-calibrated electromagnetic properties irrespective of the resolution of the physical system.

\subsection{Main Contributions}

In this work, we present a novel channel response-based calibration scheme for RT that estimates and mitigates per-path phase errors.
Our specific contributions are as follows. 
\begin{itemize}
    \item We introduce a novel \emph{phase error-aware} calibration strategy for RT.
    The proposed approach builds on the \emph{variational expectation maximization} (VEM) algorithm \cite{simeone2022machine} with a tailored choice of the variational distribution on the phase errors.
    The algorithm is computationally efficient encompassing an expectation step in closed form and a maximization step via gradient descent.
    \item Leveraging the open-source differentiable RT made available through the Sionna library \cite{hoydis2022sionna, hoydis2023sionna}, we present a variety of experiments, evaluating calibration accuracy in terms of the estimated electromagnetic parameters, as well as of predicted power maps. 
    The experiments confirm the significant advantages of the proposed phase error-aware calibration strategy as compared to existing power profile-based schemes and phase error-oblivious approaches.
\end{itemize}

The rest of the paper is organized as follows. 
Sec. \ref{sec:setting} introduces the considered communication system and the phase-error RT channel model.
In Sec. \ref{sec:baselines}, we describe the implemented calibration baselines for the phase error-oblivious (Sec. \ref{subsec:peoc}) and the power profile-based (Sec. \ref{subsec:upec}) schemes.
Sec. \ref{sec:peac} covers our proposed phase error-aware calibration scheme, which is compared to the previously introduced baselines in Sec. \ref{sec:experiments}, where we study the accuracy of the calibrated material parameters and predicted signal powers for a synthetic scenario.
Finally, Sec. \ref{sec:conclusion} concludes the paper.

\section{Setting and Problem Definition} \label{sec:setting}

In this work, we study the problem of calibrating a DT-based RT using channel frequency response observations.
This section first describes the studied communication system and the RT output before introducing a stochastic phase-error channel model that accounts for the errors of the RT-predicted phases.
We end this section by specifying the available data and the calibration objective.

\subsection{Setting}

Consider a transmitter (Tx) located at coordinates $c^\Tx \in \R^3$ and a receiver (Rx) located at coordinates $c^\Rx \in \R^3$ in a three-dimensional Cartesian space.
The transmitter and receiver are equipped with antenna arrays of $N^\Tx$ and $N^\Rx$ antennas respectively.
Using the paths computed by the RT, the DT models the channel over a bandwidth \add{$B=\abs{f^{\mathrm{max}} - f^{\mathrm{min}}}$} with end frequencies $f^{\mathrm{max}} > f^{\mathrm{min}} > 0$ and central carrier frequency $f^c = (f^{\mathrm{min}} + f^{\mathrm{max}}) / 2$ [Hz].
Assuming multi-carrier transmission, the set of $S$ subcarrier frequencies is $\{ f_s \}_{s = 1}^{S}$, where $f_s = f^{\mathrm{min}} + (s-1) \Delta f$ is the frequency of the $s\mbox{-th}$ subcarrier; $\Delta f$ [Hz] is the subcarrier spacing; and $S = \left\lfloor (f^{\mathrm{max}} - f^{\mathrm{min}})/\Delta f \right\rfloor$ is the total number of subcarriers.

As illustrated in Fig. \ref{fig:DT_RT_workflow}, the RT takes as input the geometric properties, specifying the shapes and positions of the objects in the scene, the coordinates $c = (c^\Rx, c^\Tx)$ of the transmitter and receiver devices, and the materials properties, denoted as vector $\theta$, which includes the permittivity, conductivity, permeability, scattering coefficient and cross-polarization discrimination of the materials involved \cite{itu2021effects, degli2007measurement, degli2011analysis}.
Additional RT inputs not depicted in Fig. \ref{fig:DT_RT_workflow} are the antenna patterns of the devices \cite[Ch.~2.2]{balanis2016antenna}, as well as the choice of a parameterized scattering model \cite{degli2007measurement}.
Based on this information, the RT outputs the parameters of a number $P^c$ of realizable propagation paths through specular and diffuse reflection, refraction, and diffraction \cite[Ch.~4]{molisch2012wireless}, \cite{itu2019propagation, degli2007measurement, degli2011analysis}.
Note that the RT is only able to consider a finite number of samples among the potentially infinite set diffusely scattered paths, whereas the finite set of specular components is treated exhaustively up to a pre-determined number of reflections or refractions.

\begin{figure}
    \centering
    \includegraphics[keepaspectratio, width=3.4in]{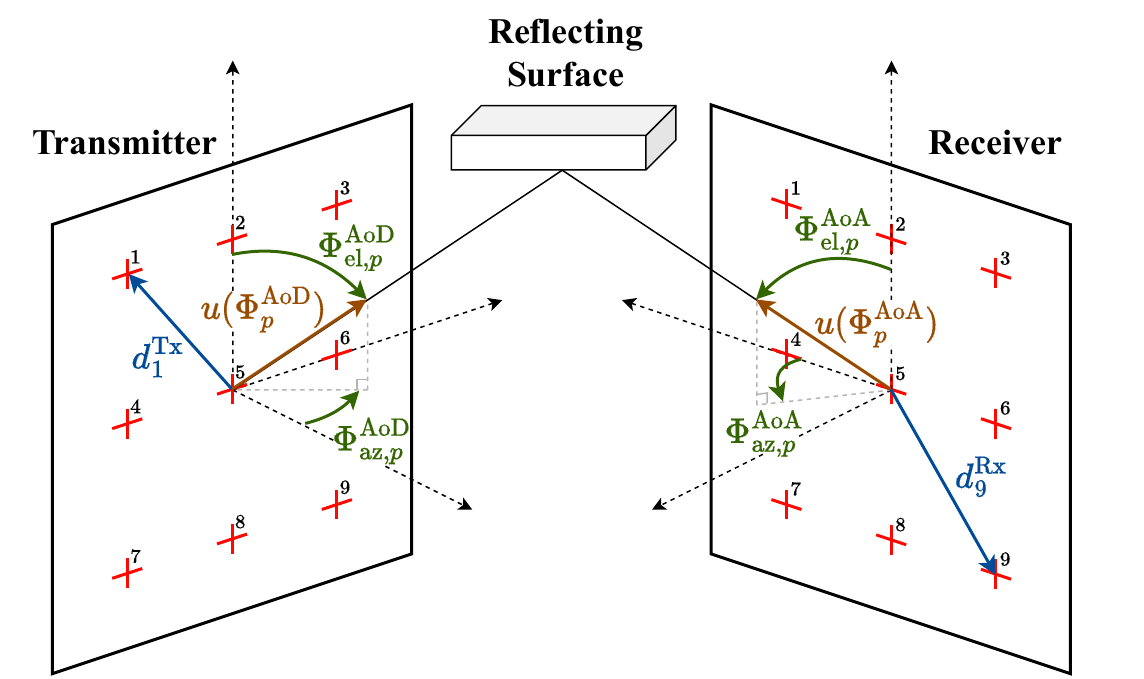}
    \caption{Illustration of the definitions of angle path parameters for planar antenna  arrays. The red crosses represent the elements of each array. The angles and directions of the departing and arriving signals, as well as the positions of the array elements, are represented in the respective local coordinates.}
    \label{fig:notation}
\end{figure}

In this work, we will only focus on the simulation of specular propagation paths, and the parameter vector $\theta$ will encompass the permittivity, conductivity and permeability of the materials in the scene.
We leave the study of diffusely scattered components, along with the calibration of scattering coefficients and cross-polarization discrimination, for future work.
Accordingly, the RT output is assumed to be deterministic and given by the function
\begin{equation}
\label{eq:rt_parameters}
    R(c | \theta) = \left\{ \alpha^c_p(\theta), \tau_p^c, \AoDc_p, \AoAc_p \right\}_{p=1}^{P^c},
\end{equation}
which maps the input coordinates $c$ and material parameters $\theta$ into the parameters of $P^c$ paths.
By \eqref{eq:rt_parameters}, each propagation path $p \in \dset{P^c}$ is described by a complex amplitude $\alpha^c_p(\theta) \in \C$, a delay $\tau_p^c > 0$ [s], a pair of angles of departure $\AoDc_p \in \angleDomain$, and a pair of angles of arrival $\AoAc_p \in \angleDomain$.
The angles of departure $\AoDc_p = (\AoDc_{\mathrm{el}, p}, \AoDc_{\mathrm{az}, p}) \in \angleDomain$ include the elevation $\AoDc_{\mathrm{el}, p}$ and azimuth $\AoDc_{\mathrm{az}, p}$ angles from the transmitter perspective, and the angles of arrival $\AoAc_p = (\AoAc_{\mathrm{el}, p}, \AoAc_{\mathrm{az}, p}) \in \angleDomain$ encompass the elevation-azimuth angles at the receiver, as illustrated in Fig. \ref{fig:notation}.
In order to keep the notation concise, in what follows, we omit the position subscript $c$ from the path parameters $R(c | \theta)$ in \eqref{eq:rt_parameters}, and from the number of paths $P^c$.

For a given receiver and transmitter position $c = (c^\Rx, c^\Tx)$, signal propagation between each receive and transmit antenna pair is assumed to follow the same path parameters $R(c | \theta)$, while differing only in terms of phase.
Accordingly, for each simulated path $p \in \dset{P}$, phase offsets between neighbouring antenna elements are modeled as a $N^\Rx \times 1$ complex-valued receive steering vector $a^\Rx(\AoA_p)$ and a $N^\Tx \times 1$ transmit steering vector $a^\Tx(\AoD_p)$, where the functions $a^\Rx(\cdot)$ and $a^\Tx(\cdot)$ depend on the shape and orientation of the receive and transmit array, respectively, and on the transmission distance \cite[Ch.~6]{balanis2016antenna}.

For instance, in the far-field regime, the steering vectors $a^\Rx(\AoA_p)$ can be approximated using a planar wavefront model as
\begin{equation}
\label{eq:steering_vector_rx}
    a^\Rx(\AoA_p) = 
    \begin{pmatrix}
        e^{j \frac{2 \pi}{v} f^c (d^{\Rx}_1)^\top u(\AoA_p)},
        \hdots,
        e^{j \frac{2 \pi}{v} f^c (d^{\Rx}_{N^\Rx})^\top u(\AoA_p)}
    \end{pmatrix}^{\top},
\end{equation}
where $v = 3 \times 10^8$ m/s denotes the speed of light, while $d^{\Rx}_n \in \R^{3 \times 1}$ is the position of the $n\mbox{-th}$ receive array element and $u(\AoA_p) \in \R^{3 \times 1}$ is a unitary-norm vector pointing in the direction specified by $\AoA_p$, both expressed in the local coordinates of the receiver (see Fig. \ref{fig:notation}).
The transmit steering vector $a^\Tx(\AoD_p)$ can also be computed in an analogous manner as
\begin{equation}
\label{eq:steering_vector_tx}
    a^\Tx(\AoD_p) =
    \begin{pmatrix}
        e^{j \frac{2 \pi}{v} f^c (d^{\Tx}_1)^\top u(\AoD_p)},
        \hdots,
        e^{j \frac{2 \pi}{v} f^c (d^{\Tx}_{N^\Tx})^\top u(\AoD_p)}
    \end{pmatrix}^{\top},
\end{equation}
where $d^{\Tx}_n \in \R^{3 \times 1}$ is the position of the $n\mbox{-th}$ transmit array element, and $u(\AoD_p) \in \R^{3 \times 1}$ denotes the direction specified by $\AoD_p$, in the local transmitter coordinate system (see Fig. \ref{fig:notation}).
Furthermore, for curved wavefronts in near-field regimes, the steering vectors can be approximated using a parabolic wavefront model \cite{do2023parabolic}.
Alternatively, it is also possible to determine the vectors $a^\Rx(\AoA_p)$ and $a^\Tx(\AoD_p)$ numerically by running the RT simulation for each antenna pair, i.e., at each coordinate $c_{n^\Rx, n^\Tx} = (c^\Rx + d^\Rx_{n^\Rx}, c^\Tx + d^\Tx_{n^\Tx})$ for $n^\Rx \in \dset{N^\Rx}$ and $n^\Tx \in \dset{N^\Tx}$ \cite{hoydis2023sionna}.

\subsection{Deterministic Channel Model}\label{subsec:deterministic_model}

Using the multipath parameters $R(c | \theta)$ produced by the RT in \eqref{eq:rt_parameters}, the DT can reconstruct the frequency response of each path component $p \in \dset{P}$ for each subcarrier frequency $f_s$ \add{as a} $N^\Rx N^\Tx \times 1$ vector of complex numbers, one for each of the transmit-receive antenna pairs.
\add{
In particular, the frequency response is given by
\begin{equation}
\label{eq:path_contribution_subcarrier}
    \alpha_p(\theta)
    e^{-j 2 \pi f_s \tau_p}
    a^\Rx(\AoA_p) \otimes
    \conj{a^\Tx(\AoD_p)},
\end{equation}
}%
where $\conj{\{ \cdot \}}$ denotes the element-wise conjugate operation, and $\otimes$ represents the Kronecker product \cite{van2000ubiquitous}.
Accordingly, each path is characterized by a magnitude $\abs{\alpha_p(\theta)}$ and by a phase $(\angle \alpha_p(\theta) - 2 \pi f_s \tau_p)$ dependent on the propagation delay $\tau_p$ and on the phase $\angle \alpha_p(\theta)$ of the complex amplitude $\alpha_p(\theta)$, in addition to the antenna-dependent contributions of the steering vectors.

Expanding from the per-subcarrier formulation in \eqref{eq:path_contribution_subcarrier}, we collect the contributions of the $p\mbox{-th}$ path across all $S$ subcarriers into a single $S N^\Rx N^\Tx \times 1$ vector $\alpha_p(\theta) a(\tau_p, \AoA_p, \AoD_p)$, where
\begin{equation}
\label{eq:time_angle_projection}
    a(\tau_p, \AoA_p, \AoD_p) = w(\tau_p) \otimes a^\Rx(\AoA_p) \otimes \conj{a^\Tx(\AoD_p)},
\end{equation}
represents the phase contributions emanating from all possible combinations of antenna pairs, represented by the  $N^\Rx N^\Tx \times 1$ vector $a^\Rx(\AoA_p) \otimes \conj{a^\Tx(\AoD_p)}$, and subcarriers, represented by the $S \times 1$ vector
\begin{equation}
    w(\tau_p) = \begin{pmatrix}
        e^{-j 2 \pi f_1 \tau_p}, ..., e^{-j 2 \pi f_S \tau_p}
    \end{pmatrix}^\top.
\end{equation} 
\add{The deterministic frequency response model assumed by the DT is then obtained by summing the contributions of each path $p \in \dset{P}$} into the $S N^\Rx N^\Tx \times 1$ vector
\begin{equation}
\label{eq:deterministic_channel_model}
    \Hhat(c | \theta) = \sum_{p=1}^{P} \alpha_p(\theta) a(\tau_p, \AoA_p, \AoD_p),
\end{equation}
\add{where the ``hat'' notation is a reminder that this is an estimate produced by the DT on the basis of the RT output.}

\subsection{Phase Error Channel Model}\label{subsec:phase_error_model}

The estimate $\Hhat( c | \theta)$ in \eqref{eq:deterministic_channel_model} hinges on the assumption that the geometric properties provided by the DT are almost identical to the physical environment of interest.
In practice, as mentioned in Sec.~\ref{sec:introduction}, even small discrepancies between the geometry assumed by the RT and the actual physical system, of a fraction of the studied wavelength, can significantly affect the accuracy of the predicted phase $(\angle \alpha_p(\theta) - 2 \pi f_s \tau_p)$ of each path $p \in \dset{P}$ \cite[Ch.~7.5]{molisch2012wireless}, resulting in random small-scale fading effects that are not captured by the deterministic model in \eqref{eq:deterministic_channel_model}.
Therefore, even for high-fidelity DT models, one cannot depend on the RT for a reliable estimate of the phases associated with each propagation path \cite[Ch.~7.5]{molisch2012wireless}, \cite{jemai2009calibration, he2018design, hoydis2023sionna}.

\begin{figure}
    \centering
    \includegraphics[keepaspectratio, width=3.4in]{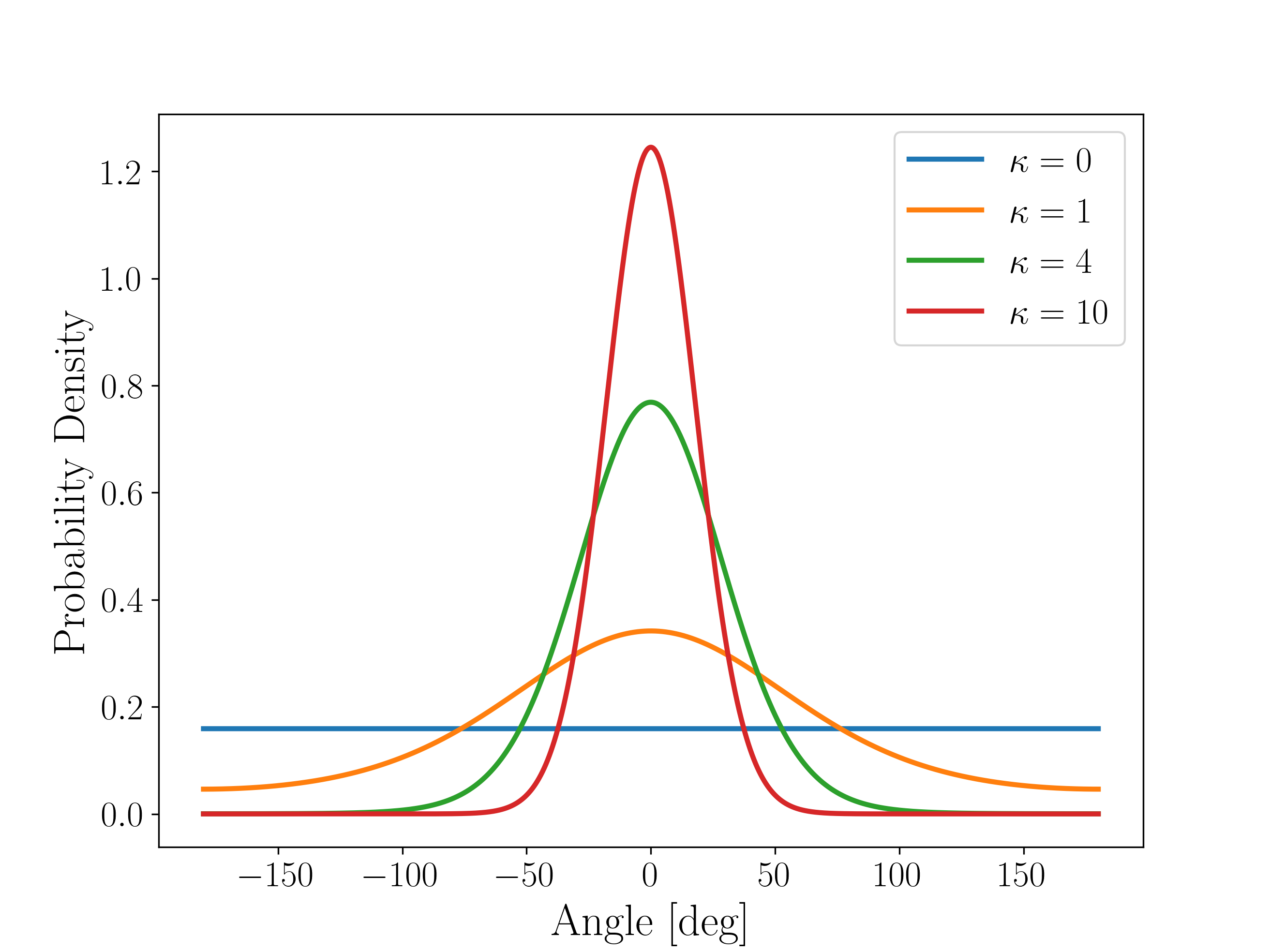}
    \caption{Von Mises probability density functions $\VM(0, \kappa)$ for different values of the concentration parameters $\kappa$.}
    \label{fig:von_mises_pdf}
\end{figure}

In order to account for the inherent phase error of the DT model in \eqref{eq:deterministic_channel_model}, we introduce a phase error term $z_p$ for each path $p \in \dset{P}$ that is distributed according to a von Mises distribution $\VM(0, \kPrior)$ with mean $0$ and concentration parameter $\kPrior \geq 0$.
\add{The von Mises distribution is the maximum entropy distribution for directional random variables with known mean direction \cite[Ch. 3.5.4]{mardia2000directional}, providing a worst-case model for randomly distributed phases where only the mean phase is assumed} \cite{badiu2019communication, wang2021outage}.
The probability density of a von Mises random variable $X \sim \VM(\mu, \kappa)$ is defined as \cite{mardia2000directional}
\begin{equation}
\label{eq:von_mises_dist}
    \VM\left(x | \mu, \kappa \right) = \frac{1}{2 \pi I_0\left( \kappa \right)} e^{\kappa \cos\left( x - \mu \right)},
\end{equation}
for $x \in \vmDomain$, with parameters $\mu \in \vmDomain$ and $\kappa \geq 0$, and where $I_0(\cdot)$ denotes the modified Bessel function of the first kind of order $0$.
The concentration parameter $\kappa$ determines the spread of the distribution as illustrated in Fig. \ref{fig:von_mises_pdf}.

Incorporating the phase errors $\{z_p\}_{p=1}^{P}$ into the deterministic model \eqref{eq:deterministic_channel_model} yields the stochastic model
\begin{equation}
\label{eq:phase_error_channel_model_analytic_form}
\begin{split}
    &\Hhat(c | \theta, \kPrior) = \sum_{p=1}^{P}
    e^{j z_p}
    \alpha_p(\theta)
    a(\tau_p, \AoA_p, \AoD_p),
\end{split}
\end{equation}
where the phase errors $\{ z_p \}_{p=1}^{P} \iid \VM(0, \kPrior)$ are assumed to be independent and identically distributed (i.i.d.).
Note that all the subcarriers and antenna pairs share the same \emph{phase error vector}
\begin{equation}
\label{eq:phaseerrorvector}
    Z = (z_1, ..., z_P)^T \iid \VM(0, \kPrior)
\end{equation}
of i.i.d. phase error terms and, by extension, the same $P\times 1$ \emph{phasor error vector}
\begin{equation}
    e^{jZ} = (e^{j z_1}, \hdots, e^{j z_P})^\top,
\end{equation}
which corresponds to the element-wise application of the exponential function to the $P \times 1$ phase-error vector $Z$.
Accordingly, the stochastic model \eqref{eq:phase_error_channel_model_analytic_form} can be written as
\begin{equation}
\label{eq:phase_error_channel_model}
    \Hhat(c | \theta, \kPrior) = \G(c , \theta) e^{jZ},
\end{equation}
with the $S N^\Rx N^\Tx \times P$ matrix
\begin{equation}
\label{eq:Gmatrix}
\begin{split}
    \G(c , \theta) =
    \begin{pmatrix}
        a(\tau_1, \AoA_1, \AoD_1), ..., a(\tau_P, \AoA_P, \AoD_P)
    \end{pmatrix}
    \Diag{ \alpha(\theta) },
\end{split}
\end{equation}
where $\alpha(\theta) = (\alpha_1(\theta), ..., \alpha_P(\theta))^\top$ is a $P \times 1$ vector representing the complex amplitude of each of the $P$ predicted paths, and $\diag{\cdot}$ denotes the diagonal matrix of the given argument.

In practice, it may be useful to assume that different paths have different concentration parameters $\kPrior$. In particular, for line of sight (LoS) paths one can set the corresponding phase error $z_p$ to zero, which corresponds to an infinite concentration value $\kPrior$. For other paths, it is common to assume a uniform phase error, which is obtained by setting $\kPrior = 0$, i.e.,  $\VM(z_p | 0, 0) = 1 / 2 \pi$ for $z_p \in \vmDomain$. In this paper, we consider for simplicity of presentation the i.i.d. model \eqref{eq:phaseerrorvector}, but it is straightforward to generalize the proposed technique to situations in which the phase errors are independent but not identically distributed. 

By decreasing the value of the concentration parameter $\kPrior$, the stochastic model \eqref{eq:phase_error_channel_model} smoothly transitions from the deterministic-phase model \eqref{eq:deterministic_channel_model} for $\kPrior \to +\infty$, which neglects phase errors, to a worst-case phase-error RT model for $\kPrior = 0$, for which the distribution of the phase error is uniform.
Current applications of RT-simulated channels mostly assume the uniform phase error setting, i.e., $\kPrior = 0$, for non-LoS paths.
Nevertheless, we keep the presentation general in order to evaluate the potential benefits of more accurate ray tracing models.

\subsection{Data-Driven Calibration of the Ray Tracer}

In addition to the geometric properties, the precision of the RT simulation also depends on the accuracy of the estimated electromagnetic properties of the materials.
These are typically more complex to acquire than their geometric counterpart \cite{royo2019overview, bedford2019modeling, lin20236g, OpenStreetMap}, and values estimated in technical reports for generic materials do not capture the variety of properties found in practical scenarios \cite{itu2021effects}.
As a result, previous works \cite{jemai2009calibration, priebe2012calibrated, he2018design, charbonnier2020calibration, kanhere2023calibration, bhatia2023tuning, hoydis2023sionna} have proposed to include the unknown electromagnetic parameters, namely the relative permittivity and conductivity, of all the material types in the scene in a vector $\theta$, which is optimized based on data obtained from the ground-truth scene at multiple locations in a process known as \emph{calibration}.

To this end, we assume the availability of a dataset 
\begin{equation}
\label{eq:dataset}
    \mathcal{D} = \left\{ \left( c_n, H_n \right) \right\}_{n=1}^{N}
\end{equation}
of $N$ \emph{channel frequency response} (CFR) observations $H_n \in \C^{S N^\Rx N^\Tx \times 1}$ for $n \in \dset{N}$, with each observation $H_n$ taken at receiver and transmitter coordinates $c_n = (c^\Rx_n, c^\Tx_n)$ for all $S$ subcarriers.
\add{Specifically, each observation $H_n$ rearranges the observed CFRs across all antenna pairs and subcarrier frequencies into a single vector following the same order as the channel models in \eqref{eq:deterministic_channel_model} and \eqref{eq:phase_error_channel_model}}.

The goal of calibration is to enhance the accuracy of the RT output in \eqref{eq:rt_parameters} for locations $c$ at which channel data are available, as well as possibly for new locations $c$ not included in the dataset, thus extending the DT simulation capacity to new receiver and/or transmitter positions.
That is, the DT aims to provide a reliable simulation of the propagation and channel conditions between any existing or new pair of receiver and transmitter devices at coordinates $c$ by leveraging the output $R(c | \theta)$ of the RT.
Such a simulator can then be used by the DT to monitor the physical deployment of interest, or to design optimal control policies without the need to collect further data by training AI models in virtual space \cite{ruah2023bayesian, jiang2023digital, salehi2023multiverse, nguyen2023probabilistic, viswanathan2020communications, ferrand2023wireless}.

\section{Calibration Baselines} \label{sec:baselines}

In this section, we review two existing classes of calibration methods.
The first approach, proposed in \cite{hoydis2023sionna}, optimizes the electromagnetic parameters $\theta$ given the channel observations $\mathcal{D}$ in \eqref{eq:dataset} by assuming the phase error-oblivious model $\Hhat(c|\theta)$ in \eqref{eq:deterministic_channel_model}. 
We refer to this channel response-based method as \emph{phase error-oblivious calibration}.
The second approach, representing the line of work \cite{jemai2009calibration, priebe2012calibrated, he2018design, charbonnier2020calibration, kanhere2023calibration, bhatia2023tuning}, calibrates the parameters $\theta$ by comparing the observed \emph{channel power profiles} estimated from the dataset $\mathcal{D}$ to the corresponding channel power profiles obtained from the RT under the uniform phase error model $\Hhat(c|\theta, 0)$ in \eqref{eq:phase_error_channel_model}, i.e., with concentration parameter $\kPrior = 0$.
We refer to this power profile-based process as \emph{uniform phase error calibration}.

\subsection{Phase Error-Oblivious Calibration} \label{subsec:peoc}

As proposed in \cite{hoydis2023sionna}, the material parameters $\theta$ can be optimized by comparing each observed CFR $H_n$ in the dataset $\mathcal{D}$ to its respective simulated CFR in $\Hhat(c_n | \theta)$ under the deterministic model \eqref{eq:deterministic_channel_model} for all observations $n \in \dset{N}$. \add{
To this end, the resulting phase error-oblivious calibration strategy addresses the least-squares problem of finding the parameter
\begin{equation}
\label{eq:error_oblivious_opt_problem}
    \theta^{*} = \argmin_{\theta} \lossPEOC(\theta | \mathcal{D})
\end{equation}
that minimizes the difference between the simulated and observed CFRs
\begin{equation}
\label{eq:loss_peoc}
    \lossPEOC(\theta | \mathcal{D}) = \sum_{n=1}^{N} \Norm{
        H_n - \Hhat(c_n | \theta)
    }^2,
\end{equation}
where $\norm{X}^2 = \conjt{X} X$ denotes the squared 2-norm for a complex vector $X$.
Problem \eqref{eq:error_oblivious_opt_problem} is tackled in \cite{hoydis2023sionna} through gradient descent.
This solution requires the differentiability of the predicted channel response $\Hhat(c_n | \theta)$ with respect to the electromagnetic parameters $\theta$, which is ensured by Sionna RT \cite{hoydis2022sionna, hoydis2023sionna} .
}

\subsection{Uniform Phase Error Calibration} \label{subsec:upec}

In earlier works \cite{jemai2009calibration, priebe2012calibrated, he2018design, charbonnier2020calibration, kanhere2023calibration, bhatia2023tuning}, RT calibration was conducted by comparing the power profiles of the measured and simulated channels over time, i.e., the power delay profiles \cite{jemai2009calibration, he2018design}, and/or over space, i.e., the power-angle-delay profiles \cite{priebe2012calibrated, charbonnier2020calibration, kanhere2023calibration, bhatia2023tuning}, while assuming uniform phase errors for the RT-predicted path components.
In this subsection, we briefly elaborate on the resulting uniform phase error calibration approach. 

Given an observation $H_n$ of the channel frequency response from the dataset $\mathcal{D}$, an estimation of its power-angle-delay profile can be obtained by projecting the CFR $H_n$ onto a channel vector matched to a given triple $(\tau, \AoD, \AoA)$ of delay $\tau \geq 0$, angle of departure $\AoD \in \angleDomain$, and angle of arrival $\AoA \in \angleDomain$.
Such vector can be defined by considering a normalized single-path component $a(\tau, \AoD, \AoA)$ from \eqref{eq:time_angle_projection} with the given path parameters $(\tau, \AoD, \AoA)$ in lieu of $(\tau_p, \AoD_p, \AoA_p)$.
Accordingly, the \emph{power-angle-delay profile} for measured CFR $H_n$ is given as
\begin{equation}
\label{eq:power_measurement}
    \powProfile_n(\tau, \AoD, \AoA) = \Abs{\frac{1}{\sqrt{L}} \conjt{a(\tau, \AoD, \AoA)} H_n}^2,
\end{equation}
with $L = S N^\Rx N^\Tx$.


To enable calibration of the RT, the measured power-angle-delay profile \eqref{eq:power_measurement} must be compared to the simulated power-angle-delay profile obtained from the RT under uniform phase errors. The latter can be obtained from the stochastic model in \eqref{eq:phase_error_channel_model} with concentration $\kPrior = 0$,  in a manner similar to \eqref{eq:power_measurement},  as
\begin{equation}
\label{eq:power_profile_def}
    \hatPowProfile(c, \tau, \AoD, \AoA | \theta) =
    \E_{Z \iid \VM(0, 0)} \left[ 
        \Abs{\frac{1}{\sqrt{L}} \conjt{a(\tau, \AoD, \AoA)} \G(c , \theta) e^{jZ}}^2
    \right],
\end{equation}
where the average is taken with respect to the uniformly distributed phases $Z = (z_1, ..., z_P)^\top$.
The average in \eqref{eq:power_profile_def} can be evaluated explicitly as (see Appendix \ref{apx:average_power_von_mises})
\begin{equation}
\label{eq:power_model}
    \hatPowProfile(c, \tau, \AoD, \AoA | \theta) =
    \Norm{\frac{1}{\sqrt{L}} \conjt{a(\tau, \AoD, \AoA)} \G(c , \theta)}^2.
\end{equation}

Optimization of the material parameters $\theta$ is carried by finding the parameter
\begin{equation}
\label{eq:upec_optimization_problem}
    \theta^{*}  = \argmin_{\theta} \lossUPEC(\theta | \mathcal{D})
\end{equation}
that minimizes the discrepancy between the measured and simulated power profiles for a set of pre-defined $M$ angle-delay triples $\{ (\tau_m, \AoD_m, \AoA_m )\}_{m=1}^{M}$ under the squared loss metric
\begin{equation}
\label{eq:upec_loss}
\begin{split}
    \lossUPEC(\theta | \mathcal{D}) =
    \sum_{n=1}^{N} \sum_{m=1}^{M} \Abs{\powProfile_n(\tau_m, \AoD_m, \AoA_m) - \hatPowProfile(c_n, \tau_m, \AoD_m, \AoA_m | \theta)}^2.
\end{split}
\end{equation}

Problem \eqref{eq:upec_optimization_problem} can be  tackled through gradient descent, where differentiability of the loss $\lossUPEC(\theta | \mathcal{D})$ with respect to the material parameters $\theta$ is ensured by the differentiability of the RT output \cite{hoydis2023sionna}. In this regard, we note that existing art has mostly relied on less efficient strategies based on  simulated annealing \cite{jemai2009calibration, priebe2012calibrated, he2018design} or exhaustive parameter search \cite{charbonnier2020calibration, bhatia2023tuning}.
\add{We refer the reader to Appendix \ref{subapx:upec_details} for a more in-depth discussion on the selection of the $M$ projection triples $\{ (\tau_m, \AoD_m, \AoA_m )\}_{m=1}^{M}$.}

\section{Phase Error-Aware Calibration} \label{sec:peac}

In this section, we introduce a novel calibration scheme that accounts for the inherent phase uncertainty associated with the RT output $R(c | \theta)$ in \eqref{eq:rt_parameters} via the stochastic channel model in \eqref{eq:phase_error_channel_model}.
The proposed \emph{phase error-aware calibration} scheme aims at extracting coherent information about the multi-path phases for each channel realization based on the available data \eqref{eq:dataset} with the goal of improving the quality of the calibrated RT parameters $\theta$.
Therefore, the approach contrasts with the phase error-oblivious strategy of \cite{hoydis2023sionna}, reviewed in the previous section, which hinges on the deterministic model in \eqref{eq:deterministic_channel_model}, as well as with existing uniform phase error calibration schemes, also described in previous section, which account only for the received power.

\subsection{Calibration Problem}

\add{The phase error-calibration scheme models the channel observations $H_n$ as noisy observations of the phase error-aware model in \eqref{eq:phase_error_channel_model} expressed as}
\begin{equation}
\label{eq:peac_measurement_model}
    H_n = \G(c_n , \theta) e^{j Z_n} + W_n,
\end{equation}
where \add{the additive noise vector $W_n \in \C^{S N^\Rx N^\Tx \times 1}$ is composed of random of i.i.d. circular Gaussian elements $\CN(0, \sigma^2)$} with zero mean and known power $\sigma^2$, and $e^{j Z_n} = (e^{j z_{n, 1}}, ..., e^{j z_{n, P}})^\top$ is the $P \times 1$ random phasor vector accounting for phase errors.
The phase errors $Z_n = (z_{n, 1}, ..., z_{n, P})^\top \iid \VM(0, \kPrior)$ are described by a prior concentration parameter $\kPrior$.
We first assume parameter $\kPrior$ to be fixed, and then we extend the proposed algorithm to enable the estimation of the prior parameter $\kPrior$ from data.

By \eqref{eq:peac_measurement_model}, the probability distribution of each observation conditioned on the phase-error vector $Z_n$ is given by
\begin{equation}
\label{eq:error_aware_conditioned_likelihood}
    P(H_n | Z_n, c_n, \theta) = \CN\left( H_n \Big| \G(c_n , \theta) e^{j Z_n}, \sigma^2 I \right),
\end{equation}
\add{where $I$ denotes the $S N^\Rx N^\Tx \times S N^\Rx N^\Tx$ identity matrix.}
The probability distribution $P(H_n | c_n, \theta)$ of each measured CFR $H_n$ is then obtained by marginalizing the joint distribution $P(H_n, Z_n | c_n, \theta, \kPrior) = P(H_n | Z_n, c_n, \theta) P(Z_n | \kPrior)$ over the distribution $P(Z_n | \kPrior)$ of the phase-error vector $Z_n$, i.e.,
\begin{equation}
\label{eq:error_aware_likelihood}
    P(H_n | c_n, \theta, \kPrior) = \int P(H_n | Z_n, c_n, \theta) P(Z_n | \kPrior) dZ_n,
\end{equation}
where integration is taken over the domain $\vmDomain^{\otimes P}$ of the phase errors.
Given the i.i.d. assumption, the phase-error vector is distributed as $P(Z_n | \kPrior) = \prod_{p=1}^{P} \VM(z_{n, p} | 0, \kPrior)$.

The cross-entropy loss with respect to the RT parameters $\theta$ is then given as
\begin{equation}
\label{eq:error_aware_cross_entropy}
\lossPEAC(\theta | \mathcal{D}) = - \sum_{n=1}^{N} \log\left( P(H_n | c_n, \theta) \right).
\end{equation}
The resulting ML calibration problem is therefore defined as the minimization
\begin{equation}
\label{eq:error_aware_opt_problem}
\begin{split}
    \theta^{*} =
    \argmin_{\theta} \left\{
        - \sum_{n=1}^{N} \log\left(
            \int P(H_n | Z_n, c_n, \theta) P(Z_n | \kPrior) dZ_n
        \right)
    \right\}.
\end{split}
\end{equation}
Problem \eqref{eq:error_aware_opt_problem} reduces to \eqref{eq:error_oblivious_opt_problem} when choosing an infinitely large prior concentration parameter $\kPrior$ for the phase noise.

\subsection{Variational Expectation Maximization Algorithm}

The direct solution of the problem \eqref{eq:error_aware_opt_problem} requires the expensive computation of the negative log-likelihood function \eqref{eq:error_aware_cross_entropy} by marginalizing over the latent variables $Z_n$ \eqref{eq:error_aware_likelihood} for all observations $n \in \dset{N}$.
To address this computational challenge, we propose a computationally efficient \emph{variational expectation maximization} (VEM) algorithm \cite[Ch.~10]{simeone2022machine} that approximates the solution of the ML problem \eqref{eq:error_aware_opt_problem} by restricting the set of possible solutions to a parametric family of distributions. 

The proposed VEM approach iterates between the problem of optimizing the parameter $\theta$ and that of estimating the phase errors $z_{n,p}$ for all paths $p\in\{1,...,P\}$ and all observations $n\in\{1,...,N\}$.
The main underlying idea is that, while a priori all phase errors are assumed to have the same distribution $\VM(0, \kPrior)$, each channel observation $H_n$ is characterized by specific per-path phase errors, which may be estimated and used to improve the calibration process.

The estimate of each phase error  $z_{n,p}$ is represented by a von Mises distribution $\VM\left( z_{n, p} | \mu_{n, p}, \kappa_{n, p} \right)$ with mean parameters $\mu_{n,p}$ and concentration parameter $ \kappa_{n,p}$.
The mean represents the nominal estimate of the phase error, and the concentration parameter quantifies the uncertainty of this estimate.
As detailed below, both parameters are estimated from the measured CFR $H_n$, as well as from the current iterate $\theta$.
Accordingly, the estimates of the phase errors for all paths $p\in \{1,...,P\}$ associated with data point $n$ are described by the \emph{variational distribution}
\begin{equation}
\label{eq:variational_distribution}
    Q\left( Z_n | \mu_n, \kappa_n \right) =
    \prod_{p=1}^{P} \VM\left( z_{n, p} | \mu_{n, p}, \kappa_{n, p} \right),
\end{equation}
which assumes the estimated phase errors to be independent (but not identically distributed).
The adjective ``variational'' indicates that distribution  $Q\left( Z_n | \mu_n, \kappa_n \right)$ depends on free parameters that can be optimized, namely the mean parameter vector $\mu_n = (\mu_{n, 1}, ..., \mu_{n, P})^\top \in \vmDomain^{\otimes P}$ and the concentration parameter vector $\kappa_n = (\kappa_{n, 1}, ..., \kappa_{n, P})^\top$ with $\kappa_{n, p} \geq 0$ for $p \in \dset{P}$.
As discussed, distribution \eqref{eq:variational_distribution} captures the coherent information about the $p\mbox{-th}$ path that can be extracted from the data.
Considering all data points $n\in\{1,...,N\}$, this information is encoded in the variational parameters $\{ \mu_n, \kappa_n \}_{n=1}^{N}$.

VEM approximates the solution of problem \eqref{eq:error_aware_opt_problem} by iteratively minimizing an upper bound on the negative log-likelihood \eqref{eq:error_aware_opt_problem} over the material parameters $\theta$ and over parameters $\{ \mu_n, \kappa_n \}_{n=1}^{N}$ of the variational distribution \eqref{eq:variational_distribution}.
This upper bound, known as the \emph{variational free energy} is given by (see, e.g., \cite{bishop2006pattern, simeone2022machine})
\begin{equation}
    \lossPEAC(\theta | \mathcal{D}) \leq \sum_{n=1}^{N} \F_n(\mu_n, \kappa_n, \theta, \kPrior),
\end{equation}
where
\begin{equation}
\label{eq:free_energy_definition}
    \F_n\left( \mu_n, \kappa_n, \theta, \kPrior \right) =
    \E_{Z_n \sim Q(Z_n | \mu_n, \kappa_n)} \left[ \log\left( \frac{Q(Z_n | \mu_n, \kappa_n)}{P(H_n, Z_n | c_n, \theta, \kPrior)} \right) \right],
\end{equation}
and $P(H_n, Z_n | c_n, \theta, \kPrior)$ is the joint distribution defined in \eqref{eq:error_aware_likelihood}.
As detailed in Appendix \ref{apx:free_energy}, using  \eqref{eq:error_aware_conditioned_likelihood} and \eqref{eq:variational_distribution}, the free energy \eqref{eq:free_energy_definition} can be written more explicitly as 
\begin{equation}
\label{eq:free_energy}
\begin{split}
\F_n\left( \mu_n, \kappa_n, \theta, \kPrior \right) =
    & \sum_{p=1}^{P} \left\{
        \log\left( \frac{I_0(\kPrior)}{I_0(\kappa_{n, p})} \right)
        + b(\kappa_{n, p}) \left(
            \kappa_{n, p}
            - \kPrior \cos(\mu_{n, p})
        \right)
    \right\} + 
    L \log\left( \pi \sigma^2 \right) \\
    & + \frac{1}{\sigma^2} \left[
        \Norm{ \G(c_n , \theta) \Diag{B(\kappa_n)} e^{j \mu_n} - H_n}^2 +
        L \left( \Abs{\alpha(\theta)}^2 \right)^\top \left( \allOnes - B^2(\kappa_n) \right)
    \right],
\end{split}
\end{equation}
where $\abs{\cdot}^2$ denotes the element-wise squared amplitude of a vector; $\allOnes = (1, ..., 1)^\top$ denotes the all-ones $P \times 1$ vector; $I_{\nu}(\cdot)$ is the modified Bessel function of the first kind of order $\nu \geq 0$; $B(\kappa_n) = (b(\kappa_{n, 1}), ..., b(\kappa_{n, P}))^\top$ is the $P \times 1$ vector of Bessel ratios
\begin{equation}
\label{eq:bessel_ratio}
    b(\kappa) = \frac{I_1(\kappa)}{I_0(\kappa)},
\end{equation}
for $\kappa \geq 0$; and $B^2(\kappa_n) = (b(\kappa_{n, 1})^2, ..., b(\kappa_{n, P})^2)^\top$ is the $P \times 1$ vector of squared Bessel ratios.
The Bessel ratio $b(\cdot)$ in \eqref{eq:bessel_ratio} is a strictly increasing and continuous function in $[0, +\infty)$, with $b(0) = 0$ and $\lim_{\kappa \to +\infty} b(\kappa) = 1$.

As in the classical expectation maximization (EM) algorithm, each iteration $i \geq 0$ of VEM comprises of two steps, an \emph{expectation step} (E-step) and a \emph{maximization step} (M-step), producing a sequence of iterates $\theta^{(i)}$ and $\{ \mu_n^{(i)}, \kappa_n^{(i)} \}_{n=1}^{N}$, with initializations  $\theta^{(0)}$ and $\{ \mu_n^{(0)}, \kappa_n^{(0)} \}_{n=1}^{N}$.
In this regard, the electromagnetic parameters $\theta^{(0)}$ may be initialized using prior knowledge from previous observations or using known properties of the materials in the scene \cite{itu2021effects}.
The E-step estimates the statistics of the phase errors  $\{ \mu_n^{(i)}, \kappa_n^{(i)} \}_{n=1}^{N}$ across all data points $(c_n, H_n) \in \mathcal{D}$ by using the current iterate $\theta^{(i)}$;  while the M-step updates the material parameters $\theta^{(i+1)}$ for fixed updated phase error parameters $\{ \mu_n^{(i+1)}, \kappa_n^{(i+1)} \}_{n=1}^{N}$.
The overall proposed phase error-aware calibration algorithm is summarized in Algorithm \ref{alg:vem}, and details are provided next.

\begin{algorithm}[h]
\setstretch{1.5}
\caption{Phase Error-Aware Calibration}
\label{alg:vem}
\begin{algorithmic}[1]
    \renewcommand{\algorithmicrequire}{\textbf{Input:}}
    \renewcommand{\algorithmicensure}{\textbf{Output:}}
    \Require
        Training data $\mathcal{D} = \{ (c_n, H_n) \}_{n=1}^{N}$;
        initial iterates $( \theta^{(0)}, \kPrior^{(0)}, \{ \mu_n^{(0)}, \kappa_n^{(0)} \}_{n=1}^{N} )$ for the material parameters and for the per-data point estimated statistics of the phase errors;
        number of iterations $I \geq 0$
    \Ensure  Calibrated parameters $ \theta^{(I)}$
    \For{$i = 0$ to $I-1$}
        \Statex \textit{\quad Expectation Step} :
        \For{$n = 1$ to $N$} \label{algline:vem_start_e_step}
            \State $\mu^{(i+1)}_n \gets \angle\left\{
                    \left( \conjt{\G(c_n , \theta^{(i)})} \G(c_n , \theta^{(i)}) \right)^{-1}
                    \left( \frac{\sigma^2 \kPrior^{(i)}}{2} \allOnes + \conjt{\G(c_n , \theta^{(i)})} H_n \right)
                \right\}$
            \For{$p = 1$ to $P$}
                \If{$\frac{L \abs{\alpha_p(\theta^{(i)})}^2}{\sigma^2} > 1$}
                    \State $\kappa^{(i+1)}_{n, p} \gets 2 \sqrt{\frac{L \abs{\alpha_p(\theta^{(i)})}^2}{\sigma^2} - 1} \sqrt{\frac{L \abs{\alpha_p(\theta^{(i)})}^2}{\sigma^2}}$
                \Else
                    \State $\kappa^{(i+1)}_{n, p} \gets 0$
                \EndIf
            \EndFor  \label{algline:vem_end_e_step}
        \EndFor
        
        \Statex \textit{\quad Maximization Step} :
        \State $\theta^{(i+1)} \gets \mathrm{GradientDescent}_\theta \left\{ \sum_{n=1}^{N} \F_n\left( \mu^{(i+1)}_n, \kappa^{(i+1)}_n, \theta, \kPrior^{(i)} \right) \right\}$  \label{algline:vem_m_step_theta}
        \If{$(N P)^{-1} \sum_{n=1}^{N} \sum_{p=1}^{P} b(\kappa^{(i+1)}_{n, p}) \cos(\mu^{(i+1)}_{n, p}) \geq 0$} \label{algline:vem_start_m_step_prior}
            \State $\kPrior^{(i+1)} \gets b^{-1} \left( \frac{1}{N P} \sum_{n=1}^{N} \sum_{p=1}^{P} b(\kappa^{(i+1)}_{n, p}) \cos(\mu^{(i+1)}_{n, p}) \right)$
        \Else
            \State $\kPrior^{(i+1)} \gets 0$
        \EndIf \label{algline:vem_end_m_step_prior}
    \EndFor
\end{algorithmic}
\end{algorithm}
 
\subsection{Expectation Step}

As summarized by lines  \ref{algline:vem_start_e_step}-\ref{algline:vem_end_e_step} of Algorithm \ref{alg:vem}, during the E-step of the proposed phase error-aware calibration scheme,  the material parameters $\theta^{(i)}$ are fixed, and minimization of the variational free energy in \eqref{eq:free_energy} is carried out with respect to the local parameters $\mu_n$ and $\kappa_n$. This is done by tackling  the minimization problem 
\begin{equation}
\label{eq:generic_vem_e_step}
    \left( \mu_n^{(i+1)}, \kappa_n^{(i+1)} \right)
    = \argmin_{\mu_n, \kappa_n} \left\{ \F_n\left( \mu_n, \kappa_n, \theta^{(i)}, \kPrior \right) \right\}.
\end{equation}
This problem can be addressed, separately for each data point $n \in \dset{N}$, using any optimization algorithms such as gradient descent.
In the following, we present an approximate analytical solution that will be tested experimentally in the next section.

As detailed in Appendix \ref{apx:vem_updates}, we propose to address a relaxed version of problem \eqref{eq:generic_vem_e_step} in which optimization is carried out over the complex phasor $e^{j\mu_n}$ while  removing the constraint that the magnitude of all entries be equal to 1. For this relaxed problem, we obtain a  stationary solution, and then we project the solution into the space of unitary magnitude vectors to recover an approximate estimate of the variational mean vector $\mu_n$. Note that we do not make any claim of optimality for this solution.

To elaborate on the obtained approximate solution, let us assume that the $L \times P$ matrix $\G(c_n , \theta^{(i)})$ in \eqref{eq:Gmatrix} is of rank $P$. 
This assumption requires that the number of channel coefficients measured for each data points, $L$, is no smaller than the number, $P$, of paths, i.e., $L \geq P$.
In this common situation, full rank implies that the space-frequency signatures of each of the $P$ paths, i.e., the columns of matrix $\G(c_n, \theta^{(i)})$, are linearly independent and hence distinguishable using linear projections.
Accordingly, this condition implies that the system has sufficient resolution in the combined space-frequency domain to resolve the different paths from the predicted channel response in the given deployment scenario.

Under this assumption, by addressing the mentioned relaxed problem and projecting back the solution into the space of phasors, we obtain the next iterate 
\begin{equation}
\label{eq:e_step_posterior_mean}
    \mu_n^{(i+1)} = \angle\biggr\{
        \Bigr( \conjt{\G(c_n , \theta^{(i)})} \G(c_n , \theta^{(i)}) \Bigl)^{-1}
        \Bigr( \frac{\sigma^2 \kPrior}{2} \allOnes + \conjt{\G(c_n , \theta^{(i)})} H_n \Bigl)
    \biggl\},
\end{equation}
where $\angle ( \cdot )$ represents the element-wise phase of a complex vector.
Furthermore, the concentration parameter vector $\kappa_{n,p}$ equals zero, i.e., $\kappa^{(i+1)}_{n,p}=0$, if the inequality $L|\alpha_p(\theta^{(i)})|^2/\sigma^2\leq1$ holds, while it satisfies the fixed-point equation
\begin{equation}
\label{eq:e_step_posterior_concentration}
    \kappa_{n,p}^{(i+1)} = \frac{2 L}{\sigma^2} \abs{\alpha_p(\theta^{(i)})}^2 b(\kappa_{n,p}^{(i+1)}).
\end{equation} otherwise.  While equation \eqref{eq:e_step_posterior_concentration} can be addressed by fixed-point iterations, an approximation can be obtained by bounding the Bessel ratio \cite{ruiz2016new}, yielding the approximate closed-form solution
\begin{equation}
    \kappa_{n, p}^{(i+1)} = 2 \sqrt{\frac{L \abs{\alpha_p(\theta^{(i)})}^2}{\sigma^2} - 1} \sqrt{\frac{L \abs{\alpha_p(\theta^{(i)})}^2}{\sigma^2}},
\end{equation}
for each path $p \in \dset{P}$ and observation $n \in \dset{N}$.

The condition \begin{equation}
\label{eq:e_step_condition_snr}
    \frac{L \abs{\alpha_p(\theta^{(i)})}^2}{\sigma^2} > 1,
\end{equation} under which the fixed-point condition \eqref{eq:e_step_posterior_concentration} applies, indicates that the signal-to-noise ratio (SNR) associated with the observation of path $p$ is larger than one.
Accordingly, if the per-path SNR, i.e., the left-hand side of \eqref{eq:e_step_condition_snr}, is smaller than one, the estimated concentration parameter $\kappa_{n,p}^{(i+1)}$  corresponds to a uniform phase error, while, if it is larger than one, the concentration parameter $\kappa_{n,p}^{(i+1)}$ is larger than zero.
For any strictly positive concentration $\kappa_{n,p}^{(i+1)} > 0$, the corresponding variational distribution $\VM(\mu_{n,p}^{(i+1)}, \kappa_{n,p}^{(i+1)})$ exhibits a preferred direction along the mean phase $\mu_{n,p}^{(i+1)}$, and, therefore, contributes to lowering the uncertainty of the phase error of path $p$ in observation $n$.


\subsection{Maximization Step}

Let us consider first the case in which the concentration parameter $\kPrior$ is fixed.
As summarized in the line \ref{algline:vem_m_step_theta} of Algorithm \ref{alg:vem}, during the M-step of the proposed scheme, the previously selected phase error parameters $\{ \mu_n^{(i+1)}, \kappa_n^{(i+1)} \}_{n=1}^{N}$, are given, and the optimal material parameter iterate $\theta^{(i+1)}$ is obtained by tackling the problem 
\begin{equation}
\label{eq:generic_vem_m_step}
    \theta^{(i+1)} = \argmin_{\theta} \left\{\sum_{n=1}^{N} \F_n\left( \mu_n^{(i+1)}, \kappa_n^{(i+1)}, \theta, \kPrior \right) \right\}.
\end{equation}
We suggest to use gradient descent for this purpose in a manner similar to the phase error-oblivious scheme in \cite{hoydis2023sionna}.


When treating the prior concentration $\kPrior$ as a parameter to be jointly optimized with $\theta$ (lines \ref{algline:vem_start_m_step_prior}-\ref{algline:vem_end_m_step_prior} of Algorithm \ref{alg:vem}), the M-step tackles the problem
\begin{equation}
\label{eq:alternative_vem_m_step}
    \left( \theta^{(i+1)}, \kPrior^{(i+1)} \right) = \argmin_{\theta, \kPrior} \left\{\sum_{n=1}^{N} \F_n\left( \mu_n^{(i+1)}, \kappa_n^{(i+1)}, \theta, \kPrior \right) \right\}
\end{equation}
in lieu of \eqref{eq:generic_vem_m_step}.
To address the minimization \eqref{eq:alternative_vem_m_step}, parameter $\theta^{(i+1)}$ may still be updated through gradient descent; while, as derived in Appendix~\ref{apx:learnable_prior}, an optimal value for the prior concentration iterate $\kPrior^{(i+1)}$ is obtained as
\begin{equation}
\label{eq:vem_prior_m_step}
    \kPrior^{(i+1)} = b^{-1} \left( \frac{1}{N P} \sum_{n=1}^{N} \sum_{p=1}^{P} b(\kappa^{(i+1)}_{n, p}) \cos(\mu^{(i+1)}_{n, p}) \right),
\end{equation}
provided that the inequality $(N P)^{-1} \sum_{n=1}^{N} \sum_{p=1}^{P} b(\kappa^{(i+1)}_{n, p}) \cos(\mu^{(i+1)}_{n, p}) \geq 0$ holds, and as $\kPrior^{(i+1)} = 0$ otherwise, where $b^{-1}(\cdot)$ denotes the inverse function of the Bessel ratio $b(\cdot)$ \cite{hill1981evaluation}.


\subsection{Complexity Analysis}

\add{
We now study the additional computational complexity incurred by the proposed method with respect to the calibration baselines introduced in Sec.~\ref{sec:baselines}.
The main new added operation is the update of the variational parameters $\{ \mu_n^{(i)}, \kappa_n^{(i)} \}$ in the E-step.
In this regard, we note that the matrix inversion during the computation of the mean parameter in \eqref{eq:e_step_posterior_mean} can be expressed as
\begin{equation}
\begin{split}
    \left( \conjt{\G(c_n , \theta^{(i)})} \G(c_n , \theta^{(i)}) \right)^{-1} =
    \Diag{ \alpha(\theta)}^{-1} A(c_n)^{-1} \Diag{ \conj{\alpha(\theta)} }^{-1},
\end{split}
\end{equation}
where the $P \times P$ matrix 
\begin{equation}
    A(c_n) =
    \begin{pmatrix}
        \conjt{a(\tau_i, \AoA_i, \AoD_i)} a(\tau_j, \AoA_j, \AoD_j)
    \end{pmatrix}_{(i, j) \in \dset{P}^2}
\end{equation}
does not depend on the material parameters $\theta^{(i)}$.
Accordingly, the inverse $A(c_n)^{-1}$ only needs to be computed once for each calibration position $c_n$ at the beginning of training, while the successive E-step iterations can be computed efficiently in $O(N \max(LP, P^2))$ operations.
All in all, our proposed method only incurs an additional complexity of $O(N P^3)$ when inverting the matrices $\{A(c_n)\}_{n=1}^N$ prior to calibration, while retaining the same computational efficiency as the baselines at each training step.
}

\section{Experiments} \label{sec:experiments}

In this section, we present numerical experiments with the aim of offering insights into the performance comparison between the existing calibration schemes summarized in Sec.~\ref{subsec:peoc} and Sec.~\ref{subsec:upec} and the proposed method introduced in the previous section over a set of synthetic examples, allowing us to control the evaluation setting.
\add{
The channel observations $\mathcal{D}$ are generated by RT with an unknown ground-truth setting.
To further validate the approach, we also consider synthetic data produced via a finite-difference time-domain (FDTD) electromagnetic wave simulator \cite{warren2016gprmax} for a small-sized scene.
}
We adopt as performance metrics the accuracy of calibration in terms of the errors on the estimated electromagnetic material parameters $\theta$ and on the estimated received signal power.
Furthermore, we evaluate the RT capacity to predict received powers for new coordinates of the receiver that were not covered by the available data.
The source code and \add{simulation assets} of all the presented experiments are available online \cite{repoRuah2023Calibrating}.

\subsection{Setting} \label{subsec:experimental_setting}

Communication between the receive and transmit devices takes place at a central carrier frequency $f^c = 6$~GHz using a bandwidth $B = S \Delta f$ [Hz] composed of $S$ evenly spaced subcarriers, with fixed spacing $\Delta f = 30$ kHz.
All antenna elements are simulated using an isotropic radiation pattern.
Antenna arrays are modelled using evenly spaced antenna elements, with a half-wavelength spacing, and their steering vectors $a^\Rx(\cdot)$ and $a^\Tx(\cdot)$ are computed using the planar wavefront assumption in the far-field regime as presented in \eqref{eq:steering_vector_rx} and \eqref{eq:steering_vector_tx}, respectively.

\begin{figure}
    \centering
    \includegraphics[keepaspectratio, width=3.4in]{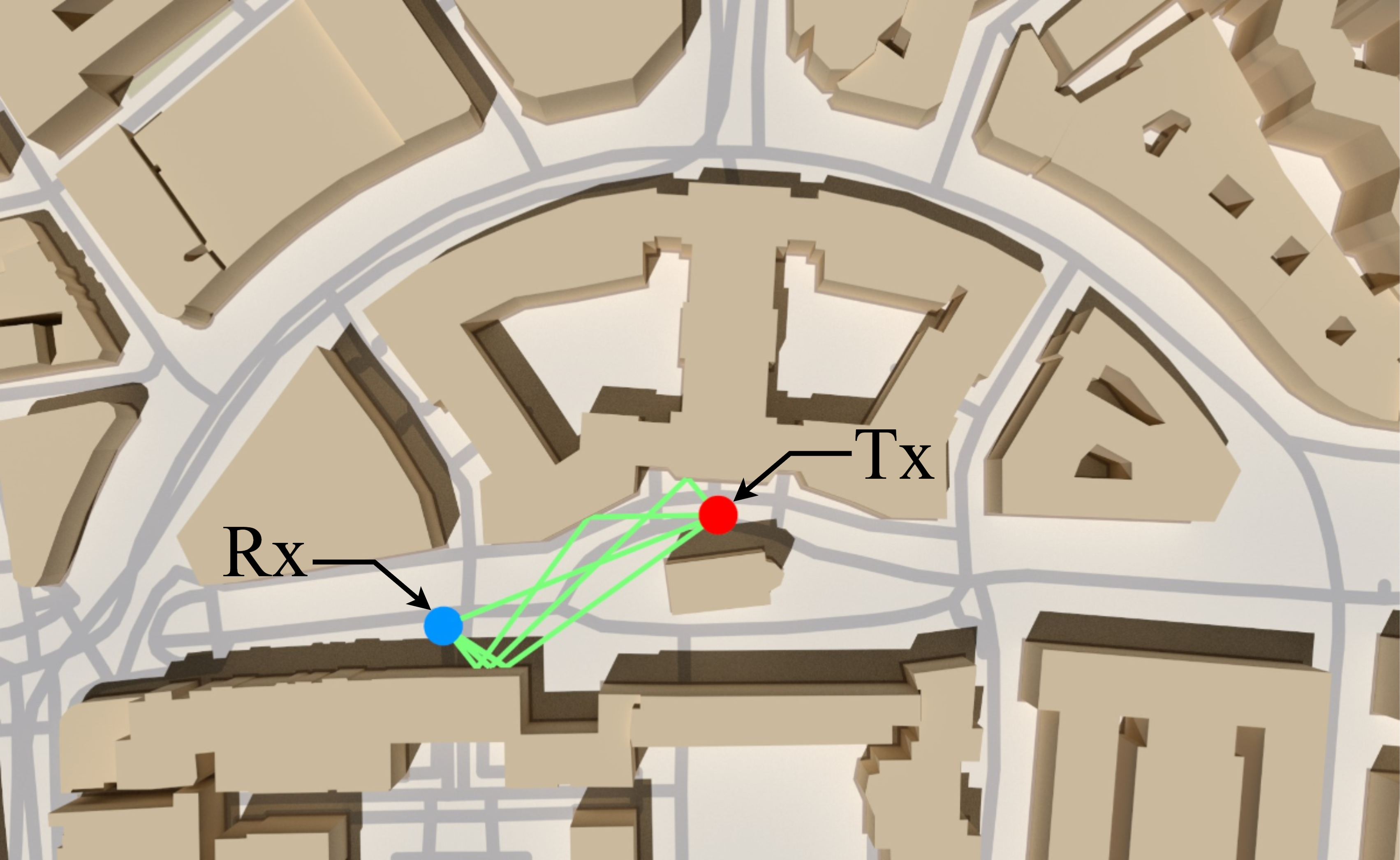}
    \caption{
    3D model of King's College London Strand campus built using data from the OpenStreetMap database \cite{OpenStreetMap}.
    As an illustration of the operation of the ray tracer (RT), we show RT-simulated wireless propagation paths (green lines) for a transmitter (red dot) and receiver (blue dot) pair.
    }
    \label{fig:the_strand}
\end{figure}

\add{
We evaluate the performance of each calibration scheme over a set of three different scenarios comprising: (\emph{i}) the toy-example presented in Fig.~\ref{fig:calibration_among_geometric_errors} (Sec.~\ref{subsec:experiments_toy_example}); (\emph{ii}) a urban scene representing the north side of King's College London Strand campus depicted in Fig. \ref{fig:the_strand} (Sec.~\ref{subsec:experiments_urban_setting}); and (\emph{iii}) a small-sized scene with a metallic blocker, as illustrated in Fig.~\ref{fig:toy_example_fdtd} (Sec.~\ref{subsec:experiments_wave_toy_example}).
}
\add{With the exception of the metallic blocker in the last scenario}, all \add{other} surfaces are assumed to be made of the same \add{concrete-like} material \add{and to share the same electromagnetic properties defined by} a single set of parameters $\theta = \{ \epsilon, \gamma \}$, where $\epsilon \geq 1$ denotes their relative (real) \emph{permittivity} and $\gamma > 0$ denotes their \emph{conductivity} [S/m].
\add{Following the ITU recommendations for metal and concrete materials \cite{itu2019propagation}, we model the metallic blocker as a perfect conductor, with permittivity $1.0$ and conductivity $10^7$ S/m, and define the ground-truth material parameters of the remaining surfaces as $\theta^{\true} = \{ \epsilon^{\true}, \gamma^{\true} \}$, with $\epsilon^{\true} = 5.31$ and $\gamma^{\true} = 0.139$ S/m.}

\add{
The channel observations $(c_n, H_n) \in \mathcal{D}$ at the receive and transmit locations $c_n = (c^\Rx_n, c^\Tx_n)$ are synthetically generated under ground-truth material parameters $\theta^{\true}$, and the channel response is evaluated via RT in scenarios (\emph{i}) and (\emph{ii}), and by solving Maxwell's equations via FDTD in scenario (\emph{iii}).
RT is implemented using the Sionna RT module \cite{hoydis2022sionna, hoydis2023sionna}, which determines the path parameters $R(c_n | \theta^{\true})$ and computes the channel response $H_n$ by using either the deterministic (Sec.~\ref{subsec:deterministic_model}) or the phase error (Sec.~\ref{subsec:phase_error_model}) channel models. 
FDTD uses the gprMax library \cite{warren2016gprmax} to simulate the electric field in time domain at the receiver, obtaining the channel response $H_n$ via an inverse Fourier transform (see Appendix~\ref{apx:fdtd_simulation} for more details).
For both simulation modalities, we add an additive Gaussian noise term with known power $\sigma^2 > 0$ to the generated channel response $H_n$.
}
Accordingly, we define the SNR of the observed data at coordinates $c$ as
\begin{equation}
\label{eq:snr}
    \snr(c) = 
    \frac{
        \pow\left(c | \theta^{\true} \right)
    }{\sigma^2},
\end{equation}
\add{
where $\pow(c | \theta^{\true})$ represents the total received power averaged across all receive-transmit antenna pairs.
}%
\add{
For RT simulations, the overall received power $\pow(c | \theta^{\true}) = \norm{\alpha(\theta^{\true})}^2$ is obtained by summing the} power across all ground-truth propagation paths, \add{while the power $\pow(c | \theta^{\true})$ in FDTD simulations is obtained by integrating the power of the channel response in time domain (see Appendix~\ref{apx:fdtd_simulation}).
}

During calibration, the ground-truth material parameters $\theta^{\true}$ are assumed to be unknown, and we leverage the given dataset $\mathcal{D}$ to calibrate the material parameters estimate $\theta = \{ \epsilon, \gamma \}$.
The geometric model available at the DT presents small-scale differences with respect to the ground-truth scene used to generate the data.
These differences are generated either by modifying the geometry of the ground-truth scene \add{in scenarios (\emph{i}) and (\emph{iii})}, as illustrated in Fig.~\ref{fig:scenario_simulation}, or by injecting \add{position errors (Sec.~\ref{subsubsec:receiver_position_mismatch}) or} phase errors \add{(Sec.~\ref{subsubsec:independent_phase_errors})} in the simulated paths responses \add{in scenario (\emph{ii})}.

Once the parameters $\theta = \{ \epsilon, \gamma \}$ are calibrated using one of the three schemes described in Sec.~\ref{sec:baselines} and Sec.~\ref{sec:peac}, we assess the relative performance of calibration in terms of: the normalized permittivity error  $\abs{\epsilon - \epsilon^{\true}}/\epsilon^{\true}$; the normalized conductivity error $\abs{\gamma - \gamma^{\true}}/\gamma^{\true}$; and the normalized received power error
\begin{equation}
\label{eq:power_error}
    \frac{
        \Abs{
            \hatpow\left(c | \theta \right)
            - \pow\left(c | \theta^{\true} \right)
        }
    }{
        \pow\left(c | \theta^{\true} \right)
    },
\end{equation}
where $\hatpow(c | \theta) = \norm{\alpha(\theta)}^2$ is the total power received at position $c$ across all propagation paths estimated using the DT model with calibrated parameters $\theta$.

For all calibration schemes, we initialize the parameters $\theta$ to be calibrated at $\epsilon^{(0)} = 3$ and $\gamma^{(0)} = 0.1$ S/m.
Throughout the experiments, the phase error-aware scheme (Sec.~\ref{sec:peac}) optimizes its prior concentration parameter $\kPrior$ as per the alternative M-step objective \eqref{eq:alternative_vem_m_step} with initialization at $\kPrior^{(0)} = 0$.
The uniform phase error scheme (Sec.~\ref{subsec:upec}) uses a set of $M=P$ projections in \eqref{eq:upec_loss} defined by the delays and angles $\{(\tau_p, \AoD_p, \AoA_p)\}_{p=1}^{P}$ of the $P$ paths computed by the RT using the DT model.


\subsection{Toy Example} \label{subsec:experiments_toy_example}

Following the discussion in Sec. \ref{sec:introduction}, we first compare the presented calibration methods for the toy example presented in Fig.~\ref{fig:calibration_among_geometric_errors}, with the receiver and transmitter modeled as single isotropic antennas located on the plane at positions $c^\Rx = (240, 0)^\top$ and $c^\Tx = (-240, 0)^\top$, respectively, and the two horizontal walls intersecting the vertical axis at positions $(0, 100)^\top$ and $(0, -180)^\top$, where the coordinates represent multiples of the carrier wavelength $\lambda^c \approx 5$ cm.
To model a two-dimensional scenario, we discard the paths that reflect on the floor, and focus exclusively on the two non-line of sight specular paths displayed in Fig.~\ref{fig:calibration_among_geometric_errors}, which are caused by reflections on the two horizontal walls.
Accordingly, the two propagation paths interfere constructively at the receiver coordinates, with a time of arrival difference of $\Delta \tau \approx 13.3$ ns.

We consider a set of $N=50$ synthetically generated channel observations $\{ H_n \}_{n=1}^{N} \iid \CN(\Hhat(c | \theta^{\true}), \sigma^2)$ with $\sigma^2 > 0$, where $\Hhat(c | \theta^{\true})$ is obtained from the deterministic model in \eqref{eq:deterministic_channel_model} by running the RT at the receive and transmit locations $c = (c^\Rx, c^\Tx)$.
During calibration, we adopt the geometric model available at the DT which, as illustrated in Fig.~\ref{fig:scenario_simulation}, presents a small-scale discrepancy with respect to the ground-truth geometry in the form of a shift of the position of the lower wall by an amount of $\Delta d = 0.4 \lambda^c$ m.
Accordingly, the lower wall intercepts the vertical axis at coordinates $(0, -180.4)$ in the DT model, and the two simulated paths interfere destructively at the receiver.

\begin{figure}
    \centering
    \begin{subfigure}{3.4in}
        \centering
        \hspace{0.4in}\includegraphics[keepaspectratio=true, width=3in]{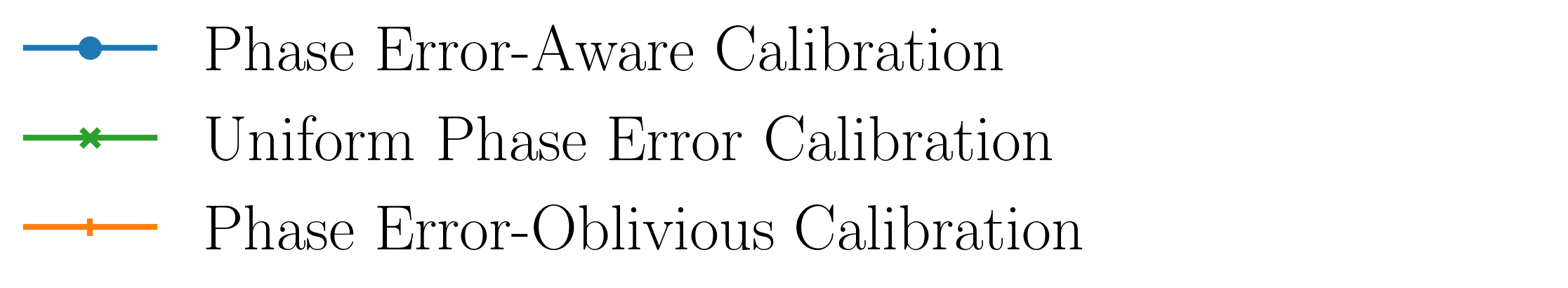}
    \end{subfigure}
    \centering
    \begin{subfigure}{3.4in}
        \centering
        \includegraphics[keepaspectratio=true, width=3.4in]{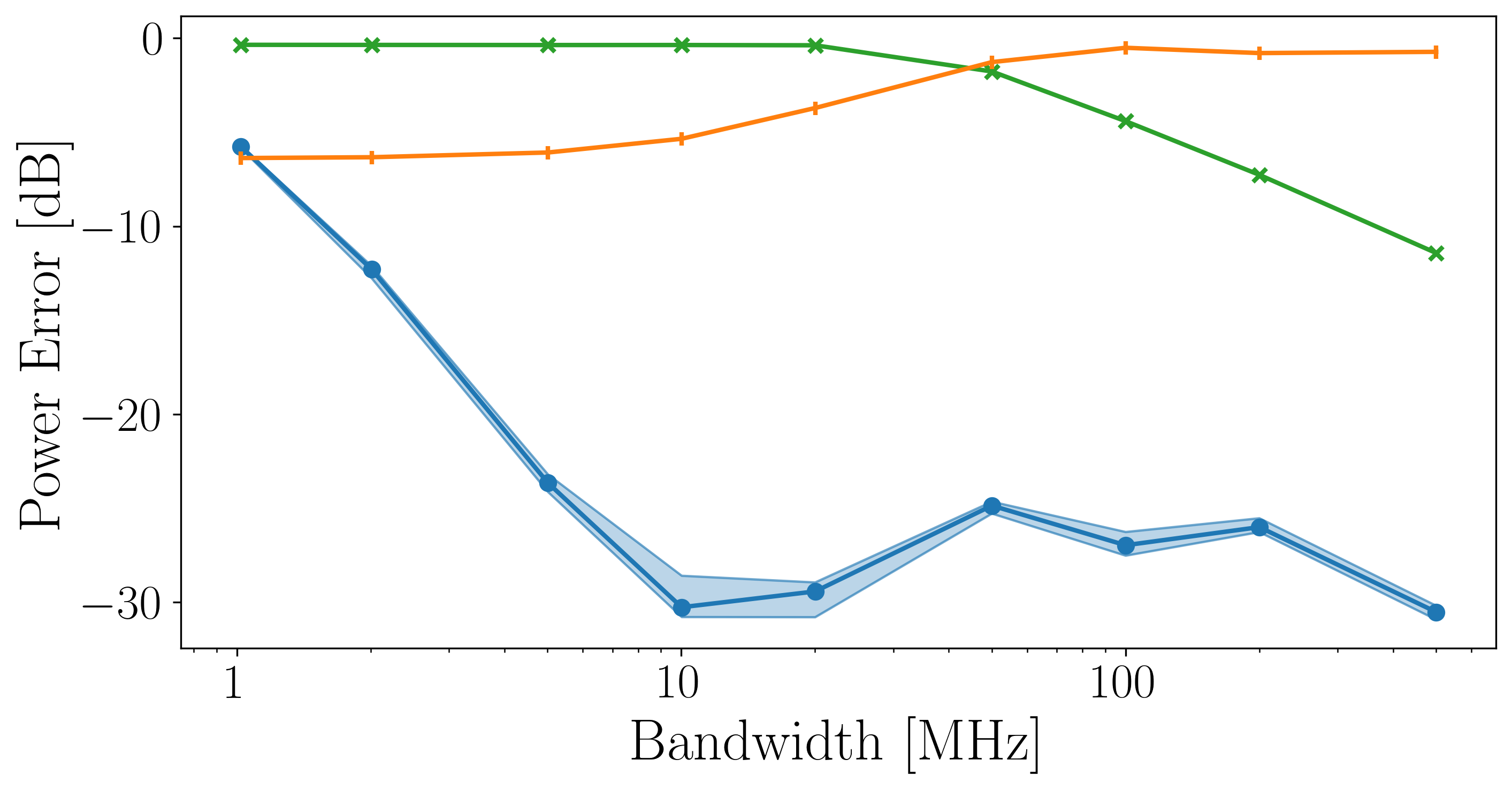}
    \end{subfigure}
    \caption{
    Relative power estimation errors at the position $c$ used to collect data for phase error-oblivious calibration (Sec. \ref{subsec:peoc}), uniform phase error calibration (Sec. \ref{subsec:upec}), and for the proposed phase error-aware calibration (Sec. \ref{sec:peac}) as a function of the available bandwidth $B$, with a signal-to-noise (SNR) ratio of $20$~dB.
    Lines represent the median error across ten independent channel observation and calibration runs.
    Shaded areas represent the first and third quartiles.
    }
    \label{fig:calibration_error_vs_bandwidth}
\end{figure}

We plot the power estimation errors of all schemes as a function of the bandwidth $B$ in Fig. \ref{fig:calibration_error_vs_bandwidth} for an SNR equal to $20$ dB.
To scale the bandwidth $B$, the subcarrier spacing $\Delta f = 30$ kHz is kept constant, while the number of subcarriers $S$ is increased.
Confirming the insights in Fig.~\ref{fig:calibration_among_geometric_errors}, the performance of the power profile-based scheme starts improving once the system bandwidth approaches $B = 1 / \Delta \tau = 75$ MHz, i.e., once its temporal resolution is high enough to separate the contribution of each path based on their arrival times at the receiver.  
Accordingly, the relative power estimation error of the uniform phase error scheme decreases from $0$ dB at bandwidths lower than $B < 20$ MHz, to $-11$ dB at bandwidth $B = 500$ MHz.
As the number of subcarriers increases for larger bandwidths, the phase error-oblivious scheme is affected by a larger number of phase errors in the path responses simulated using the DT model, yielding an error increase from $-6$ dB at lower bandwidths $B < 5$ MHz, to $-1$ dB at $B = 500$ MHz.
On the contrary, owing to the larger number of subcarriers at higher bandwidths, the phase error-aware scheme has access to more information to estimate the phase errors, outperforming the phase error-oblivious and uniform phase error schemes for bandwidths of $B = 2$ MHz and above.
The relative power error of the phase error-aware scheme stabilizes around $-27$ dB for bandwidths larger than $B > 50$ MHz, which correspond to time resolutions at which the system can resolve the two propagation paths.

\subsection{Urban Setting} \label{subsec:experiments_urban_setting}

Throughout the rest of this section, we study a more realistic scenario that simulates radio propagation \add{via RT} on the north side of King's College London Strand campus, whose 3D representation, depicted in Fig. \ref{fig:the_strand}, is obtained using geo-spatial data from the OpenStreetMap database \cite{OpenStreetMap}.
We now consider $8 \times 8$ two-dimensional antenna arrays at both receive and transmit locations, and we fix the number of subcarriers to $S = 64$.
Non-line of sight propagation paths are simulated through specular reflection only, with up to ten reflections per path.
We leave the evaluations under diffracted, refracted and diffusely reflected paths for future work.


To enable calibration, we use synthetically generated channel observations for a single pair of receiver and transmitter devices at coordinates $c_{\rmcal} = (c^{\Rx}_\rmcal, c^{\Tx}_\rmcal)$ depicted in Fig.~\ref{fig:the_strand}, with the receiver and transmitter at heights $1$ and $5$ meters, respectively.
Specifically, we generate a synthetic dataset $\mathcal{D} = \{ (c_{\rmcal}, H_n) \}_{n=1}^{N}$ of $N=50$ channel observations by running the RT in the same 3D model available at the DT with ground-truth material parameters $\theta^\true$.
We account for the discrepancy between the ground-truth (simulated) deployment and the virtual geometry assumed by the DT in one of the following ways.


\subsubsection{Receiver position mismatch}\label{subsubsec:receiver_position_mismatch}

A first approach is to directly model small geometric discrepancies by assuming that the position of the receiver is subject to uncertainty.
Each channel observation $H_n$ for $n \in \dset{N}$ is obtained by placing the receiver at the position 
\begin{equation}
\label{eq:rx_pos_error}
    \Tilde{c}^\Rx_n = c^\Rx_\rmcal + \eta^{\true} \lambda^c u_n,
\end{equation}
where $\lambda^c \approx 5$ cm is the carrier wavelength, $u_n \in \R^{3 \times 1}$ is a uniformly sampled direction in the unit sphere, i.e., $\norm{u_n} = 1$, and $\eta^{\true} \in [0, 0.5]$ dictates the level of geometric discrepancy, with $\eta^{\true} = 0$ indicating a perfect geometric model at the DT.
Accordingly, the position $c^\Rx_\rmcal$ assumed by the DT differs from the true position $\Tilde{c}^\Rx_n$ of observation $H_n$ due to the random displacement $\eta^{\true} \lambda^c u_n$.

 

\subsubsection{Independent phase errors}\label{subsubsec:independent_phase_errors}

The stochastic phase errors caused by the geometric displacement \eqref{eq:rx_pos_error} are generally correlated across different paths, since they stem from the same uncertainty.
In order to capture a worst case-scenario in terms of phase errors caused by a geometric mismatch between the DT and the PT, we also study a setting with independent phase errors.
To this end, observations $\{H_n\}_{n=1}^{N}$ are generated as independent samples from the phase-error model $\Hhat(c_{\rmcal} | \theta^{\true}, \kPrior^{\true})$ in \eqref{eq:peac_measurement_model}.
The ground-truth concentration parameter $\kPrior^{\true}$ controls the level of geometric discrepancy, with $\kPrior^{\true} = 0$ corresponding to the most challenging case of uniform phase errors, and $\kPrior^{\true} \to +\infty$ corresponding to a perfect DT model.

For both data generation methods, the SNR is defined as in \eqref{eq:snr}, with the caveat that the power $\pow(c | \theta^{\true})$ at the numerator is averaged over the positions given by the random displacements in \eqref{eq:rx_pos_error} for the former method.

\begin{figure*}
    \centering
    \begin{subfigure}{\textwidth}
        \centering
        \includegraphics[width=\textwidth]{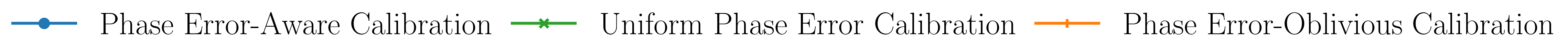}
    \end{subfigure}
    \begin{subfigure}{0.49\textwidth}
        \centering
        \includegraphics[width=\textwidth]{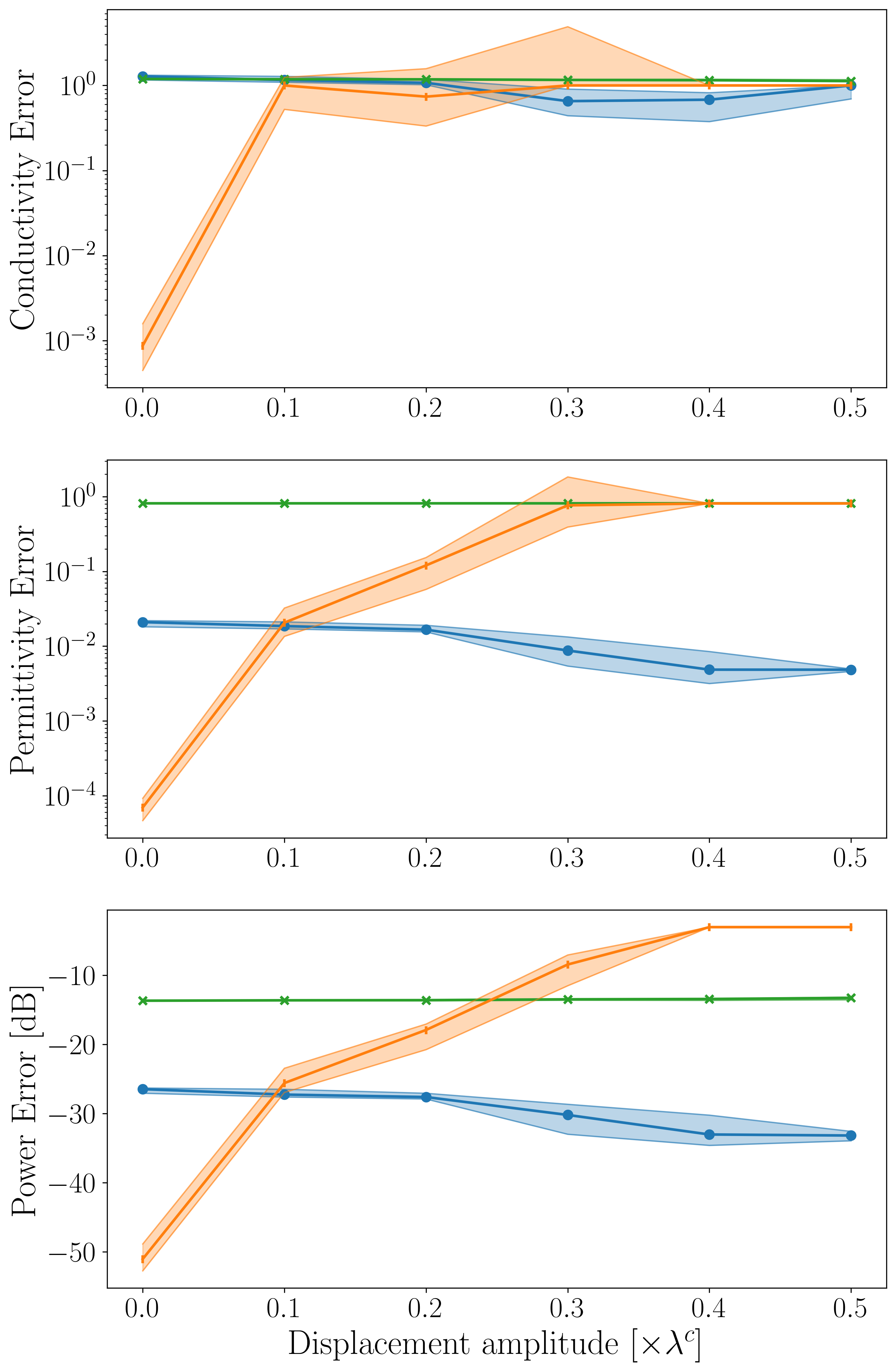}
        \vspace{-0.6cm}
        \caption{}
        \label{fig:calibration_error_vs_rx_displacement}
    \end{subfigure}
    \begin{subfigure}{0.49\textwidth}
        \centering
        \includegraphics[width=\textwidth]{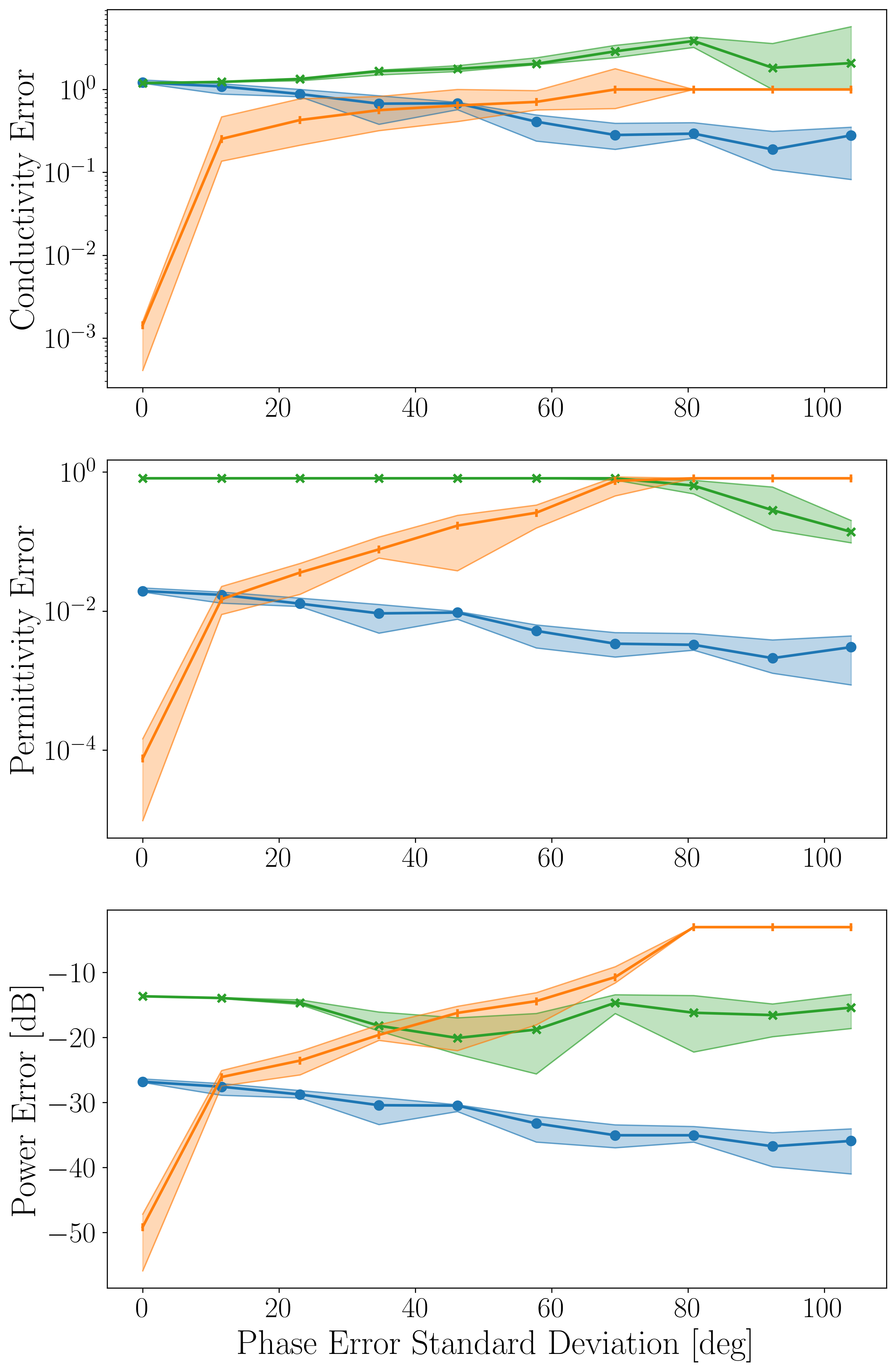}
        \vspace{-0.6cm}
        \caption{}
        \label{fig:calibration_error_vs_std}
    \end{subfigure}
    \caption{
    Calibration relative errors in terms of estimation of the electromagnetic properties (top) and of power estimation at the receiver collecting data (bottom) for phase error-oblivious calibration (Sec. \ref{subsec:peoc}), uniform phase error calibration (Sec. \ref{subsec:upec}), and for the proposed phase error-aware calibration (Sec. \ref{sec:peac}) as a function of the magnitude of the random receiver displacements (a), and as a function of the standard deviation of the independently sampled phase errors (b).
    Lines represent the median error across ten independent channel observation and calibration runs, with a signal-to-noise ratio (SNR) of $20$ dB.
    Shaded areas represent the first and third quartiles.
    }
    \label{fig:calibration_error}
\end{figure*}

\subsection{Calibration Performance} \label{subsec:experiments_calibration_performance}

We start by comparing the three considered calibration schemes for geometric discrepancies simulated through receiver position uncertainty (Fig.~\ref{fig:calibration_error_vs_rx_displacement}), and through independent phase errors (Fig.~\ref{fig:calibration_error_vs_std}), for an SNR of $20$ dB.
Using the three metrics described in Sec.~\ref{subsec:experimental_setting}, Fig.~\ref{fig:calibration_error} displays the performance of each scheme in terms of normalized errors of the predicted permittivity $\epsilon$, conductivity $\gamma$, and received power $\hatpow(c_{\rmcal} | \theta)$ at coordinates $c_{\rmcal}$.
The standard deviation in Fig.~\ref{fig:calibration_error_vs_std} is a function of the ground-truth concentration parameter $\kPrior^{\true}$, and is chosen here as it is easier to interpret as compared to the concentration parameter.
Note that, in the uniform phase error case, the standard deviation equals $ \pi / \sqrt{3} \mbox{ rad} \approx 104 \mbox{ deg}$.

A first general observation is that the three considered calibration schemes display similar relative performance gaps for both discrepancy models.
This motivates the adoption of the worst case of independently and uniformly distributed ($\kPrior^{\true} = 0$) phase errors at each path for the following experiments.
Furthermore, from the power error plots in the bottom row of Fig. \ref{fig:calibration_error}, we observe that the proposed phase error-aware scheme scheme outperforms the uniform phase error baseline across all considered random receiver displacements amplitudes and phase error standard deviation levels, while also performing better than the phase error-oblivious scheme for displacements amplitudes above $\eta^{\true} = 0.1$ and phase error standard deviations of $20$ degrees and higher.
In particular, in the most common regime of uniform phase errors, i.e., at a standard deviation of approximately $104$ degrees, the proposed scheme outperforms the two baselines in terms of power error, with a performance gap of about $20$ dB when compared to the uniform phase error scheme, and about $32$ dB when compared to the phase error-oblivious scheme.
We also observe from the two top rows of Fig~\ref{fig:calibration_error} that an accurate estimate of the true material parameters $\theta$ is not a pre-requisite to achieve reliable power estimation, as multiple combinations of permittivity $\epsilon$ and conductivity $\gamma$ values can result in the same predicted power.

\begin{figure}
    \centering
    \begin{subfigure}{3.4in}
        \centering
        \hspace{0.4in}\includegraphics[keepaspectratio=true, width=3in]{legend_single_column}
    \end{subfigure}
    \centering
    \begin{subfigure}{3.4in}
        \centering
        \includegraphics[keepaspectratio=true, width=3.4in]{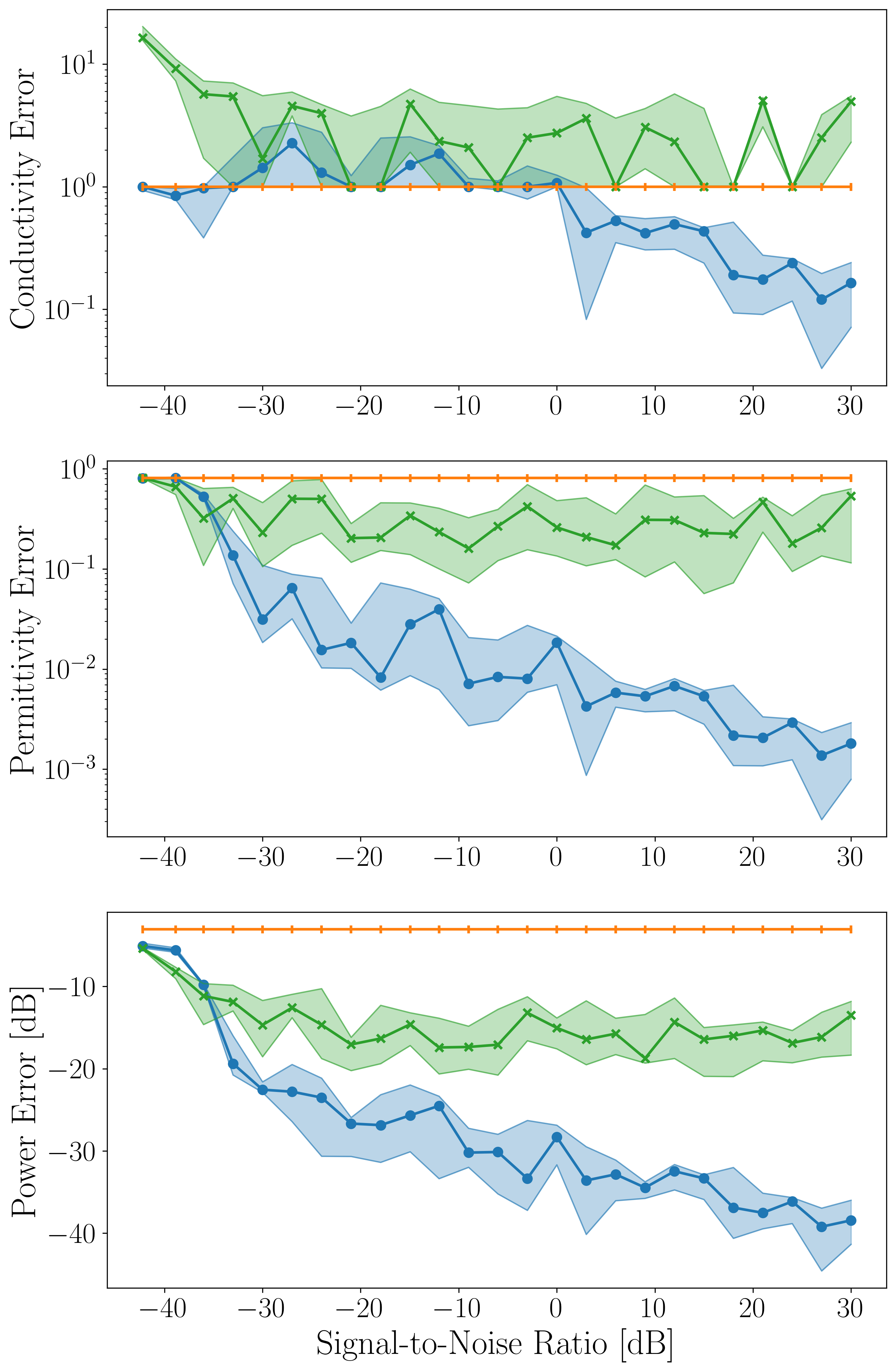}
    \end{subfigure}
    \caption{
    Calibration relative errors in terms of estimation of the electromagnetic properties (top) and of power estimation at the receiver collecting data (bottom) for phase error-oblivious calibration (Sec. \ref{subsec:peoc}), uniform phase error calibration (Sec. \ref{subsec:upec}), and for the proposed phase error-aware calibration (Sec. \ref{sec:peac}) as a function of the SNR for uniformly sampled independent phase errors ($\kPrior^{\true} = 0$).
    Lines represent the median error across ten independent channel observation and calibration runs.
    Shaded areas represent the first and third quartiles.
    }
    \label{fig:calibration_error_vs_snr}
\end{figure}

Another general observation is that the proposed phase error-aware scheme is able to benefit from the increased diversity in the data set produced by larger random displacements, in Fig.~\ref{fig:calibration_error_vs_rx_displacement}, or larger phase error standard deviation, in Fig.~\ref{fig:calibration_error_vs_std}.
In fact, the figure illustrates how phase error-aware calibration yields improved estimates of material parameters as the displacement amplitude and the phase error standard deviations increase, which, in turn, also improves the power estimation.
In this regard, the diversity in the calibration data set resulting from randomness in the geometry or path phases has the effect of mitigating the inherent limited resolution arising from a limited bandwidth by allowing the observation of different randomized path combinations in the $N$ channel observations.
For example, with a single observation, i.e., $N = 1$ (not shown), the power error remains around $-31 \pm 1$ dB across all phase error standard deviations, while with $N = 50$, as illustrated in Fig.~\ref{fig:calibration_error_vs_std}, the power error drops from $-27$ dB when there are no phase errors to $-36$ dB for uniform phase errors.

Fig.~\ref{fig:calibration_error_vs_snr} displays the calibration performance of all schemes under uniform ground-truth phase errors as a function of the SNR.
The performance gains of the proposed phase error-aware scheme range from around $13$ dB at an SNR of $0$ dB to around $25$ dB at an SNR of $30$ dB in terms of power error.
Moreover, the phase error-oblivious calibration scheme is unable to correctly calibrate its material parameters under uniform phase errors, yielding inaccurate material and power estimates at the calibration location $c_{\rmcal}$ for all studied SNR levels, while the best benchmark in terms of power estimation is given by the uniform phase error scheme.

\subsection{Generalization Performance for Received Power Prediction}

We now study the implications of calibration on the operations of the DT for the urban scenario in Sec.~\ref{subsec:experiments_urban_setting}.
Specifically, we consider the task of predicting received power at positions of the receiver that are different from that used to collect the calibration data set. This allows the DT to build power maps, which can be used for functionalities such as network optimization and resource allocation. 

To this end, we compare the average signal power predicted by the RT using the calibrated parameters $\theta$ to the average ground-truth signal power obtained by using the ground-truth parameters $\theta^{\true}$ for all receiver positions within coverage of the transmitter.
Calibration data $\mathcal{D}$ is generated using independently sampled phase errors at each path, as presented in Sec.~\ref{subsubsec:independent_phase_errors}, with concentration parameter $\kPrior^{\true} \geq 0$.
We emphasize that the transmitter's position $c^{\Tx}_{\rmcal}$ is fixed across calibration and testing, while the position of the receiver $c^{\Rx}_{\rmcal}$ displayed in Fig. \ref{fig:the_strand} describes the conditions used to collect the calibration data.

\begin{figure*}
    \centering
    \begin{subfigure}{0.03\textwidth}
        \raisebox{1.25cm}{\rotatebox[origin=t]{90}{\small{No Phase Error}}}
    \end{subfigure}
    \begin{subfigure}{0.31\textwidth}
        \captionsetup{margin={-1cm, 0cm}}
        \centering
        \hspace{-1cm}\small{Phase Error-Aware Calibration}
        \includegraphics[width=\textwidth]{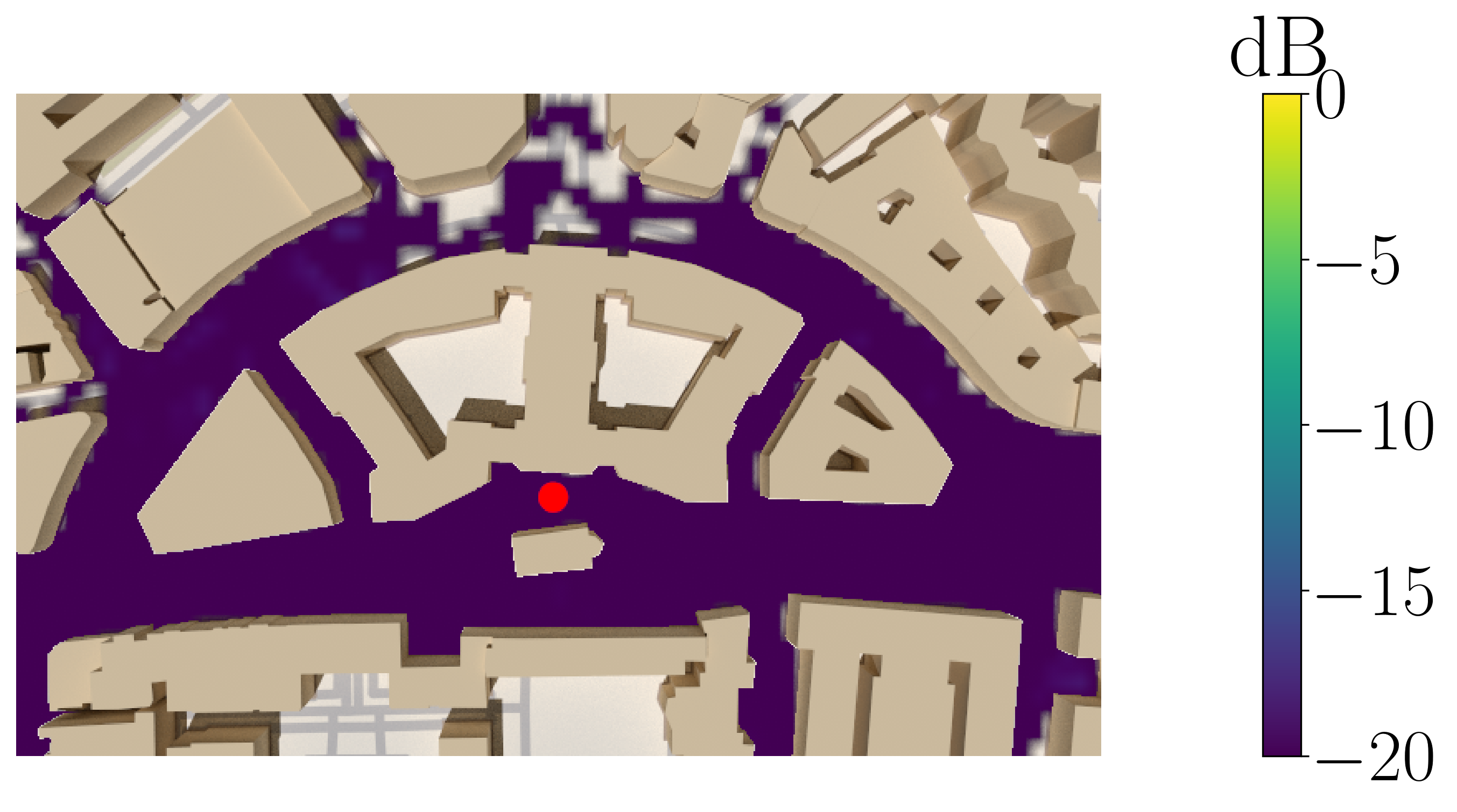}
        \vspace{-0.7cm}
        \caption{}
        \label{fig:power_map_peac_no_phase_error}
    \end{subfigure}
    \begin{subfigure}{0.31\textwidth}
        \captionsetup{margin={-1cm, 0cm}}
        \centering
        \hspace{-1cm}\small{Phase Error-Oblivious Calibration}
        \includegraphics[width=\textwidth]{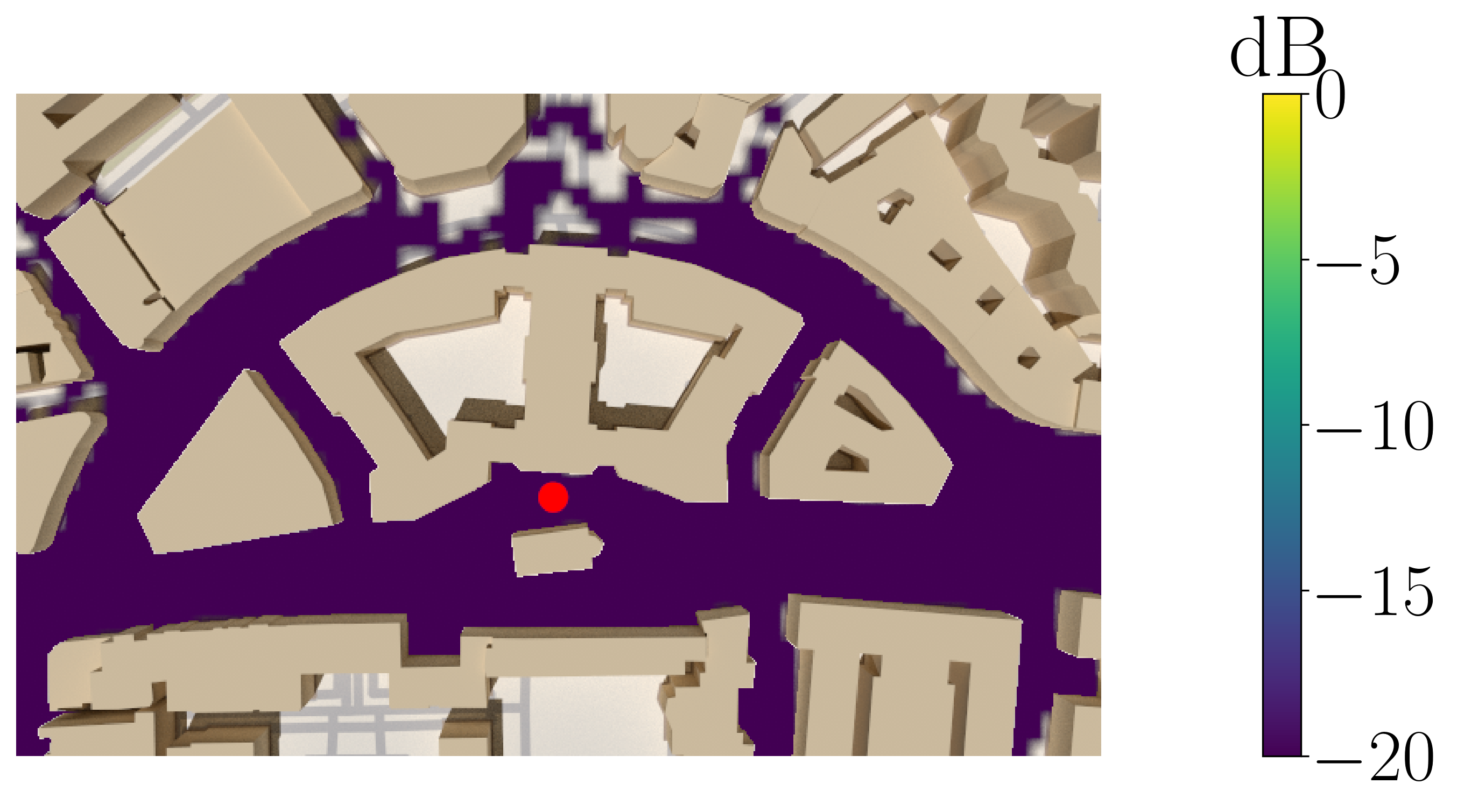}
        \vspace{-0.7cm}
        \caption{}
        \label{fig:power_map_peoc_no_phase_error}
    \end{subfigure}
    \begin{subfigure}{0.31\textwidth}
        \captionsetup{margin={-1cm, 0cm}}
        \centering
        \hspace{-1cm}\small{Uniform Phase Error Calibration}
        \includegraphics[width=\textwidth]{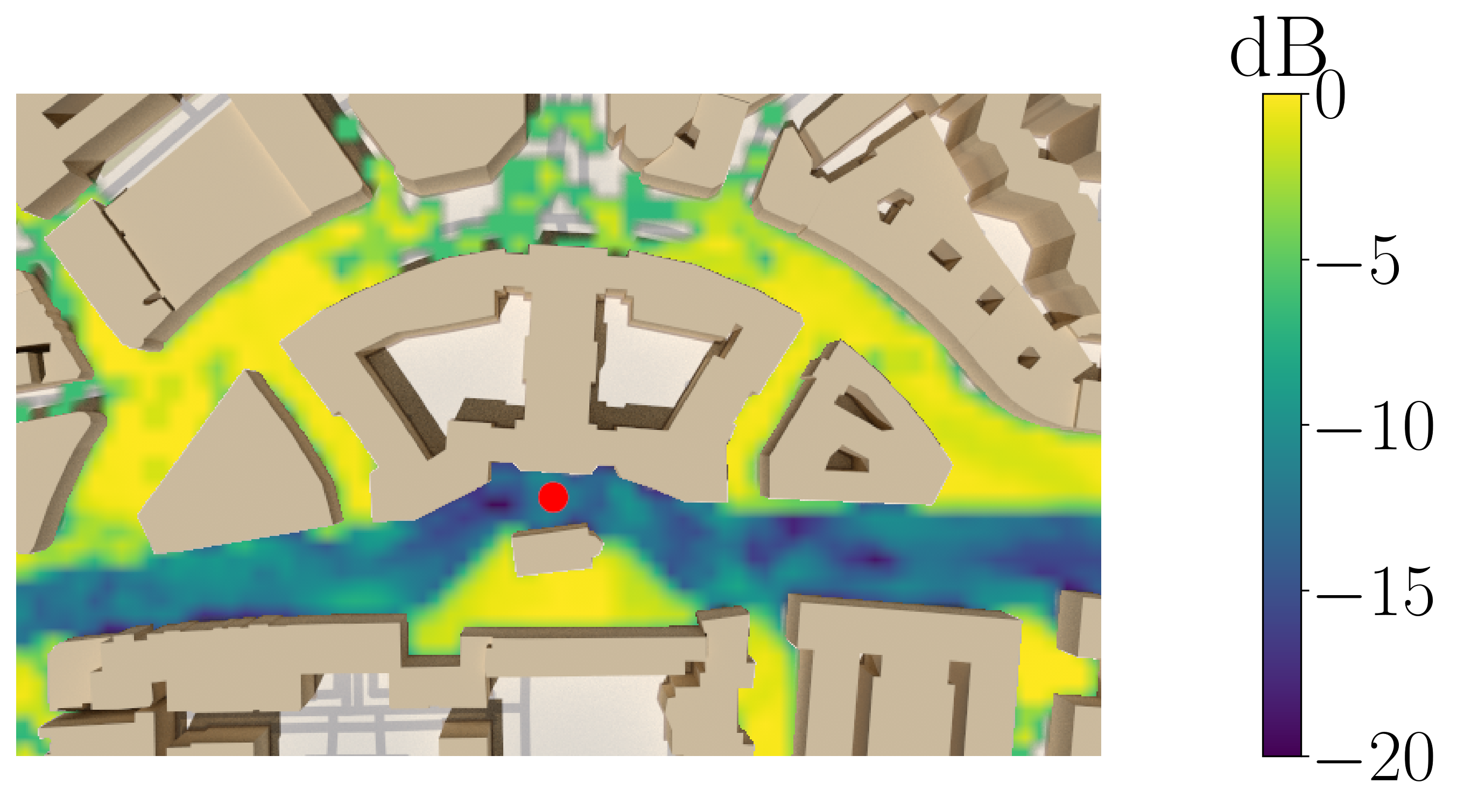}
        \vspace{-0.7cm}
        \caption{}
        \label{fig:power_map_upec_no_phase_error}
    \end{subfigure}
    \hfill
    \begin{subfigure}{0.03\textwidth}
        \raisebox{1.25cm}{\rotatebox[origin=t]{90}{\small{Uniform Phase Error}}}
    \end{subfigure}
    \begin{subfigure}{0.31\textwidth}
        \captionsetup{margin={-1cm, 0cm}}
        \centering
        \includegraphics[width=\textwidth]{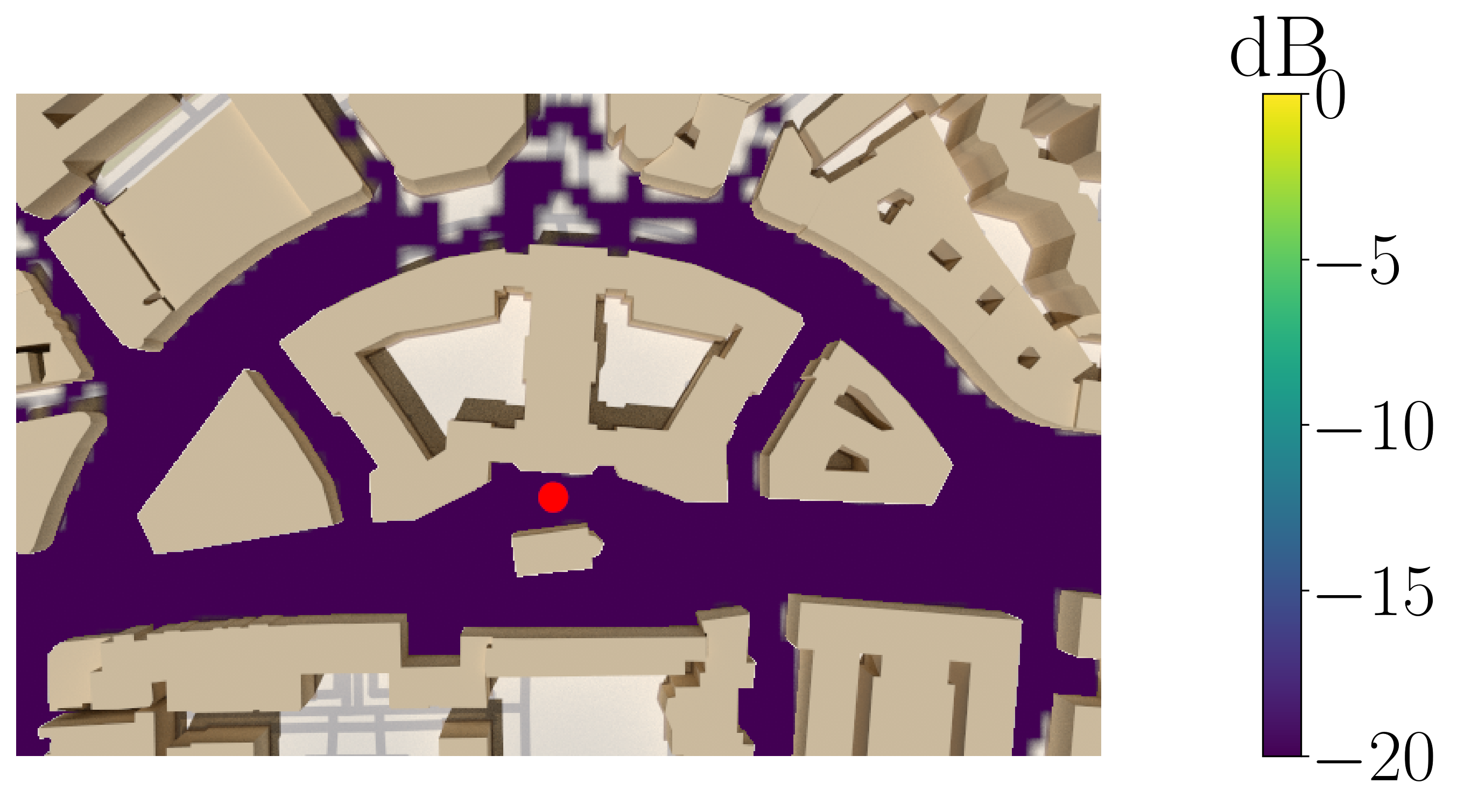}
        \vspace{-0.7cm}
        \caption{}
        \label{fig:power_map_peac_uniform_phase_error}
    \end{subfigure}
    \begin{subfigure}{0.31\textwidth}
        \captionsetup{margin={-1cm, 0cm}}
        \centering
        \includegraphics[width=\textwidth]{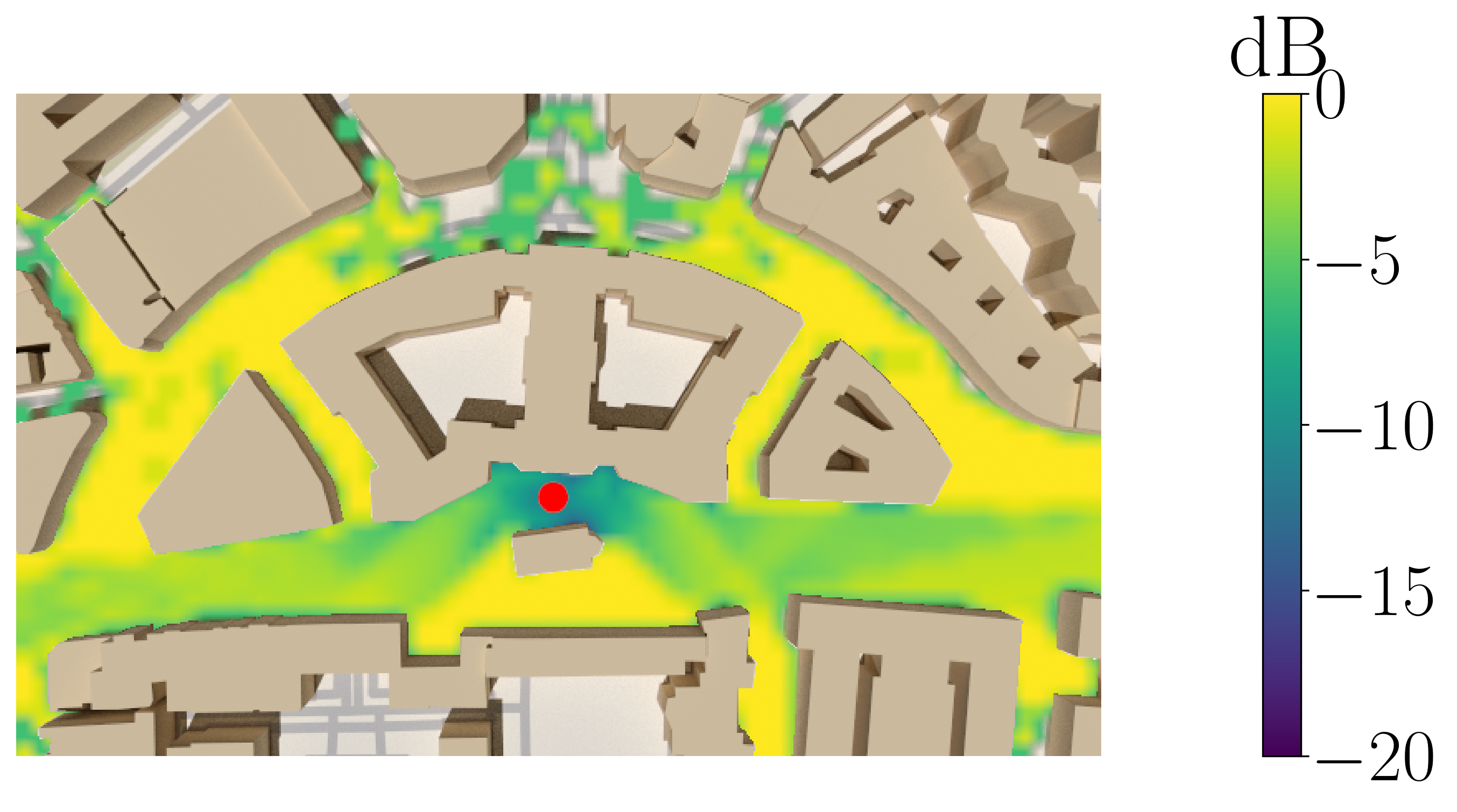}
        \vspace{-0.7cm}
        \caption{}
        \label{fig:power_map_peoc_uniform_phase_error}
    \end{subfigure}
    \begin{subfigure}{0.31\textwidth}
        \captionsetup{margin={-1cm, 0cm}}
        \centering
        \includegraphics[width=\textwidth]{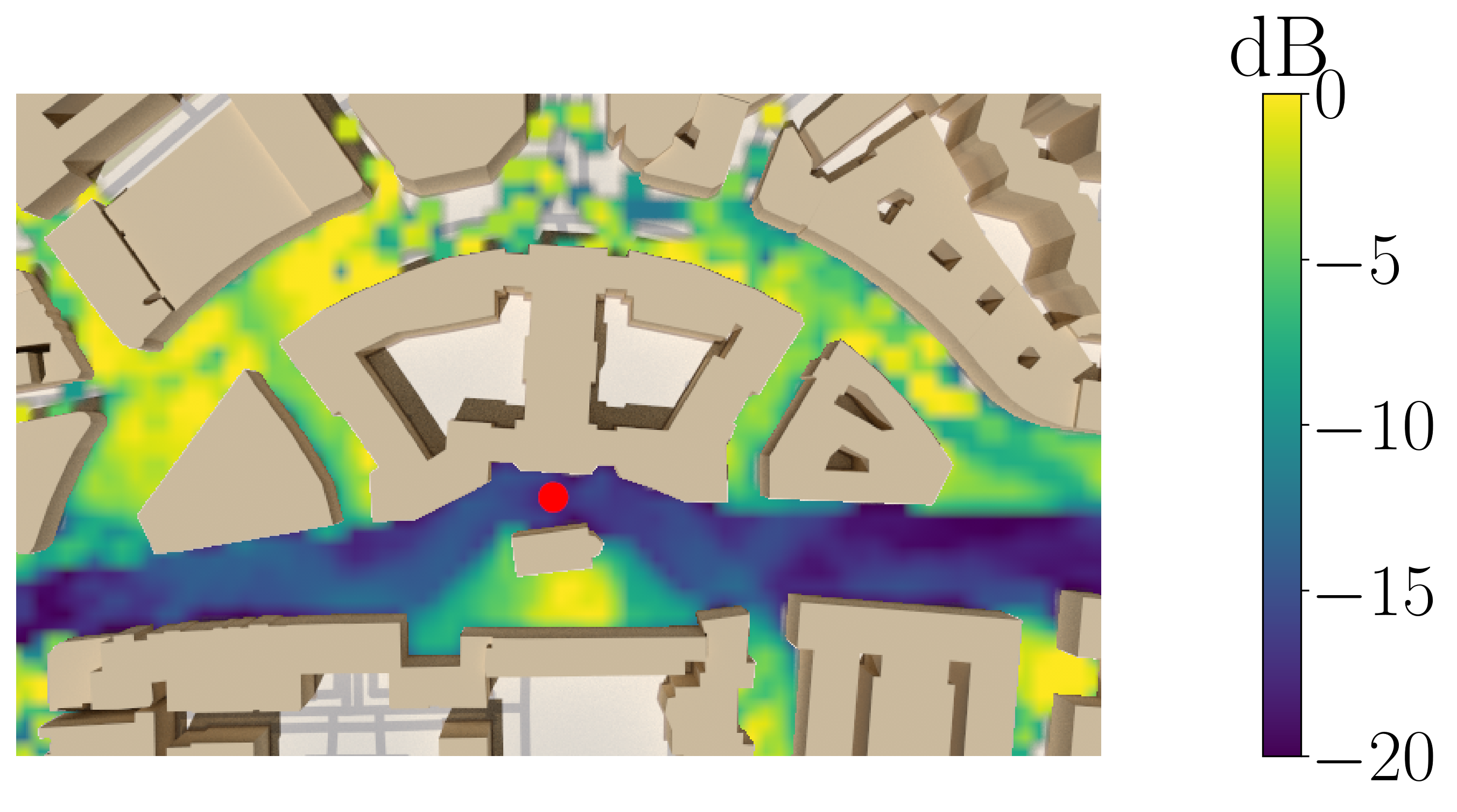}
        \vspace{-0.7cm}
        \caption{}
        \label{fig:power_map_upec_uniform_phase_error}
    \end{subfigure}
    \caption{
    Relative estimation errors (in dB) for the average predicted power of the signal sent by a single transmitter (red dot) under phase error-aware calibration ((a) and (d)), phase error-oblivious calibration ((b) and (e)), and uniform phase error calibration ((c) and (f)) in the absence of phase errors $\kPrior^{\true} \to +\infty$ ((a), (b) and (c)), and for uniform phase errors $\kPrior^{\true} = 0$ ((d), (e) and (f)).
    The estimation errors at each position are averaged across ten independent channel observation and calibration procedures, with a signal-to-noise ratio (SNR) of $20$ dB during calibration.
    }
    \label{fig:power_maps}
\end{figure*}

Given a new receiver position $c^{\Rx}_{\new}$, the ground-truth received power $\pow(c_{\new} | \theta^{\true})$ and the RT-predicted signal power $\hatpow(c_{\new} | \theta)$ in \eqref{eq:power_error} are computed at receiver and transmitter position $c_{\new} = (c^{\Rx}_{\new}, c^{\Tx}_{\rmcal})$.
Accordingly, Fig. \ref{fig:power_maps} displays the normalized power prediction error \eqref{eq:power_error} between predicted and true received powers in dB as a function of the position of the receiver.
For each row in Fig. \ref{fig:power_maps}, the relative power error is averaged across a set of ten independent channel observation and calibration procedures with SNR equal to $20$ dB.
The top row in Fig. \ref{fig:power_maps} shows the power prediction error in the absence of phase errors during calibration, i.e., $\kPrior^{\true} \to +\infty$, while the bottom row displays the power errors under uniformly distributed phase errors, i.e., $\kPrior^{\true} = 0$.

\begin{figure*}
    \centering
    \begin{subfigure}{\textwidth}
        \centering
        \includegraphics[width=\textwidth]{legend_double_column}
    \end{subfigure}
    \begin{subfigure}{0.49\textwidth}
        \centering
        \includegraphics[width=\textwidth]{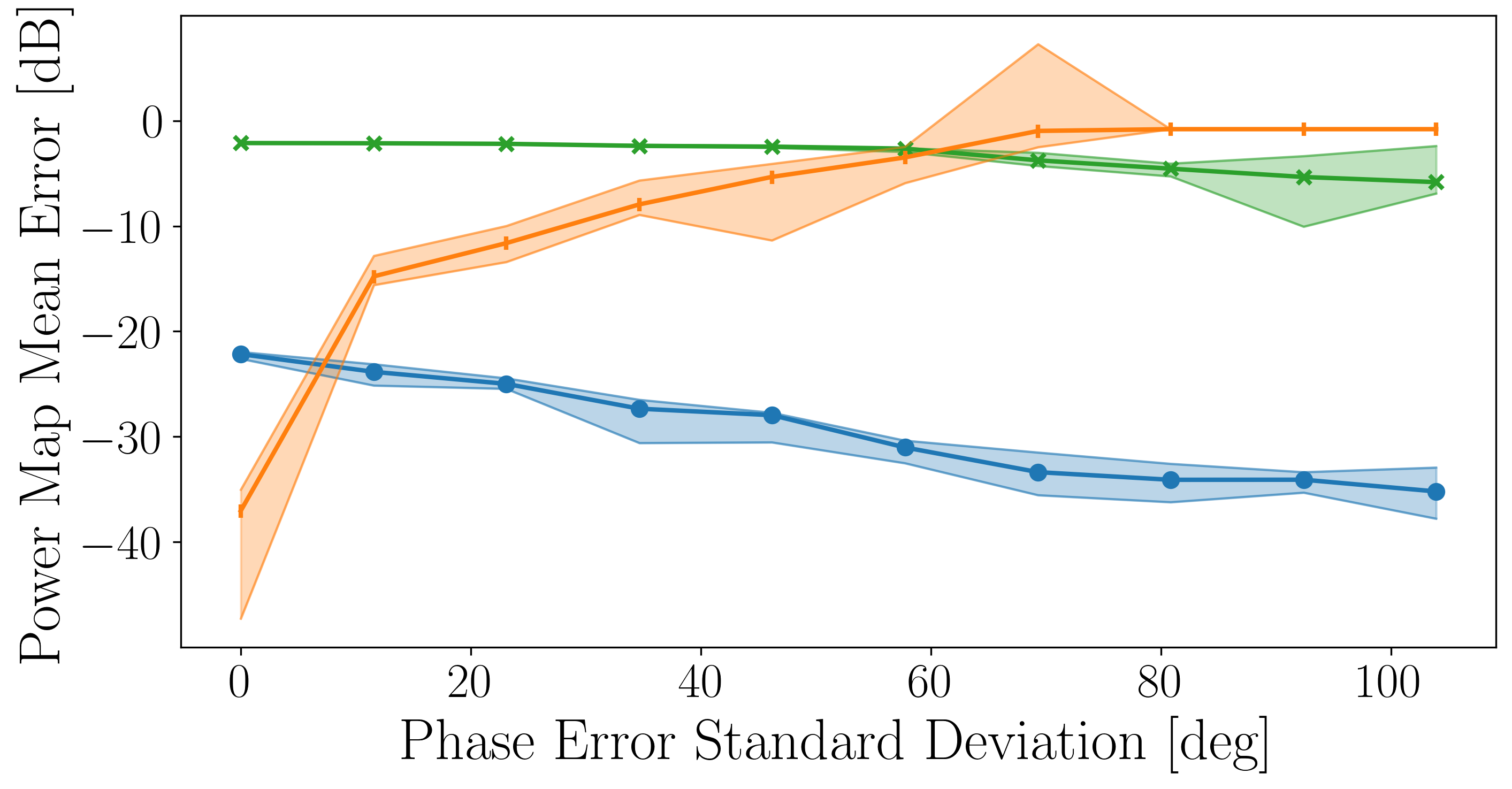}
        \vspace{-0.7cm}
        \caption{}
        \label{fig:power_map_generalization_error_vs_std}
    \end{subfigure}
    \begin{subfigure}{0.49\textwidth}
        \centering
        \includegraphics[width=\textwidth]{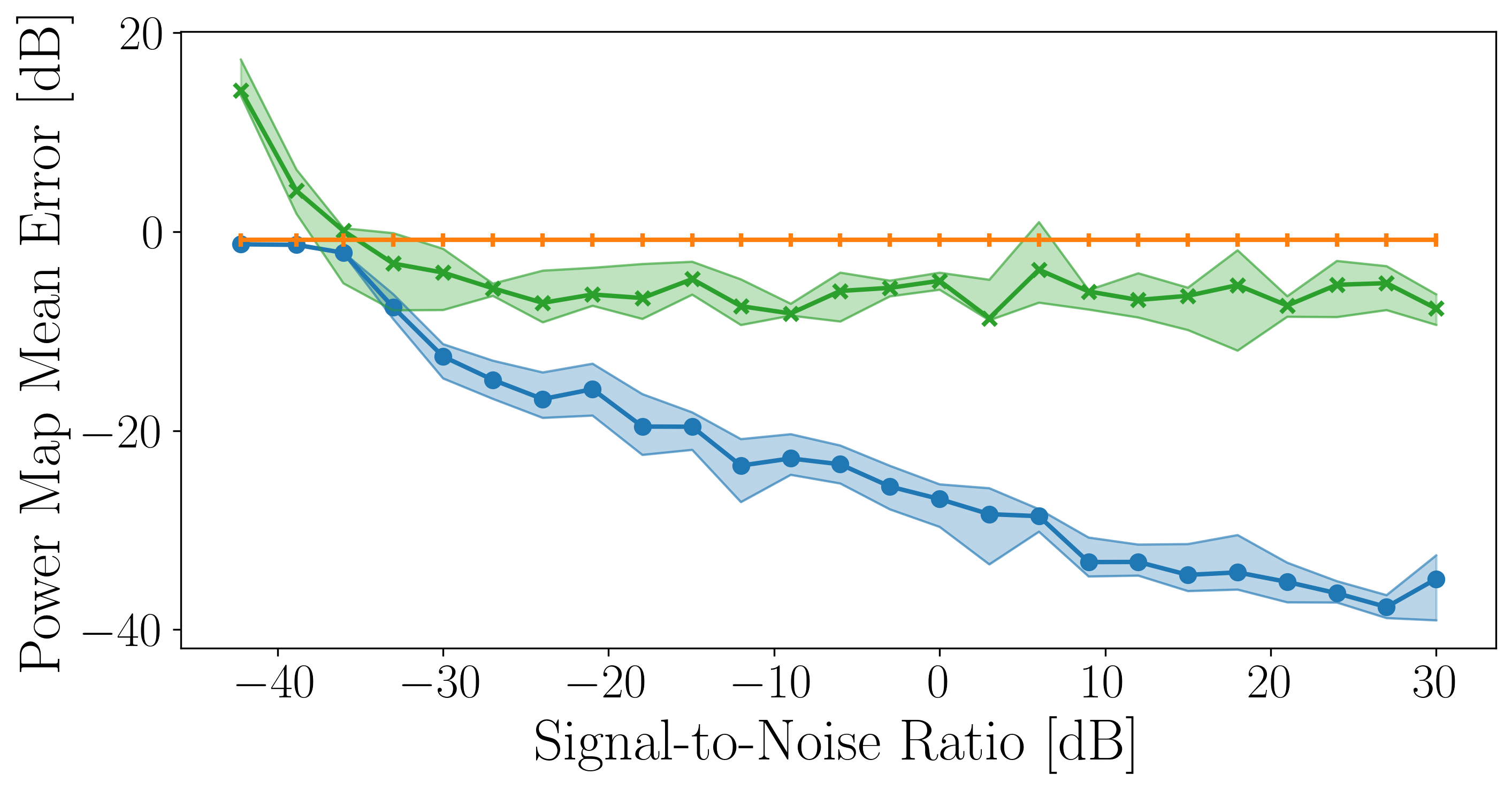}
        \vspace{-0.7cm}
        \caption{}
        \label{fig:power_map_generalization_error_vs_snr}
    \end{subfigure}
    \caption{
    Average estimation error of the predicted signal powers across all new positions $c_{\new}$ within coverage of the transmitter under ground-truth conditions as a function of the standard deviation of the phase error (a), and as a function of the signal-to-noise ratio (SNR) (b); for the phase error aware, phase error oblivious, and uniform phase error calibration schemes.
    Lines represent the median error across ten independent channel observation and calibration runs, with a $20$ dB SNR during calibration in (a), and a uniform phase error during calibration in (b).
    Shaded areas represent the first and third quartiles.
    }
    \label{fig:power_map_generalization_error}
\end{figure*}

In line with the observations made on the calibration performance of the phase error-oblivious calibration strategy in Sec.~\ref{subsec:experiments_calibration_performance}, the relative prediction error under this scheme is satisfactory, below $-20$ dB, for zero phase error (top row), while, for uniform phase errors (bottom row), the estimation error rises above $-5$ dB in zones with LoS coverage and around $0$ dB in most locations relying exclusively on non-LoS components, where propagation conditions are more sensitive to calibration inaccuracies.
In contrast, the uniform phase error calibration baseline in Fig. \ref{fig:power_map_upec_uniform_phase_error} is able to keep the power estimation error below $-15$ dB in locations with LoS components, and around $-3$ dB in locations covered exclusively by non-LoS components.
Finally, the estimation errors of the proposed phase error-aware scheme in Fig. \ref{fig:power_map_peac_no_phase_error} and \ref{fig:power_map_peac_uniform_phase_error} are below $-20$ dB across all covered receiver's positions, even in the worst case of uniform phase error.

The relative power predition error is further studied in Fig. \ref{fig:power_map_generalization_error}, for which the relative error is averaged over all covered receiver's positions and plotted as a function of the phase error standard deviation in Fig. \ref{fig:power_map_generalization_error_vs_std}, and as a function of the SNR in Fig. \ref{fig:power_map_generalization_error_vs_snr}.
The figure confirms that the proposed scheme outperforms both baselines in terms of its capacity to predict the received power for new positions across a different phase error and SNR levels, as long as the phase error is not negligible.

\subsection{FDTD Wave Simulation Example}\label{subsec:experiments_wave_toy_example}

\begin{figure}
    \centering
    \includegraphics[keepaspectratio, width=3.4in]{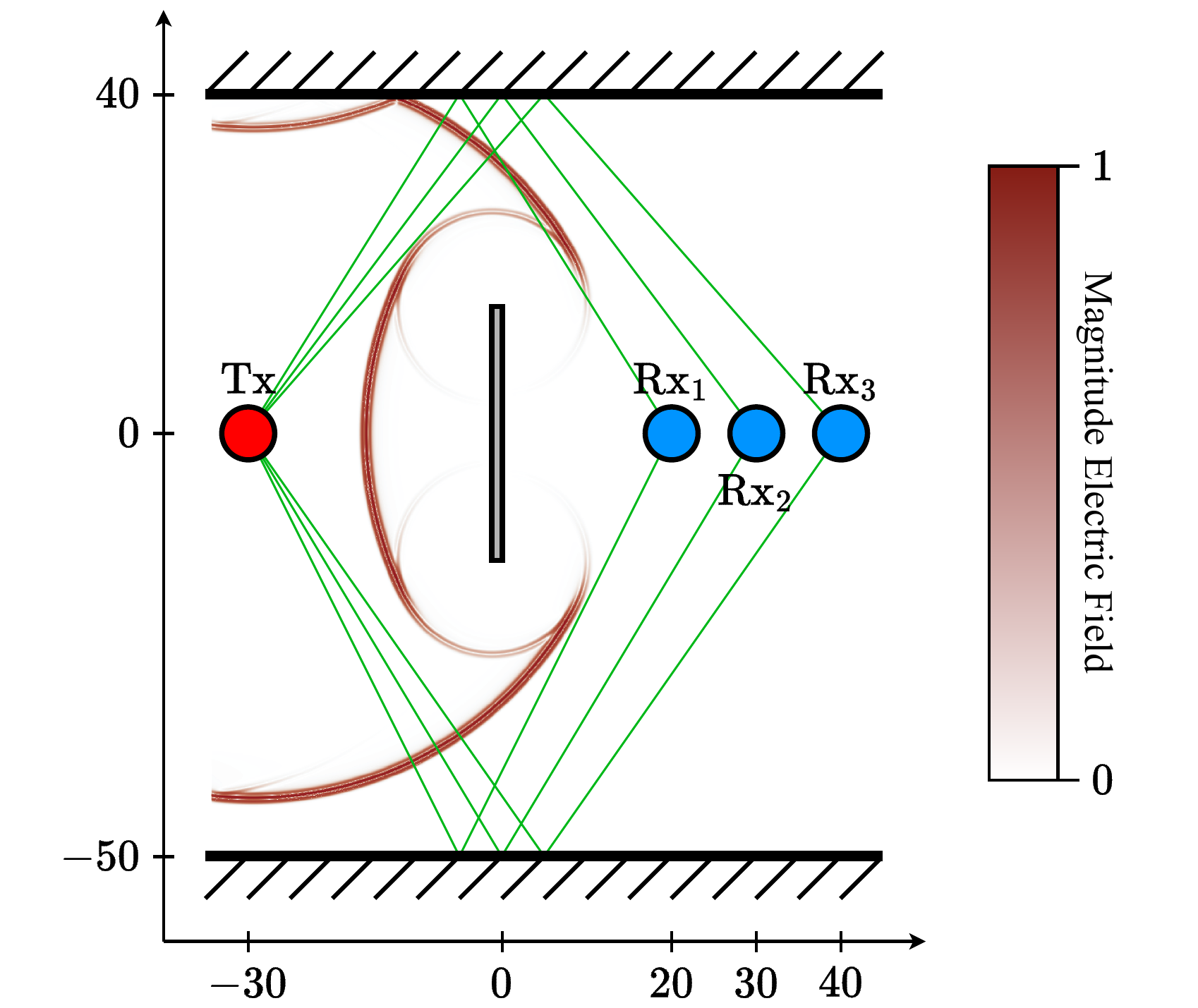}
    \caption{
    \add{
    Two-dimensional scenario used for finite-difference time-domain (FDTD) wave propagation simulation.
    A metallic blocker is located near the center of the scene and two concrete walls are present on both ends of the vertical axis.
    The single-reflection propagation paths between the transmitter ($\Tx$) and the three receivers ($\Rx_1$, $\Rx_2$, $\Rx_3$) are displayed as thin green lines.
    A snapshot of the normalized magnitude of the simulated electric field is displayed in the background.
    Axis values represent multiples of the carrier's wavelength $\lambda^c \approx 5$ cm.
    }
    }
    \label{fig:toy_example_fdtd}
\end{figure}

\add{
Stepping away from RT-generated data, we now experiment with channel observations $\mathcal{D}$ generated by solving Maxwell's equations via FDTD simulations using the gprMax library \cite{warren2016gprmax}.
Given the computational complexity of FDTD simulations, we restrict the analysis to the small-sized two-dimensional (2D) setting depicted in Fig.~\ref{fig:toy_example_fdtd}, which consists of single-antenna devices with one transmitter at position $c^\Tx = (-30, 0)$, and three receivers at positions $c^\Rx_1 = (20, 0)$, $c^\Rx_2 = (30, 0)$ and $c^\Rx_3 = (40, 0)$, where coordinates are expressed as multiples of the carrier's wavelength $\lambda^c \approx 5$ cm.
A metallic blocker of length $30 \lambda^c$ and width $1 \lambda^c$ is located near the center of the scene, and two horizontal walls sharing the same concrete-like material parameters $\theta^\true$ intersect the vertical axis at locations $(0, 40)$ and $(0, -50)$.
Using only the first two receivers at locations $c^\Rx_1$ and $c^\Rx_2$, we generate a dataset $\mathcal{D}$ of $100$ observations comprising $50$ observations at each location $c_1 = (c^\Rx_1, c^\Tx)$ and $c_2 = (c^\Rx_2, c^\Tx)$, with an SNR equal to $20$ dB.
The last transmit-receive pair at $c_3 = (c^\Rx_3, c^\Tx)$ is kept aside to test the generalization performance of the calibration schemes.
Implementation details of the FDTD simulation and the conversion of time-domain electric fields to channel frequency responses can be found in Appendix~\ref{apx:fdtd_simulation}.
}

\begin{figure}
    \centering
    \begin{subfigure}{3.4in}
        \centering
        \hspace{1.5cm}\includegraphics[width=3.4in]{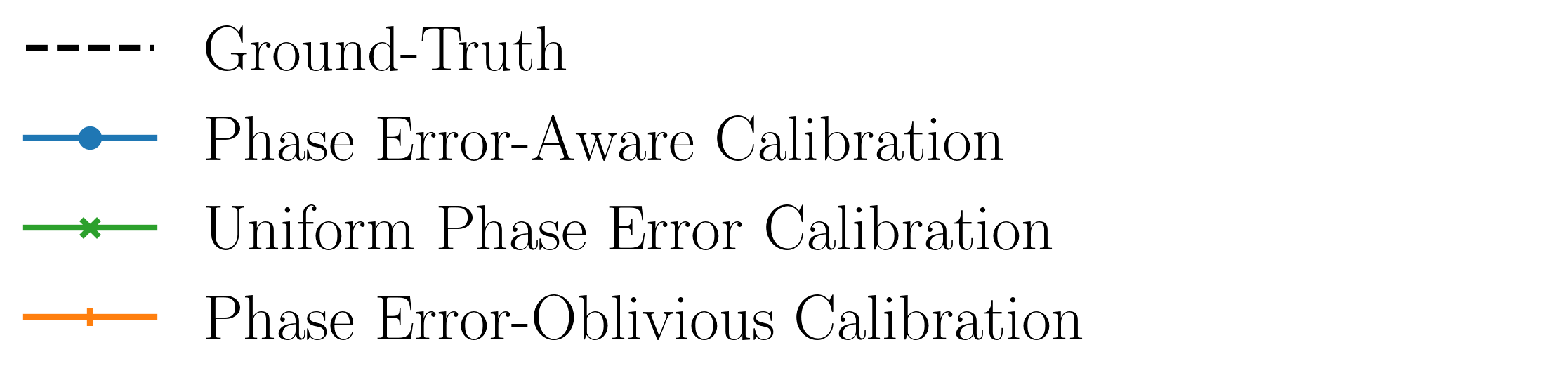}
    \end{subfigure}
    \centering
    \begin{subfigure}{3.4in}
        \centering
        \includegraphics[width=3.4in]{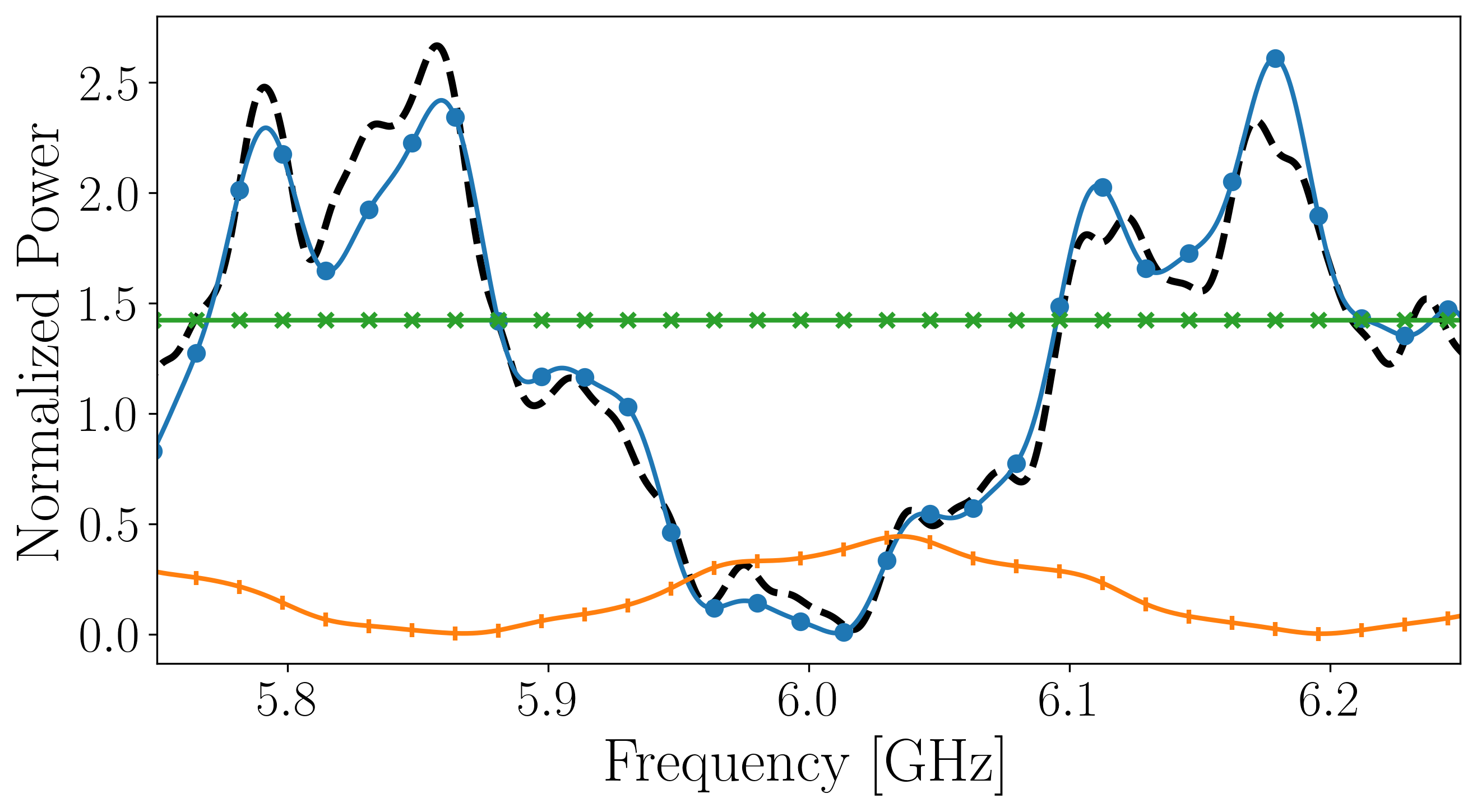}
    \end{subfigure}
    \caption{
    \add{
    Signal power as a function of the frequency $f$ at location $c_1$ for the scenario in Sec.~\ref{subsec:experiments_wave_toy_example}.
    The dashed-line represents the ground-truth power of the FDTD simulation, while solid lines represent the powers predicted using the calibrated materials and phase-error assumptions of the phase error-aware, phase error-oblivious, and uniform phase-error calibration schemes.
    Lines represent the median power value across ten independent channel observation and calibration runs, with an SNR equal to $20$ dB and a bandwidth $B = 500$ MHz.
    }
    }
    \label{fig:fdtd_freq_power_profile}
\end{figure}

\begin{figure*}
    \centering
    \begin{subfigure}{\textwidth}
        \centering
        \includegraphics[width=\textwidth]{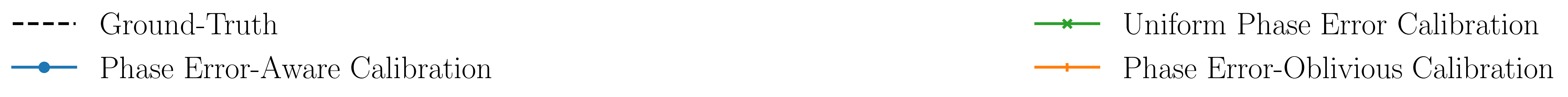}
    \end{subfigure}
    \begin{subfigure}{0.32\textwidth}
        \centering
        \includegraphics[width=\textwidth]{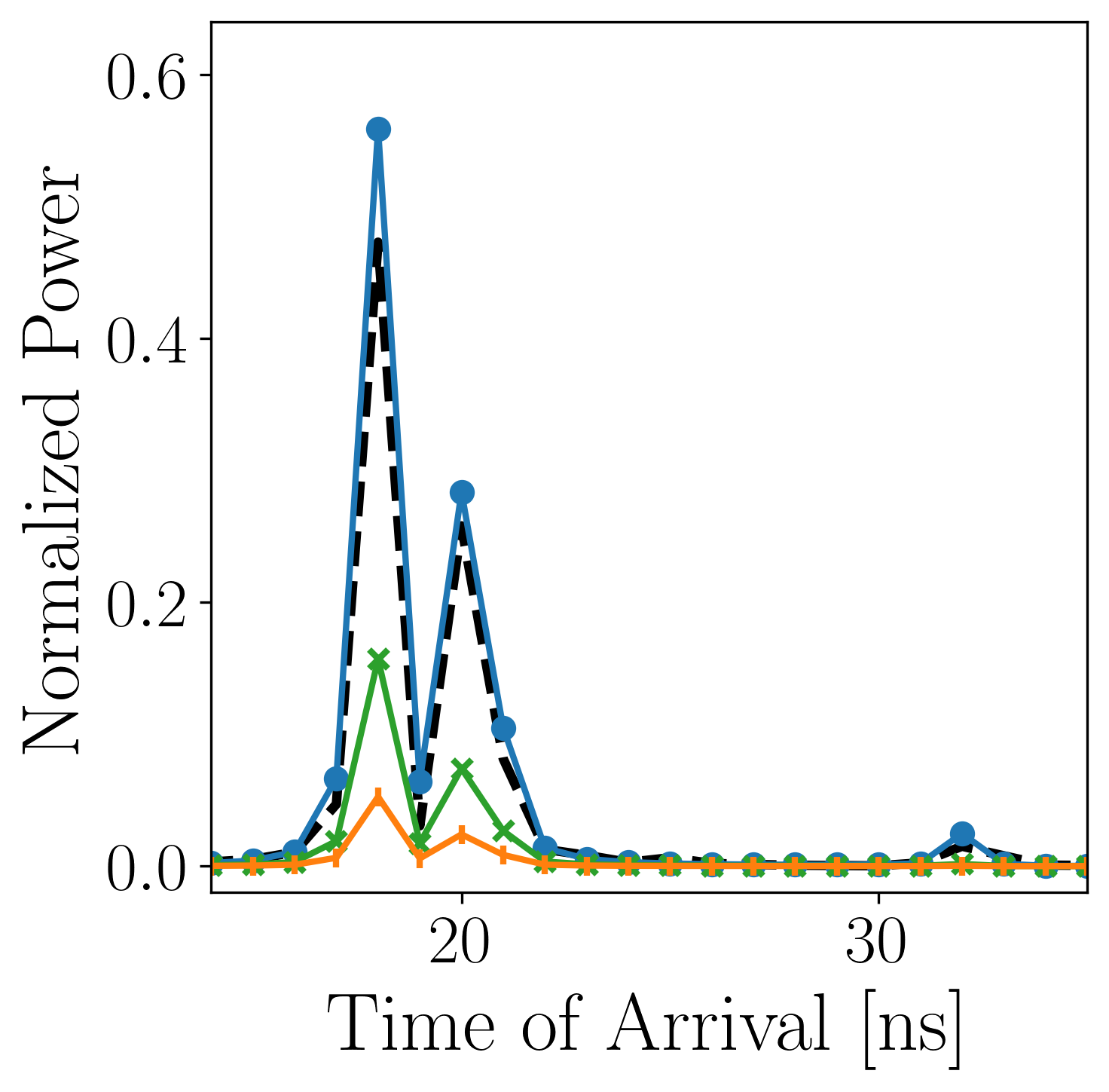}
        \vspace{-0.6cm}
        \caption{}
    \end{subfigure}
    \begin{subfigure}{0.32\textwidth}
        \centering
        \includegraphics[width=\textwidth]{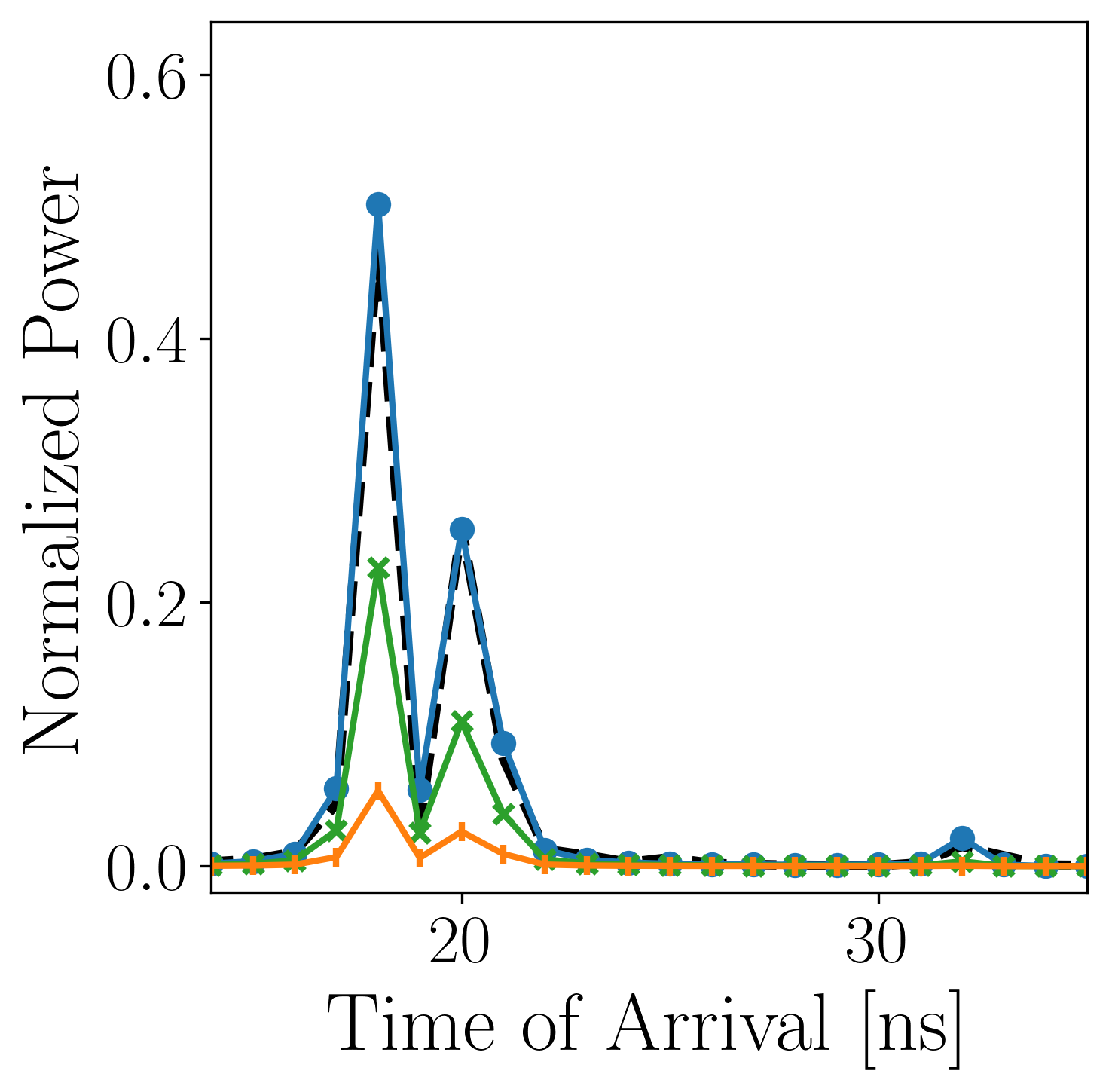}
        \vspace{-0.6cm}
        \caption{}
    \end{subfigure}
    \begin{subfigure}{0.32\textwidth}
        \centering
        \includegraphics[width=\textwidth]{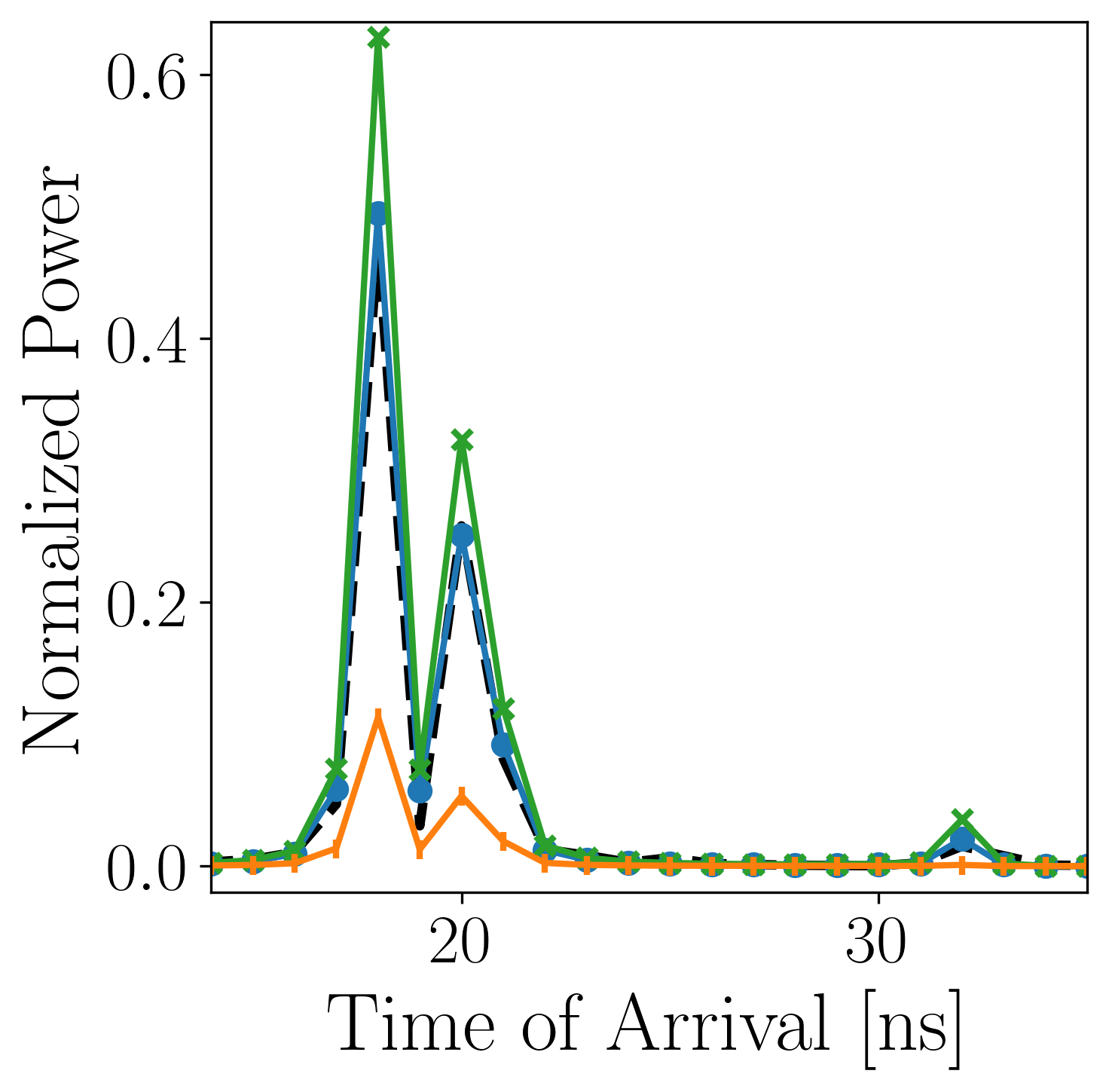}
        \vspace{-0.6cm}
        \caption{}
    \end{subfigure}
    \caption{
    \add{Power-delay profiles for the ground-truth FDTD-simulated channel (dashed line) and the RT-simulated channels (solid lines) using material parameters calibrated with the phase error-aware, phase error-oblivious, and uniform phase-error calibration schemes for bandwidths (a) $B = 100$ MHz, (b) $B = 200$ MHz, and (c) $B = 500$ MHz.
    Lines represent the median power value across ten independent channel observation and calibration runs, with an SNR equal to $20$ dB.}
    }
    \label{fig:fdtd_time_power_profile}
\end{figure*}

\add{
The material parameters of the blocker are assumed to be known, and we focus exclusively on learning the parameters $\theta$ of the two horizontal walls during calibration.
To simulate the mismatch between the ground-truth geometry used to generate the data and the available geometry at the DT, the lower wall in the DT model is shifted down by $\Delta d = 0.25 \lambda^c$ m, intersecting the vertical axis at $(0, -50.25)$.
We experiment with three different levels of available bandwidth $B$ at $100$ MHz, $200$ MHz, and $500$ MHz, and report the results in Fig.~\ref{fig:fdtd_freq_power_profile} and Fig.~\ref{fig:fdtd_time_power_profile}.
Throughout the calibration procedures, the RT at the DT is set to only simulate specular paths, with up to three reflections per path.
In this regard, we note that, while specularly reflected paths are present in both RT and FDTD simulation modalities, paths that diffract on the edges of the metallic blocker are only present in the FDTD simulation.
}

\add{
Using a bandwidth $B = 500$ MHz to calibrate the material parameters $\theta$, we display the predicted signal power $\hatpow(c_1, f | \theta, \mu_1, \kappa_1)$ at location $c_1$ as a function of the frequency $f$ in Fig.~\ref{fig:fdtd_freq_power_profile}, where the predicted power is averaged across the phase errors assumed by each scheme.
Accordingly, $\hatpow(c_1, f | \theta, \mu_1, \kappa_1)$ is computed via \eqref{eq:apx_signal_power}, with phase-error statistics set to $\mu_{1, p} = 0$ and $\kappa_{1, p} = +\infty$, i.e., deterministic phases, for the phase error-oblivious scheme; to $\mu_{1, p} = 0$ and $\kappa_{1, p} = 0$, i.e., uniform phases, for the uniform phase error scheme; and to the calibrated parameters $\mu_1$ and $\kappa_1$ in \eqref{eq:e_step_posterior_mean} and \eqref{eq:e_step_posterior_concentration} for phase error-aware scheme.
The received power $\hatpow(c_1, f | \theta, \mu_1, \kappa_1)$ is plotted in Fig.~\ref{fig:fdtd_freq_power_profile} upon normalization by the ground-truth average power $\pow(c_1 | \theta^\true)$ of the FDTD simulation in \eqref{eq:apx_fdtd_power}.
}

\add{
Due to the uniform phase assumption, the uniform phase-error scheme produces a constant power across all frequencies, which corresponds to the average predicted power $\hatpow(c | \theta) = \norm{\alpha(\theta)}^2$ across all paths.
Accordingly, this scheme tends to over-estimate the available average power at location $c_1$, which is roughly $1.5$ times higher than the true power $\pow(c_1 | \theta^\true)$ of the simulated channel.
The phase error-oblivious scheme is unable to cope with the mismatched geometry at the DT, resulting in wrongly calibrated material parameters $\theta$ that lead to a very small received power.
In contrast, the proposed phase error-aware scheme is able to correctly predict the phase shifts $\mu_1$ due to errors in the geometry with high concentration $\kappa_1$, resulting in a predicted power curve that closely matches the ground-truth received power.
}

\add{
Finally, we also evaluate the capacity of each scheme to yield estimates $\theta$ that enable accurate power prediction at new locations.
To this end, we use the material parameters $\theta$ calibrated using data from locations $\{c_1, c_2\}$, and plot the power-delay profiles $\hatpow(c_3, \tau | \theta) = \abs{h(c_3, \tau | \theta)}^2$ at the held-out location $c_3$ in Fig.~\ref{fig:fdtd_time_power_profile}, where $h(c_3, \tau | \theta)$ denotes the inverse Fourier transform of the deterministic channel model $\Hhat(c_3 | \theta)$ in \eqref{eq:deterministic_channel_model}.
}

\add{
Fig.\ref{fig:fdtd_time_power_profile} demonstrates that the power-delay profiles produced by the phase error-aware calibration scheme follows the ground-truth profile closely, outperforming both baselines across all calibration bandwidth levels $B$.
The uniform phase-error scheme is able to retrieve some of the power of the two main paths, but it under-estimates their values at lower bandwidths $B = 100$ MHz and $B = 200$ MHz, while over-estimating the powers at bandwidth $B = 500$ MHz.
Furthermore, the phase error-oblivious scheme is unable to converge to a meaningful material parameter $\theta$ in all settings, resulting in a power profile that greatly under-estimates the received power.
}

\section{Conclusion} \label{sec:conclusion}

In this work, we have proposed a new phase error-aware method to calibrate the material properties of RT simulations based on multi-carrier channel response observations.
The proposed scheme estimates RT-predicted phase-errors from data, accounting for small-scale discrepancies between the assumed three-dimensional DT model and the ground-truth geometry.
The phase-error estimates are used to correct the predicted interference pattern among propagation paths during calibration of the material parameters.
This is demonstrated to yield more precise conductivity and permittivity estimates, as well as more accurate signal power predictions, especially in the low-resolution regime characterized by low-bandwidth conditions.
We conducted quantitative experiments for a single-material scene, showcasing the advantages of this approach against current state-of-the-art power profile-based and channel response-based schemes for practical conditions characterized by unreliable RT phase predictions.

Future work may investigate the simulation of \add{diffracted and} diffusely scattered propagation paths, along with the calibration of scattering coefficients and cross-polarization discrimination; the validation of the proposed approach on real-world measurements; the impact of antenna patterns on phase uncertainty; and the extension of the current study to a more diverse set of materials, such as position-dependent material properties through neural network parameterization \cite{hoydis2023learning}.


\appendices

\section{Discussion on Calibration Baselines}\label{apx:calibration_baselines}

\subsection{Phase Error-Oblivious Calibration as Observation Noise}\label{subapx:peoc_details}

\add{
The phase error-oblivious calibration scheme in Sec.~\ref{subsec:peoc}  assumes that each channel observation $H_n$ is a noisy realization of the deterministic channel model $\Hhat(c|\theta)$ in \eqref{eq:deterministic_channel_model} given by
\begin{equation}
    H_n = \Hhat(c_n | \theta) + W_n,
\end{equation}
where the entries of the stochastic vector $W_n \in \C^{S N^\Rx N^\Tx \times 1}$ are i.i.d. circular Gaussian $\CN(0, \sigma^2)$ variables with known noise power $\sigma^2$.
Accordingly, the distribution $P(H_n | c_n, \theta)$ of each observation $H_n$ is modelled as a circular Gaussian distribution
\begin{equation}
    P(H_n | c_n, \theta) = \CN(H_n | \Hhat(c_n | \theta), \sigma^2 I)
\end{equation}
with mean $\Hhat(c_n | \theta)$.
The phase error-oblivious calibration process addresses the maximum likelihood (ML) problem of finding the parameter $\theta^{*}$ that minimizes the cross-entropy loss
\begin{equation}
    \frac{\lossPEOC(\theta | \mathcal{D})}{\sigma^2} = - \sum_{n=1}^{N} \log\left( P(H_n | c_n, \theta) \right)
\end{equation}
for the available data $\mathcal{D}$, which amounts the least squares problem presented in \eqref{eq:error_oblivious_opt_problem}.
}

\subsection{Selection of Projections in Uniform Phase-Error Calibration}\label{subapx:upec_details}

\add{
The definition of criterion \eqref{eq:upec_loss} requires the choice of the $M$ reference triples $\{ (\tau_m, \AoD_m, \AoA_m )\}_{m=1}^{M}$.
These may be selected from the predicted path triples $\{ (\tau_p, \AoD_p, \AoA_p )\}_{p=1}^{P}$, or as uniformly spaced values in a suitable space.
For instance, one may consider evenly spaced delays $\tau_m$ between the minimal and maximal predicted times of arrival, while selecting the set of angles of departure $\AoD_m$ and arrival $\AoA_m$ as (approximately) evenly spaced points on a sphere via a spherical Fibonacci lattice \cite{keinert2015spherical}.
In practice, one may even restrict the points of the spherical lattice to the surface spanned by the estimated elevation and azimuth angles of departure or arrival, or to the relevant angles for the given antenna patterns.
}

\add{
These choices of angle-delay triples are in line with previous works, which have proposed to evenly spread delay points across a given delay-window \cite{he2018design}, or to select angle-delay points close to the angles and times of arrival of the predicted paths \cite{jemai2009calibration, priebe2012calibrated, charbonnier2020calibration, kanhere2023calibration, bhatia2023tuning}.
That said, we note that the calibration approaches in \cite{priebe2012calibrated, charbonnier2020calibration, kanhere2023calibration, bhatia2023tuning} first carry out a pre-processing step of the measured CFRs, which assumes the time-angle resolution of the system to be sufficient to distinguish the different path contributions in the measured data.
Accordingly, such schemes first pair the predicted rays to estimated measured rays by matching the predicted time of arrival, angle of departure and angle of arrival of each simulated path to its nearest peak in the measured power-angle-delay profile.
The corresponding loss function is then computed by comparing the predicted and measured power of each pair. 
}

\section{Average Signal Power under Von Mises Distributed Phases} \label{apx:average_power_von_mises}

Under the phase error model \eqref{eq:phase_error_channel_model} with electromagnetic parameters $\theta$ and \add{phase-errors $Z_n = (z_{n, 1}, ..., z_{n, P})^\top$ distributed as $z_{n, p} \sim \VM(\mu_{n, p}, \kappa_{n, p})$ for $\mu_{n, p} \in [-\pi, \pi)$ and $\kappa_{n, p} \geq 0$}, the average signal power at a location $c_n \in \R^{3} \times \R^{3}$ per subcarrier and antenna pair is computed as
\add{
\begin{equation}
\label{eq:apx_signal_power_def}
    \hatpow\left(c_n | \theta, \mu_n, \kappa_n \right) =
        \frac{1}{L} \E_{Z_n \sim \VM(\mu_n, \kappa_n)} \left[ \Norm{\G(c_n , \theta) e^{j Z_n} }^2 \right],
\end{equation}
where $L = S N^\Rx N^\Tx$, $\mu_n = (\mu_{n, 1}, ..., \mu_{n, P})^\top$, and $\kappa_n = (\kappa_{n, 1}, ..., \kappa_{n, P})^\top$}.
Using the invariance of the trace to cyclic permutations\cite{lang1987linear}, the expectation term in \eqref{eq:apx_signal_power_def} can be evaluated as
\begin{subequations}
\label{eq:apx_trace_trick}
\begin{align}
    \E\left[ \Norm{ \G(c_n , \theta) e^{j Z_n} }^2 \right]
    &= \E \left[
        \Tr \left\{
            \conjt{\left( e^{j Z_n} \right)} \conjt{\G(c_n , \theta)} \G(c_n , \theta) e^{j Z_n}
        \right\}
    \right] \label{eq:apx_trace_trick_trace_scalar} \\
    &= \Norm{ \G(c_n , \theta) \E\left[ e^{j Z_n} \right] }^2
        + \Tr \left\{
            \G(c_n , \theta) \cov\left[ e^{j Z_n} \right] \conjt{\G(c_n , \theta)}
        \right\}, \label{eq:apx_trace_trick_result}
\end{align}
\end{subequations}
where we have indicated $\E[\cdot]$ for \add{$\E_{Z_n \sim \VM(\mu_n, \kappa_n)}[\cdot]$}, $\cov(e^{j Z_n})$ denotes the $P \times P$ covariance matrix of the random phasor $e^{j Z_n}$ for \add{$Z_n \sim \VM(\mu_n, \kappa_n)$}, and $\Tr\{ \cdot \}$ denotes the trace of a matrix.
The mean and variance terms in \eqref{eq:apx_trace_trick_result} are then evaluated as \cite{mardia2000directional}
\add{
\begin{equation}
\label{eq:apx_von_mises_circular_mean}
\begin{split}
\E\left[ e^{j Z_n} \right]
    &= \begin{pmatrix}
        b\left( \kappa_{n, 1} \right) e^{j \mu_{n, 1}}
        \hdots
        b\left( \kappa_{n, P} \right) e^{j \mu_{n, P}}
    \end{pmatrix}^\top \\
    &= \Diag{B(\kappa_n)} e^{j \mu_n},
\end{split}
\end{equation}
and
\begin{equation}
\label{eq:apx_von_mises_circular_cov}
\begin{split}
\cov(e^{j Z_n})
    &= \Diag{1 - b(\kappa_{n, 1})^2, ...,  1 - b(\kappa_{n, P})^2} \\
    &= \Diag{\allOnes - B^2(\kappa_n)},
\end{split}
\end{equation}
}%
where $\allOnes = (1, ..., 1)^\top$ denotes the $P \times 1$ all-ones vector, \add{$B(\kappa_n) = (b(\kappa_{n, 1}), ..., b(\kappa_{n, P}))^\top$ is the vector of Bessel ratios $b(\kappa)$ defined in \eqref{eq:bessel_ratio} for $\kappa \geq 0$, and $B^2(\kappa_n) = (b(\kappa_{n, 1})^2, ..., b(\kappa_{n, P})^2)^\top$ is the squared Bessel ratios vector.}
\add{
Moreover, since the covariance matrix $\cov(e^{j Z_n})$ is diagonal, the trace term in \eqref{eq:apx_trace_trick} can be efficiently computed as
\begin{equation}
\label{eq:apx_tr_cov_vm}
    \Tr \left\{\G(c_n , \theta) \cov(e^{j Z_n}) \conjt{\G(c_n , \theta)}\right\} =
    L \left( \Abs{\alpha(\theta)}^2 \right)^\top \left( \allOnes - B^2(\kappa_n) \right)
\end{equation}
All in all, the average signal power expression in \eqref{eq:apx_trace_trick} can be computed as
\begin{equation}
\label{eq:apx_signal_power}
\hatpow\left(c_n | \theta, \mu_n, \kappa_n \right) = 
    \frac{1}{L} \Norm{ \G(c_n , \theta) \Diag{B(\kappa_n)} e^{j \mu_n} }^2
    + \left( \Abs{\alpha(\theta)}^2 \right)^\top \left( \allOnes - B^2(\kappa_n) \right),
\end{equation}
where $\alpha(\theta)$ is the $P \times 1$ vector of complex paths coefficients in \eqref{eq:Gmatrix}
}

\add{For i.i.d. phase errors $\{ z_{n, p} \}_{p=1}^{P} \iid \VM(0, \kPrior)$ with prior concentration parameter $\kPrior \geq 0$, the signal power expression in \eqref{eq:apx_trace_trick} simplifies to}
\begin{equation}
\label{eq:apx_signal_power_prior}
\hatpow\left(c_n | \theta, \kPrior \right) =
    \frac{b(\kPrior)^2}{L} \Norm{ \G(c_n , \theta) \allOnes }^2
    + \left( 1 - b(\kPrior)^2 \right) \Norm{\alpha(\theta)}^2.
\end{equation}
The result in \eqref{eq:apx_signal_power_prior} writes the overall received power as the sum of the average power predicted by the RT in the absence of phase errors, i.e., with $\kPrior= +\infty$ and hence $b(\kPrior)=1$, and of the average power under uniform phase errors for each propagation path, i.e., with $\kPrior=0$ and thus $b(\kPrior)=0$.

The result presented in \eqref{eq:apx_signal_power_prior} can be extended to any projection $\G^{'}(c_n , \theta) = \conjt{a} \G(c_n , \theta)$ of the $L \times P$ matrix $\G(c_n , \theta)$ by an $L \times 1$ vectors $a$ as
\begin{equation}
\begin{split}
\hatpow\left(c_n | \theta, \kPrior \right)
    &= \frac{1}{L} \E_{Z_n \iid \VM(0, \kPrior)} \left[ \Norm{\G^{'}(c_n , \theta) e^{j Z_n} }^2 \right] \\
    &= b(\kPrior)^2 \frac{\abs{ \conjt{a} \G(c_n , \theta) \allOnes }^2}{L} + \left( 1 - b(\kPrior)^2 \right) \Norm{\conjt{a} \G(c_n , \theta)}^2.
\end{split}
\end{equation}

\section{Closed-Form Expression of the Free Energy} \label{apx:free_energy}

Following the notation in Sec. \ref{sec:peac}, this appendix summarizes the main steps to derive a closed-form expression for the variational free energy in \eqref{eq:free_energy}.
Given that phase errors are assumed to be independent, both under the prior $P(Z_n | \kPrior) = \prod_{p=1}^{P} \VM(z_{n, p} | 0, \kPrior)$ and under the variational posterior distribution $Q(Z_n | \mu_n, \kappa_n) = \prod_{p=1}^{P} \VM( z_{n, p} | \mu_{n, p}, \kappa_{n, p} )$, the free energy decomposes as
\begin{equation}
\label{eq:apx_free_energy_decomposition}
\begin{split}
\F_n\left( \mu_n, \kappa_n, \theta, \kPrior \right) =
    & \sum_{p=1}^{P} \left\{
        \mathcal{H}\left( \mu_{n, p}, \kappa_{n, p} \big| \big| 0, \kPrior \right) -
        \mathcal{H}\left( \mu_{n, p}, \kappa_{n, p} \right)
    \right\} \\
    & + \E_{Z_n \sim Q(Z_n | \mu_n, \kappa_n)} \left[
        -\log\left( P(H_n | Z_n, c_n, \theta) \right)
    \right],
\end{split}
\end{equation}
where
\begin{equation}
\label{eq:apx_von_mises_entropy}
\begin{split}
\mathcal{H}\left( \mu, \kappa \right)
    &= \E_{z \sim \VM(\mu, \kappa)} \left[
        -\log\left( \VM\left( z | \mu, \kappa \right) \right)
    \right] \\
    &= \log\left( 2 \pi I_0(\kappa) \right) - \kappa b(\kappa)
\end{split}
\end{equation}
denotes the differential entropy of the von Mises distribution $\VM(\mu, \kappa)$ \cite{lund2000entropy}; and
\begin{equation}
\label{eq:apx_von_mises_cross_entropy}
\begin{split}
\mathcal{H}\left( \mu, \kappa \big| \big| \mu_0, \kPrior \right)
    &= \E_{z \sim \VM(\mu, \kappa)} \left[
        -\log\left( \VM\left( z | \mu_0, \kPrior \right) \right)
    \right] \\
    &= \log\left( 2\pi I_0(\kPrior) \right) - \kPrior b(\kappa) \cos(\mu - \mu_0)
\end{split}
\end{equation}
is the cross-entropy between von Mises distributions $\VM(\mu, \kappa)$ and $\VM(\mu_0, \kPrior)$ \cite{lund2000entropy}.
Using the distribution $P(H_n | Z_n, c_n, \theta)$ in \eqref{eq:error_aware_conditioned_likelihood}, the remaining expectation term in \eqref{eq:apx_free_energy_decomposition} can be expressed as 
\begin{subequations}
\label{eq:apx_free_energy_expectation_term}
\begin{align}
    & \E_{Z_n \sim Q(Z_n | \mu_n, \kappa_n)}\left[
        -\log\left( P(H_n | Z_n, c_n, \theta)  \right)
    \right] \nonumber \\
    & = \log\left( \pi^{L} \sigma^{2L} \right) +
    \frac{1}{\sigma^2} \E\left[
        \Norm{ H_n - \G(c_n , \theta) e^{j Z_n} }^2
    \right] \\
    & = L \log\left( \pi \sigma^2 \right) + \frac{1}{\sigma^2} \biggr[
        \Norm{ \G(c_n , \theta) \E\left[ e^{j Z_n} \right] - H_n }^2
        + \Tr \left\{
            \G(c_n , \theta) \cov(e^{j Z_n}) \conjt{\G(c_n , \theta)}
        \right\}
    \biggl] \label{eq:apx_free_energy_trace_trick_usage} 
\end{align}
\end{subequations}
with $\E[ \cdot ]$ as a shorthand for $E_{Z_n \sim Q(Z_n | \mu_n, \kappa_n)}[ \cdot ]$, \add{and where $\E[ e^{j Z_n} ]$ is the $P \times 1$ mean vector in \eqref{eq:apx_von_mises_circular_mean} and $\cov(e^{j Z_n})$ is the $P \times P$ the covariance matrix in \eqref{eq:apx_von_mises_circular_cov} for the stochastic phasor $e^{j Z_n}$ with $Z_n \sim Q(Z_n | \mu_n, \kappa_n)$.}
\add{Plugging the trace term expression in \eqref{eq:apx_tr_cov_vm} into \eqref{eq:apx_free_energy_expectation_term}}, the free energy in \eqref{eq:free_energy} can be expressed as
\begin{equation}
\label{eq:apx_free_energy_von_mises}
\begin{split}
\F_n\left( \mu_n, \kappa_n, \theta, \kPrior \right) =
    & \sum_{p=1}^{P} \left\{
        \log\left( \frac{I_0(\kPrior)}{I_0(\kappa_{n, p})} \right)
        + b(\kappa_{n, p}) \left(
            \kappa_{n, p}
            - \kPrior \cos(\mu_{n, p})
        \right)
    \right\} + 
    L \log\left( \pi \sigma^2 \right) \\
    & + \frac{1}{\sigma^2} \left[
        \Norm{ \G(c_n , \theta) \Diag{B(\kappa_n)} e^{j \mu_n} - H_n}^2 +
        L \left( \Abs{\alpha(\theta)}^2 \right)^\top \left( \allOnes - B^2(\kappa_n) \right)
    \right].
\end{split}
\end{equation}

\section{Derivation of the VEM Updates} \label{apx:vem_updates}

In this appendix, we derive an approximation of a stationary point of the free energy to be used as closed-form updates for the E-step of the VEM algorithm presented in Sec. \ref{sec:peac}.
In what follows, we adopt a Jacobian convention for partial derivatives, such that, for $y = (y_1, ..., y_{N_y})^\top \in \C^{N_y \times 1}$ and $x = (x_1, ..., x_{N_x})^\top \in \R^{N_x \times 1}$, $\frac{\partial y_i}{\partial x_j}$ is the element at the $i\mbox{-th}$ row and $j\mbox{-th}$ column of the Jacobian matrix $\frac{\partial y}{\partial x} \in \C^{N_y \times N_x}$ for $i \in \dset{N_y}$ and $j \in \dset{N_x}$.
We first evaluate the gradients with respect to the variational parameters $\mu_n$ and $\kappa_n$ for $n \in \dset{N}$, and then elaborate on the approximation of a stationary point with respect to these parameters.

\subsection{Gradients}

\subsubsection{Gradient for the Variational Means}

From the variational free energy expression in \eqref{eq:free_energy}, with fixed parameters $\theta$, we evaluate the gradient with respect to the variational mean parameter vector $\mu_n$ as
\begin{equation}
\label{eq:apx_gradient_free_energy_variational_mean}
\begin{split}
\frac{\partial \F_n\left( \mu_n, \kappa_n, \theta, \kPrior \right)}{\partial \mu_n}
    &= -\frac{\partial \sum_{p=1}^{P} b(\kappa_{n, p}) \kPrior \cos(\mu_{n, p})}{\partial \mu_n} +
        \frac{1}{\sigma^2} \frac{\partial \norm{ \G(c_n , \theta) \Diag{B(\kappa_n)} e^{j \mu_n} - H_n}^2}{\partial \mu_n} \\
    &= \Im\left\{ C(\mu_n, \kappa_n) \right\} \Diag{B(\kappa_n)},
\end{split}
\end{equation}
with $1 \times P$ row-vector
\begin{equation}
\label{eq:apx_vem_c_term}
\begin{split}
C(\mu_n, \kappa_n) =
    &\kPrior \left( e^{j \mu_n} \right)^\top
    + \frac{2}{\sigma^2} \conjt{H_n} \G(c_n , \theta) \Diag{e^{j \mu_n}} \\
    & -\frac{2}{\sigma^2} B(\kappa_n)^\top \conjt{\left( \G(c_n , \theta) \Diag{e^{j \mu_n}} \right)} \left( \G(c_n , \theta) \Diag{e^{j \mu_n}} \right),
\end{split}
\end{equation}
where $\Im\{\cdot\}$ denotes the element-wise imaginary part of a complex vector or matrix.

\subsubsection{Gradient for the Variational Concentration Parameters}

The free energy gradient with respect to the concentration parameter vector $\kappa_n$ is decomposed as
\begin{equation}
\label{eq:apx_gradient_free_energy_variational_concentration}
\begin{split}
\frac{\partial \F_n\left( \mu_n, \kappa_n, \theta, \kPrior \right)}{\partial \kappa_n} 
    &= -\frac{\partial \sum_{p=1}^{P} \log{I_0(\kappa_{n, p})}}{\partial \kappa_n}
        + \frac{\partial \sum_{p=1}^{P} b(\kappa_{n, p}) \left( \kappa_{n, p} - \kPrior \cos(\mu_{n, p}) \right)}{\partial \kappa_n} +\\
        &\quad + \frac{1}{\sigma^2} \left[
            \frac{\partial \norm{ \G(c_n , \theta) \Diag{e^{j \mu_n}} B(\kappa_n) - H_n}^2}{\partial \kappa_n} +
            L \left( \Abs{\alpha(\theta)}^2 \right)^\top \frac{\partial \left( \allOnes - B^2(\kappa_n) \right)}{\partial \kappa_n}
        \right] \\
    &= \left(
            \kappa_n^\top 
            - \frac{2 L}{\sigma^2} B(\kappa_n)^\top \Diag{\Abs{\alpha(\theta)}^2}
            - \Re \left\{ C(\mu_n, \kappa_n) \right\}
        \right) \frac{\partial B(\kappa_n)}{\partial \kappa_n},
\end{split}
\end{equation}
with row-vector $C(\mu_n, \kappa_n)$ defined in \eqref{eq:apx_vem_c_term}.

\subsection{Approximating the Stationary Points via Phasor Relaxation}



In what follows, we address the subset of stationary points $(\mu_n^{*}, \kappa_n^{*})$ for the variational parameters that satisfy the equality $C(\mu_n^{*}, \kappa_n^{*}) = 0$. Note that this may be a strict subset of the set of all stationary points, and that it may potentially be empty. To proceed, we relax the problem by optimizing over an arbitrary  $P \times 1$  complex vector  $\rho_n$ in lieu of the phasor $e^{j \mu_n}$ in \eqref{eq:apx_vem_c_term}. 
The stationarity condition of mean gradient \eqref{eq:apx_gradient_free_energy_variational_mean} is verified for all solutions satisfying the equality  $C(\rho_n^{*}, \kappa_n^{*}) = 0$. Note that we have replaced $\mu_n$ by $\rho_n$ in order to highlight the adopted change of variables for optimization. Considering also the stationarity condition for the concentration gradient \eqref{eq:apx_gradient_free_energy_variational_concentration}, we obtain the system of equations
\begin{equation}
\label{eq:apx_subset_stationary_point_e_step}
\left\{
\begin{array}{ll}
    C(\rho_n^{*}, \kappa_n^{*}) = 0 \\
    \left( \kappa_n^{*} - \frac{2 L}{\sigma^2} \Diag{\Abs{\alpha(\theta)}^2} B(\kappa_n^{*}) \right) \frac{\partial B(\kappa_n)}{\partial \kappa_n}\evalgrad{\kappa_n^{*}} = 0.
\end{array}
\right.
\end{equation}

Noting that the elements in the main diagonal of the matrix $\partial B(\kappa_n) / \partial \kappa_n$ are all non-null, the second equation in \eqref{eq:apx_subset_stationary_point_e_step} simplifies into $P$ independent scalar equations $f_p(\kappa_{n, p}^{*}) = 0$, with
\begin{equation}
    f_p(\kappa_{n, p}) = \kappa_{n, p} - \frac{2 L}{\sigma^2} \Abs{\alpha_p(\theta)}^2 b(\kappa_{n, p}),
\end{equation}
for $p \in \dset{P}$.
While $\kappa_{n, p}^{*} = 0$ is a trivial solution to $f_p(\kappa_{n, p}^{*}) = 0$, a strictly positive solution $\kappa_{n, p}^{*} > 0$ can be found if and only if we have
\begin{equation}
\label{eq:apx_condition_kappa_positive}
    \frac{L \abs{\alpha_p}^2}{\sigma^2} > 1.
\end{equation}
This statement can be proven by studying the sign of $f_p(\kappa_{n, p})$ for $\kappa_{n, p} \in [0, +\infty)$.
When inequality \eqref{eq:apx_condition_kappa_positive} is verified, a solution $\kappa_{n, p}^{*} > 0$ to $f_p(\kappa_{n, p}^{*}) = 0$ can be bounded using the Bessel ratio bounds in \cite{ruiz2016new} as
\begin{equation}
\label{eq:apx_fixed_point_bounds}
    2 \sqrt{\frac{L \abs{\alpha_p}^2}{\sigma^2} - 1} \sqrt{\frac{L \abs{\alpha_p}^2}{\sigma^2}} < 
    \kappa_{n, p}^{*}
    < 2 \sqrt{\frac{L \abs{\alpha_p}^2}{\sigma^2} - \frac{1}{2}} \sqrt{\frac{L \abs{\alpha_p}^2}{\sigma^2}},
\end{equation}
for $p \in \dset{P}$.
In practice, we use the lower bound in \eqref{eq:apx_fixed_point_bounds} to approximate $\kappa_{n, p}^{*}$ for $p \in \dset{P}$.
Note that a tighter bound in \eqref{eq:apx_fixed_point_bounds} can be obtained using sharper bounds of the Bessel ratio in \cite{ruiz2016new}.

Furthermore, assuming the $L \times P$ matrix $\G(c_n , \theta)$ is of rank $P$,  the first equation in \eqref{eq:apx_subset_stationary_point_e_step} yields
\begin{equation}
\label{eq:apx_pre_solution_variation_mean}
    \Diag{B(\kappa_n^{*})} \rho^{*}_n =
    \left( \conjt{\G(c_n , \theta)} \G(c_n , \theta) \right)^{-1} \left( \frac{\sigma^2 \kPrior}{2} \allOnes + \conjt{\G(c_n , \theta)} H_n \right),
\end{equation}
where existence of the inverse $(\conjt{\G(c_n , \theta)} \G(c_n , \theta))^{-1}$ stems from the full-column rank assumption on $\G(c_n , \theta)$.
By \eqref{eq:apx_pre_solution_variation_mean}, if all entries in $B(\kappa_n^{*})$ are strictly positive, the element-wise phase of vector $\rho^{*}_n$ satisfies
\begin{equation}
\label{eq:apx_stationary_point_variational_mean}
    \angle\left\{ \rho^{*}_n \right\} = \angle\biggr\{
        \Bigr( \conjt{\G(c_n , \theta)} \G(c_n , \theta) \Bigl)^{-1}
        \Bigr( \frac{\sigma^2 \kPrior}{2} \allOnes + \conjt{\G(c_n , \theta)} H_n \Bigl)
    \biggl\}.
\end{equation}
Finally, we project this solution back into the space of phasors, producing the proposed approximate solution \eqref{eq:e_step_posterior_mean}.

\section{Optimizing the Prior Concentration Parameter} \label{apx:learnable_prior}

From \eqref{eq:alternative_vem_m_step}, given that the function $\sum_{n=1}^{N} \F_n(\mu_n, \kappa_n, \theta, \kPrior)$ is convex with respect to $\kPrior$, an optimal solution $\kPrior^{*}$ can be computed by evaluating a stationary point $\kPrior^{*}$ as
\begin{equation}
\label{eq:apx_stationary_point_kprior}
    \frac{\partial \sum_{n=1}^{N} \F_n\left( \mu_n, \kappa_n, \theta, \kPrior \right)}{\partial \kPrior}\evalgrad{\kPrior^{*}} = 0,
\end{equation}
where
\begin{equation}
\label{eq:apx_gradient_free_energy_kprior}
    \frac{\partial \sum_{n=1}^{N} \F_n\left( \mu_n, \kappa_n, \theta, \kPrior \right)}{\partial \kPrior} =
    \sum_{n=1}^{N} \sum_{p=1}^{P} \left\{
        b(\kPrior) - 
        b(\kappa_{n, p}) \cos(\mu_{n, p})
    \right\}.
\end{equation}
This yields
\begin{equation}
\label{eq:apx_prior_k_m_step}
    \kPrior^{*} = b^{-1} \left( \frac{1}{N P} \sum_{n=1}^{N} \sum_{p=1}^{P} b(\kappa_{n, p}) \cos(\mu_{n, p}) \right),
\end{equation}
where $b^{-1}(\cdot)$ denotes the inverse function of the Bessel ratio $b(\cdot)$ \cite{hill1981evaluation}.
Given that $b^{-1}(\cdot)$ is defined on $[0, 1)$, the result in \eqref{eq:apx_prior_k_m_step} holds only if $0 \leq (NP)^{-1} \sum_{n=1}^{N} \sum_{p=1}^{P} b(\kappa_{n, p}) \cos(\mu_{n, p}) < 1$.
While the upper bound is verified for any set of parameters $\{\mu_n, \kappa_n\}_{n=1}^{N}$, the lower bound can be exceeded if some of the variational mean estimates verify $\abs{\mu_{n, p}} > \pi / 2$ for $\mu_{n, p} \in [-\pi, \pi)$.
For such cases, i.e., when $\sum_{n=1}^{N} \sum_{p=1}^{P} b(\kappa_{n, p}) \cos(\mu_{n, p}) < 0$, the free energy derivative in \eqref{eq:apx_gradient_free_energy_kprior} is always positive, and $\sum_{n=1}^{N} \F_n( \mu_n, \kappa_n, \theta, \kPrior )$ becomes a strictly increasing function with respect to $\kPrior$.
Accordingly, we extend the result in \eqref{eq:apx_prior_k_m_step} to $\kPrior^{*} = 0$ when $\\ (NP)^{-1} \sum_{n=1}^{N} \sum_{p=1}^{P} b(\kappa_{n, p}) \cos(\mu_{n, p}) < 0$.

\section{FDTD Wave Propagation Simulation Implementation} \label{apx:fdtd_simulation}

\add{
FDTD simulations are carried using the gprMax library \cite{warren2016gprmax} for the two-dimensional (2D) scenario displayed in Fig.~\ref{fig:toy_example_fdtd}.
The transmit and the receive antennas are set to the same polarization, which is perpendicular to the 2D-plane of the simulation.
We denote as $e^\Rx_n(t) \in \R$ the amplitude of the electric field along the given polarization for time $t \geq 0$ at the $n$-th receiver with $n \in \{1, 2, 3\}$.
To avoid numerical dispersion errors, both axes in the scene are finely discretized into steps of $\delta = \lambda^c / 20$ m, where the wavelength $\lambda^c \approx 5$ cm corresponds to the central frequency $f^c = 6$ GHz.
We set a current source at the transmitter equal to
\begin{equation}
i^\Tx(t) = \left(
    2 (\pi f^c)^2 \left(t - \frac{\sqrt{2}}{f^c} \right)^2 - 1
\right) e^{-(\pi f^c)^2 \left(t - \frac{\sqrt{2}}{f^c} \right)^2},
\end{equation}
which is proportional to a Ricker impulse with spectral power centered at the carrier frequency $f^c$.
The resulting transmitter acts as a dipole of size $\delta$ perpendicular to the plane of interest, which radiates energy uniformly across all 2D directions \cite{costen2009comparison}.
By solving the wave equations via finite differences, the electric field is propagated across the scene, and we record the fields $\{ e^\Rx_n(t) \}_{n=1}^{3}$ at the receive locations every $1.18$ ps in the interval $t \in [0, t_{\mathrm{max}}]$, where $t_{\mathrm{max}} = 100$ ns.
}

\add{
Using wave superposition, the electric field $e^\Tx(t, r)$ generated in the 2D plane at a distance $r \gg \lambda^c$ of the transmitter (i.e., at the far field) can be equivalently stated as a sum of Hertzian dipoles $e^\Tx(t, r) = \int_{f=-\infty}^{+\infty} E^{\Tx}_0(f) (e^{-j k_f r} / r) e^{j 2 \pi f t} df$ with 
\begin{equation}
    E^{\Tx}_0(f) = j \frac{Z_0 k_f \delta}{4 \pi} I^\Tx(f),
\end{equation}
where $k_f = 2 \pi f / v$ is the wave number associated to frequency $f$, $Z_0 \approx 377$~$\Omega$ is the impedance of free space, and $I^\Tx(f) = \mathcal{F}\{ i^\Tx \}(f)$ is the Fourier transform of $i^\Tx(t)$ \cite[Ch.4.2.6]{balanis2016antenna}.
We similarly decompose the received fields $e^\Rx_n(t)$ into a sum of harmonic contributions weighted by their respective Fourier transform $E^\Rx_n(f) = \mathcal{F}\{ e^\Rx_n \}(f)$.
In the following, we assume that the receive and transmit antennas share the same impedance and have a uniform pattern in the 2D plane, with the impedance of the transmission lines matched to their respective antennas.
Accordingly, the channel response at location $c_n = (c^\Rx_n, c^\Tx)$ for frequencies $f \in \R$ in the spectral support of the transmitted impulse (i.e., for $\abs{I^\Tx(f)} \neq 0$) is obtained as \cite{fugen2006capability}
\begin{equation}
\label{eq:apx_fdtd_cfr}
    H\left( c_n, f \right) = \frac{\lambda}{4 \pi} \sqrt{G^\Tx G^\Rx} \frac{E^\Rx_n(f)}{E^{\Tx}_0(f)},
\end{equation}
where $\lambda = v / f$ is the wavelength and $G^\Tx$ and $G^\Rx$ denote the antenna gains at the receive and transmit devices, respectively.
The corresponding time-domain channel response $h(c_n, \tau) = \mathcal{F}^{-1}\{ H( c_n, f) \}$ is then obtained via the inverse Fourier transform; and the total received power is computed as the integral
\begin{equation}
\label{eq:apx_fdtd_power}
    \pow(c_n) = \int_{\tau=0}^{t_{\mathrm{max}}} \Abs{h(c_n, \tau)}^2 d\tau.
\end{equation}
}

\bibliographystyle{IEEEtran}
\bibliography{refs}

\end{document}